\documentclass[3p,times,numbers,sort]{elsarticle} 

\journal{Elsevier}

\author[ad1]{Steven N. Rodriguez\corref{cor1}}

\author[ad1]{Athanasios P. Iliopoulos}

\author[ad2]{Kevin T. Carlberg}

\author[ad2]{\\Steven L. Brunton} 

\author[ad1]{John C. Steuben}

\author[ad1]{John G. Michopoulos}

\address[ad1]{Computational Multiphysics Systems Laboratory, U. S. Naval Research Laboratory, Washington DC, United States}

\address[ad2]{Department of Mechanical Engineering, University of Washington, Seattle, WA, United States}

\usepackage{lineno,hyperref}
\modulolinenumbers[5]
\usepackage{bm}
\usepackage{amsmath}
\usepackage{amsfonts}
\usepackage{amsthm}
\usepackage{amssymb}
\usepackage{xcolor}
\usepackage{enumitem}
\usepackage{graphicx}
\usepackage{enumitem}
\usepackage[linesnumbered,ruled]{algorithm2e}
\usepackage{algpseudocode}
\usepackage{setspace}
\usepackage{mathtools}
\usepackage{amsthm}
\usepackage{subcaption}
\usepackage{ragged2e} 
\usepackage{natbib}
\usepackage{pdflscape}
\usepackage{fancyhdr}
\usepackage{hhline}
\usepackage{multirow}
\usepackage{array}
\usepackage{booktabs}

\DeclareMathAlphabet\mathbfcal{OMS}{cmsy}{b}{n}
\DeclareMathOperator*{\argmin}{arg\,min}
\DeclareMathOperator*{\argmax}{arg\,max}
\newcommand{\code}{\texttt}
\newcommand*{\argminl}{\argmin\limits}

\newtheorem{definition}{Definition}
\newtheorem*{erb}{Error bounds}
\algdef{SE}[FUNCTION]{Function}{EndFunction}%
[2]{\algorithmicfunction\ \textproc{#1}\ifthenelse{\equal{#2}{}}{}{(#2)}}%
{\algorithmicend\ \algorithmicfunction}%

\SetCommentSty{mycommfont}
\SetKwInput{KwInput}{Input}                
\SetKwInput{KwOutput}{Output}              
\SetKwFunction{KwFn}{\textproc{Function}}

\fancypagestyle{plain}{%
	\fancyhf{} 
	\cfoot{\vspace{0.5cm}  Distribution A: Approved for public release; Distribution unlimited.}\rfoot{\thepage}
	\renewcommand{\headrulewidth}{0pt}
}

\title{Projection-tree reduced order modeling for fast $N$-body computations}

\begin{document}
	
\begin{abstract}

This work presents a data-driven reduced-order modeling framework to accelerate the computations of $N$-body dynamical systems and their pair-wise interactions. The proposed framework differs from traditional acceleration methods, like the Barnes--Hut method, which requires online tree building of the state space, or the fast-multipole method, which requires rigorous \textit{a priori} analysis of governing kernels and online tree building. Our approach combines Barnes-Hut hierarchical decomposition, dimensional compression via the least-squares Petrov--Galerkin (LSPG) projection, and hyper-reduction by way of the Gauss-Newton with approximated tensor (GNAT) approach. The resulting \textit{projection-tree} reduced order model (PTROM) enables a drastic reduction in operational count complexity by constructing sparse hyper-reduced pairwise interactions of the $N$-body dynamical system. As a result, the presented framework is capable of achieving an operational count complexity that is independent of $N$, the number of bodies in the numerical domain. Capabilities of the PTROM method are demonstrated on the two-dimensional fluid-dynamic Biot-Savart kernel within a parametric and reproductive setting. Results show the PTROM is capable of achieving over 2000$\times$ wall-time speed-up with respect to the full-order model, where the speed-up increases with $N$. The resulting solution delivers quantities of interest with errors that are less than 0.1\% with respect to full-order model.

\end{abstract}


\fancypagestyle{pprintTitle}{%
	\lhead{} \chead{}\rhead{}
	\lfoot{\textit{Preprint submitted to Elsevier}}\cfoot{\vspace{0.5cm} Distribution A: Approved for public release; Distribution unlimited.}}\rfoot{}
\renewcommand{\headrulewidth}{0.0pt}

\pagestyle{plain}{%
	\lhead{}\chead{}\rhead{}
	\lfoot{\footnotesize \textit{Preprint submitted to Elsevier}}\cfoot{ \vspace{0.5cm}  Distribution A: Approved for public release; Distribution unlimited. }\rfoot{\thepage}
	\renewcommand{\headrulewidth}{0.0pt}
}

\maketitle

\section{Introduction} \label{Section_Introduction}

Lagrangian and discrete $N$-body computational modeling, or meshless computational modeling, of dynamical systems are ubiquitous across different disciplines in science and engineering. For instance, vortex methods, such as the free-vortex wake method (FVM) \cite{rodriguez2018phd, rodriguez2019strongly, rodriguez2020strongly, rodriguez2020stability, rodriguez2020investigating, sebastian2012development, gaertner2015modeling}, the vortex panel and particle methods  \cite{jeon2014unsteady, colmenares2015computational, kebbie2018fast, eldredge2007numerical,eldredge2002vortex}, are commonly used in the aerospace community to capture the near-wake dynamics of rotorcraft and fixed-wing aircraft. The smooth particle hydrodynamic (SPH) method provides effective modeling in fluids \cite{tartakovsky2005smoothed}, additive manufacturing \cite{russell2018numerical}, and has even been used to model cosmological shock waves \cite{pfrommer2006detecting} and dark matter halos \cite{mocz2015numerical}. The discrete element method (DEM) has been used to simulate the thermomechanical states of additive manufacturing \cite{steuben2016discrete}, and has also been used to simulate complex granular flow \cite{jing2007DEMfundamentals}. The molecular dynamics (MD) method \cite{hansson2002moleculardynamics, noe2015kinetic, wang2019machine, doerr2016htmd}, which is akin to DEM, has been used to model the physical movements of atoms and molecules. These computational methods are often employed due to their many benefits, such as bypassing Eulerian grid-based artificial numerical dissipation \cite{leishman2006principles}, tracking individual particle time histories \cite{liu2010smoothed}, enabling constitutive behavior not available in grid-based (continuum) methods \cite{steuben2016discrete}, and modeling of multiphase multiphysics, free-surface flow, and splash with complex geometries \cite{shadloo2016smoothed}. Unfortunately, discrete computational $N$-body methods suffer from poor operational count complexity (OCC) associated with pairwise interactions that generally scale quadratically, $\mathcal{O}(N^2)$, or super-linearly, $\mathcal{O}(cN)$, where $N$ is the number of bodies in the computational domain, $c\in \mathbb{N}$ and $c: N \mapsto m(N)$, such that $m$ is a mapping that dictates the neighbor particle count for each $N$-body, which tends to be much higher than neighboring nodes in traditional mesh-based methods.

Many research efforts have been dedicated to reducing the cost of computational methods based on $N$-body pairwise interactions  \cite{ying2004kernel}, where the most successful and general effort has been the fast multipole method (FMM) \cite{greengard1987fast}, which was named one of the top ten algorithms of the $20^{\textup{th}}$ century \cite{cipra2000best, dongarra2000guest}. The FMM computes a multipole expansion of the field potential and executes a hierarchical decomposition to separate near-field from far-field particles. The influence of far-field particles is then approximated by lower order terms, and the far-field particles are grouped together to form fewer but stronger particles in the far field domain. The FMM approximation can at best reduce OCC to a linear scaling, with respect to the number of particles in the domain \cite{ying2004kernel}. However, FMM has drawbacks in application. For instance, the FMM can be very difficult to implement in three-dimensions, can be kernel dependent (i.e.~not all Greens functions are FMM adaptable), and the multipole expansion computation of the field potential is costly \cite{gnedin2019hierarchical, ying2004kernel}. Kernel independent variants of the FMM exist, which aim to reduce the cost of the field potential computation independent of the kernel type, for example see \cite{martinsson2007accelerated,ying2006kernel,jiang2016n}. However, for dynamical simulations the FMM and its variants are ultimately bounded by OCC that depends on multiple online updates of the hierarchical decomposition, and depend on potential field computations over all particles in the domain, i.e.~FMM-based methods are at best $\mathcal{O}(N)$ or ``$N-dependent$". 

In contrast, acceleration methods based on Lagrangian counterparts, i.e.~Eulerian grid-based methods, have achieved complexity reduction independent of the number of nodal degrees-of-freedom in the computational domain, i.e.~$N$-independent. This $N$-independent complexity reduction has been achieved by means of data-driven projection reduced order modeling (PROM) for applications in multi-query loops, i.e., optimization, control, uncertainty quantification, and inverse problems, where a non-exhaustive list of these developments are included in \cite{carlberg2013gnat, carlberg2015preserving, carlberg2017galerkin, farhat2014dimensional, farhat2015structure, rowley2005model, rowley2004model, antoulas2005approximation, benner2017model, brunton2015closed}. The objective of PROM is to perform dimensional compression by learning about a dynamical system's solution manifold and determine a corresponding low-dimensional embedding (mappings from high-dimensions to low-dimensions) where the system's governing equation can be computed. Finding the dynamical system's low-dimensional embedding is  performed \emph{a priori}, during an off-line training stage, which scales with the intrinsic dimension of the solution manifold under consideration, i.e. offline stages are $N$-dependent.  In PROM, finding the system's low-dimensional embedding is enabled by projection methods such as Galerkin \cite{sirovich1987turbulence, carlberg2017galerkin}, Petrov-Galerkin \cite{bui2008model,amsallem2012stabilization}, or least-square Petrov-Galerkin projection \cite{carlberg2013gnat, carlberg2017galerkin}. Once the low-dimensional embedding has been discovered, the PROM can deployed on-line to perform rapid $N$-independent computations for multiquery setting, such as design optimization, control, uncertainty quantification and inverse problems. 

Traditionally in PROM, the embedding is generated by an affine subspace approximation, a reduced basis, of the entire solution manifold (a data-driven global basis function). A non-exhaustive list of methods that construct these subspace approximations include the proper orthogonal decomposition (POD) \cite{sirovich1987turbulence,holmes2012turbulence, rowley2004model}, balanced POD \cite{rowley2005model,willcox2002balanced}, symplectic POD \cite{peng2016symplectic, guo2019data,afkham2017structure}, reduced basis method \cite{grepl2007efficient, buffa2012priori}, and dynamic mode decomposition (DMD) \cite{schmid2010dynamic, tu2013dynamic,kutz2016dynamic, erichson2019randomized, noack2016recursive}. \citet{taira2017modal} provide a great overview and review of popular projection techniques widely used in PROMs. Other methods exist to alleviate the strong affine approximation over the solution manifold, which can be nonlinear, by finding local affine approximations (akin to linearly discretizing over the solution manifold), such as the local reduced order basis method \cite{amsallem2012nonlinear}. Recent works have also taken advantage of new developments in convolutional deep neural networks to determine optimal nonlinear, global, and  low-dimensional embeddings to overcome strong affine approximations and Kolmogorov width limitations \cite{lee2020model}. \citet{brunton2019data} provide a great overview and introduction to these machine-learning and PROM techniques used in nonlinear dynamical systems. From a general and overhead perspective, PROM can be perceived as a data-driven Ritz method, where the basis function is determined from a dynamical system's data \textit{a posteriori} and is used to project the full-order system of equations onto a low-dimensional embedding that approximates the dynamical system behavior. 

For nonlinear dynamical systems, identifying a low-dimensional embedding is often not enough to enable $N$-independent computations, due to persistent high dimensional, higher-order, non-linear, and parametrized dependencies of the underlying system of equations. As a result, nonlinear PROMs are often accompanied by additional layers of complexity reduction, known as hyper-reduction, which alleviate the embedded system from the persistent high-dimensional dependence after projection is performed. In essence, hyper-reduction enables the computation of dynamical system's evolution in the low-dimensional embedding over a sparse set of sampled points in the numerical domain. A non-exhaustive list of these hyper-reduction methods include the emperical interpolation method (EIM) \cite{barrault2004empirical}, discrete EIM (DEIM) \cite{chaturantabut2010nonlinear}, unassembled DEIM (UDEIM) \cite{tiso2013discrete},  Gauss-Newton with approximated tensors (GNAT) \cite{carlberg2013gnat}, and energy conserving sampling and weighing (ECSW) \cite{farhat2014dimensional, farhat2015structure}.

To the best of the authors' knowledge, PROM approaches developed for Eulerian grid-based methods have not been cast into an effective $N$-independent Lagrangian framework for $N$-body problems until the work presented herein. This paper presents an OCC-reducing framework that employs hierarchical decomposition to reduce pairwise interaction operation counts, projection based dimensionality reduction, and hyper-reduction to perform a sparse set of pairwise interactions. The method presented in this work can be perceived as a data-driven kernel independent acceleration algorithm, where hierarchical decomposition occurs \emph{once} offline and the kernel is approximated via a sparse representation of LSPG projection, all of which delivers an $N$-independent acceleration framework for $N$-body problems.

The remainder of this paper is organized as follows. Section \ref{Section_ProblemFormulation} presents the problem formulation. Specifically, the two-dimensional Biot--Savart kernel is presented in the form of a parametric ordinary differential equation (ODE). Section \ref{Section_PROM} introduces projection-based reduced order modeling, specifically the least-squares Petrov-Galerkin projection. Section \ref{Section_QuadTreeDecomp} introduces the Barnes--Hut quad-tree (two-dimensional) hierarchical decomposition method. The main contribution of this work, i.e.~the PTROM, is presented in Section \ref{Section_PTROM}. Applications of the presented PTROM framework in the form of reproductive and parametric studies are presented Section \ref{Section_Results}. Finally, conclusions and future directions are offered in Section \ref{Section_Conclusion}.

\section{Problem Formulation} \label{Section_ProblemFormulation}
Development of the PTROM in this work is rooted in the Biot--Savart kernel, which is often used to model vorticity transport in fluid-dynamics as an $N$-body problem. The Biot--Savart kernel is often used as the underlying theoretical foundation for many Lagrangian computational fluid-dynamics frameworks, such as the vorticity transport model (VTM) \cite{brown2005efficient}, FVM \cite{sebastian2012development, rodriguez2017jert, rodriguez2019strongly, rodriguez2020strongly}, and many others \cite{jeon2014unsteady, colmenares2015computational, kebbie2018fast, akoz2018unsteady}. The overarching goal of this work is to develop the mathematical foundations of the PTROM using the Biot--Savart kernel, but to maintain a general structure of the formulation, such that any other $N$-body kernel could be substituted in the presented framework. In this paper, the Biot--Savart kernel and its $N$-body pair-wise full-order model (FOM) is presented for a two-dimensional study in the form of a time-continuous ordinary differential equation (ODE). For a particle, $i$,  the ODE formulation is defined by

\begin{equation}
\frac{d \bm{\chi}_i}{dt}=\sum_{j\neq i}^{N} \bm{k}(\bm{\chi}_i,\bm{\chi}_j,t;\boldsymbol{\mu}_j),\:\:\:\bm{\chi}_i(0; \boldsymbol{\mu})=\bm{\chi}_i^0(\boldsymbol{\mu}),
\label{pairwise_ode}
\end{equation}
where $\bm{\chi}_i=\bm{\chi}_i(t;\bm{\mu}):=\left\{ \chi_i, \psi_i,0\right\}^T, i =1, \ldots, N$, is the position vector of the $i^\textup{th}$ particle and $\chi$ and $\psi$ are the two-dimensional Cartesian coordinates.  Here, $t\in [0, T_f]$ denotes time with the final time $T_f\in \mathbb{R}_{+}$, $\bm{\mu}$ is the parameter container of all particles (i.e., contains circulation or density), $N$ is the number of particles, the second equality in Eq.~\ref{pairwise_ode} defines the initial conditions, and superscript $T$ denotes the transpose operations. Here, $\bm{k}$ is the Biot--Savart kernel, 
\begin{equation}
\sum_{j\neq i}^N\bm{k}(\bm{\chi}_i,\bm{\chi}_j,t;\boldsymbol{\mu}_j)= \sum_{j\neq i}^N\frac{{\Gamma}_j}{2\pi}\frac{\hat{\bm{e}}_3 \times [\bm{\chi}_i-\bm{\chi}_j]}{\lVert \bm{\chi}_i-\bm{ \chi}_j\rVert^2 + \delta_k}=\{k_{i,\chi},k_{i,\psi},0\},
\label{BS_kernel}
\end{equation}
where $\bm{\mu}_j \leftarrow \bm{\Gamma_j}$ is the circulation, and $\hat{\bm{e}}_3=\{0,0,1\}$ is the out-of-plane unit normal vector of the $j^{\textup{th}}$ body, and $\delta_k$ is the de-singularization constant where $\delta_k \in \mathbb{R}_+$. The kernel outputs are denoted by $\bm{k}(\bm{\chi}_i,\bm{\chi}_j,t;\boldsymbol{\mu}_j):=\{k_{i,\chi}, k_{i,\psi},0\}$ where the subscripts $\chi$ and $\psi$ denote the $\chi$ and $\psi$ output components. For the current two-dimensional study, it is convenient to introduce $\bm{x}:=\left\{ x_1, \ldots, x_{2N} \right\}^T=\left\{ \chi_1, \ldots, \chi_N, \psi_1,\ldots, \psi_N \right\}^T$ that collects the particle positions in a vector such that $\bm{x} \in \mathbb{R}^{N_d}$, with $N_d=dN$, where $d=2$ refers to our two-dimensional system, and where  $\left\{ \chi_i, \psi_i \right\}^T \mapsto \left\{{x}_i, {x}_{i+N}\right\}^T$. Similarly, we introduce $\bm{f}:=\left\{f_{1},\ldots, f_{2N}  \right\}^T=\left\{k_{1,\chi},\ldots, k_{N,\chi}, k_{1,\psi}, \ldots, k_{N,\psi}  \right\}^T$ that collects 
the particle velocity in a vector such that $\bm{f}\in \mathbb{R}^{N_d}$, and $\left\{ k_{i,\chi}, k_{i,\psi} \right\}^T \mapsto \left\{{f}_i, {f}_{i+N}\right\}^T$. As a result the particle-wise ODE formulation in Eq.~\ref{pairwise_ode} can be rewritten into a traditional vector ODE form: 
\begin{equation}
\frac{d \bm{x}}{dt}= \bm{f}\left(\bm{x},t;\bm{\mu}\right),\:\:\:\bm{x}(0; \boldsymbol{\mu})=\bm{x}^0(\boldsymbol{\mu}),
\label{govparticle}
\end{equation}
where $\bm{x}: [0, T_F] \times \mathcal{D} \rightarrow \mathbb{R}^{N_d}$ denotes the time-dependent parameterized state, which is implicitly defined as the solution to the full $N$-body pair-wise interaction problem in Eq. \refeq{govparticle}, with parameters $\boldsymbol{\mu}\in \mathcal{D}$. Here, $\mathcal{D}\subseteq \mathbb{R}^{n_{\mu}}$ denotes the parameter space of $n_{\mu}$ parameters, and $\bm{x}^{0}:\mathcal{D}\rightarrow\mathbb{R}^{N_d}$ is the parametrized initial condition. Finally, $\bm{f}: \mathbb{R}^{N_d} \times [0, T_f] \times \mathcal{D} \rightarrow \mathbb{R}^{N_d}$ which denotes the vector of velocity components generated by the kernel $\bm{k}$.

\begin{figure*}[t]
	\centering
		\includegraphics[scale=0.55, trim=3cm 0cm 50 0cm]{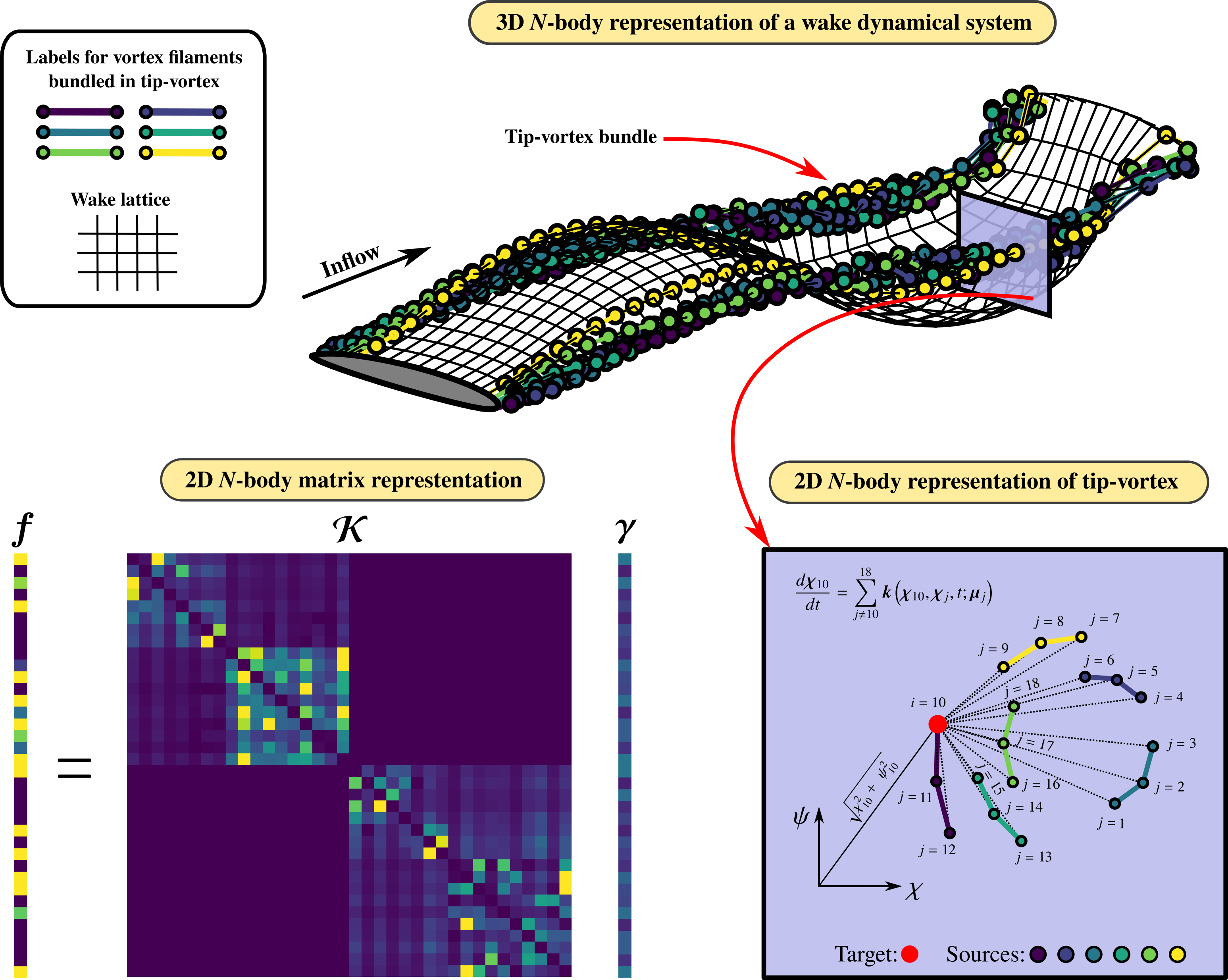}
	\caption[]
	{Illustration of a wake past an elliptical wing posed as an $N$-body dynamical systems by the Biot--Savart kernel. The three-dimensional dynamics are projected on a two-dimensional plane and a matrix of pair-wise interactions.} 
	\label{PairWiseSchematic}
\end{figure*}

It is also useful to arrange the Biot-Savart law pair-wise interaction in block-matrix form. Let 
\begin{equation}
\boldsymbol{\kappa}_{ij}=\left\{\kappa_{ij1},\kappa_{ij2}\right\}^T=\frac{\hat{\boldsymbol{e}}_3 \times [\boldsymbol{\chi}_i-\boldsymbol{\chi}_j]}{\lVert \boldsymbol{\chi}_i-\boldsymbol{ \chi}_j\rVert^2 + \delta_k} \:\:\:\:\: \textup{and} \:\:\:\: \boldsymbol{\gamma}=\frac{1}{2\pi}\left\{\Gamma_1,\ldots,\Gamma_N, \Gamma_1,\ldots,\Gamma_N\right\}^T,
\label{BS_MatrixFormulation}
\end{equation}
where $\bm{\gamma}\in\mathbb{R}^{N_d}$. Next, a block matrix is formed such that 
\begin{align}
\boldsymbol{\mathcal{K}}=\left(
\begin{matrix}
\boldsymbol{\mathcal{K}^{ul}} & \mathbf{0} \\
\mathbf{0} & \boldsymbol{\mathcal{K}^{br}}
\end{matrix} \right),
\end{align}
where
\begin{align}
\boldsymbol{\mathcal{K}^{ul}} \equiv \left\{\mathcal{K}_{ij}^{ul}\right\}=\left\{\kappa_{ij1}\right\}\:\:\:\: \textup{and} \:\:\:\:
\boldsymbol{\mathcal{K}^{br}} \equiv \left\{\mathcal{K}_{ij}^{br}\right\}=\left\{\kappa_{ij2}\right\}
\end{align}
and $\bm{\mathcal{K}}\in \mathbb{R}^{N_d\times N_d}$. The two-dimensional Biot-Savart kernel can then be written as
\begin{equation}
\bm{f}(\bm{x}, t;\boldsymbol{\mu})=\bm{\mathcal{K}}\bm{\gamma},
\label{pairwise_matrix}
\end{equation}
which helps contextualize the pair-wise interactions with the ``target-source" relationship. Figure ~\ref{PairWiseSchematic} illustrates an example of the target-source relationship of a tip-vortex shed off of an elliptical wing formed in with the Biot--Savart kernel, where the three-dimensional dynamics have been projected on a two-dimensional plane. Through-out this paper, the pre-defined circulation of the particles will serve as parametric variables, as will be shown in Section \ref{Section_Results}. 

The two-dimensional dynamical system presented in Eq.~\refeq{govparticle} can be discretized in time by a $\breve{k}$-step linear multistep scheme, where $\breve{k}$ denotes the number of steps in the multistep scheme and $\breve{k} \in \mathbb{N}$, in residual form as:
\begin{equation}
\bm{r}^n(\bm{x}^n;\boldsymbol{\mu})=\mathbf{0}, \:\: n=1 \ldots N_t,
\end{equation}
where the superscript $n$ designates the value of a variable at time step $n\in\mathbb{N}(N_t)$, $N_t\in\mathbb{N}$ denotes the final number of time steps taken, and $\mathbb{N}(N_t):=\{1 \ldots N_t \}$. The time-discrete residual ${\bm{r}}: \mathbb{R}^{N_d} \times \mathcal{D} \rightarrow \mathbb{R}^{N_d}$ is defined as 
\begin{equation}
{\bm{r}}^n: (\boldsymbol{\xi}^n; \boldsymbol{\nu}) \mapsto \alpha_0\boldsymbol{\xi}^n -\Delta t \beta_0\bm{f}(\boldsymbol{\xi}^n, t^n; \boldsymbol{\nu}) + \sum_{j=1}^{\breve{k}} \alpha_j \bm{x}^{n-j}- \Delta t \sum_{j=1}^{\breve{k}} \beta_j \bm{f}(\bm{x}^{n-j},t^{n-j}, \boldsymbol{\nu}),
\label{linKstep}
\end{equation}
where the current work employs the implicit trapezoidal rule, such that $\breve{k}=1;\: \alpha_{0}=1,\:\alpha_{1}=-1; \: \beta_{0}=\beta_{1}=1/2$. Furthermore, the time step is denoted by $\Delta t \in \mathbb{R}_{+}$ and is considered uniform. Here, $\bm{x}^{\breve{k}}$ denotes the numerical approximation to $\bm{x}(\breve{k}\Delta t; \boldsymbol{\mu})$, and $\boldsymbol{\xi}^n \in \mathbb{R}^{N_d}$ is the unknown state vector that is implicitly solved to explicitly update the state, i.e. $\bm{x}^n=\boldsymbol{\xi}^n$. Finally, the implicit trapezoidal integration employed herein is solved via an inexact Newton method, such that the Jacobian is updated every $p_{\textup{it}}$ time-steps, where $p_{\textup{it}}:= cn$ and $c \in \mathbb{N}$. It is also important to note that the current work introduces an inexact kernel Jacobian to compute the residual Jacobian, where only the diagonal block entries are computed, i.e. $\partial \bm{f} / \partial \boldsymbol{\xi}$. This inexact approach is performed to avoid computing a fully-populated matrix where the off-diagonal blocks of the kernel Jacobian provide neglible contributions to the residual Jacobian.

\section{Projection-based reduced order modeling} \label{Section_PROM}

To enable rapid computations of the $N$-body problem in a low-dimensional embedding, the presented PTROM performs dimensional compression of Eq.~\ref{linKstep} via the least-squares Petrov--Galerkin (LSPG) projection \cite{carlberg2013gnat,carlberg2017galerkin}. Specifically, the PTROM seeks an approximate solution, $\tilde{\bm{x}} \approx \bm{x}$, of the form 
\begin{equation}
\tilde{\bm{x}}(t;\boldsymbol{\mu}) = \bm{x}_{\textup{ref}}(\boldsymbol{\mu})+\bm{g}(\hat{\boldsymbol{x}}(t;\boldsymbol{\mu}))
\end{equation}
where $\tilde{\bm{x}}: \mathbb{R}_{+} \times \mathcal{D} \rightarrow \bm{x}_{\textup{ref}} + \mathcal{X}$ and $\mathcal{X}:=\{g(\hat{\boldsymbol{\xi}})\:|\:\hat{\boldsymbol{\xi}} \in \mathbb{R}^M \}$ denotes some trial manifold. Here $\bm{x}_{\textup{ref}}: \mathcal{D}\rightarrow\mathbb{R}^{N_d}$ denotes some parameterized reference state and $\bm{g}: \hat{\boldsymbol{\xi}} \mapsto \bm{g}(\hat{\boldsymbol{\xi}})$ with $\bm{g}:\mathbb{R}^M\rightarrow\mathbb{R}^{N_d}$ and $M\le N$ denotes a parameterization function that projects or maps the low-dimensional generalized coordinates $\hat{\bm{x}}: \mathbb{R}_{+} \times \mathcal{D}\rightarrow \mathbb{R}^M$ to the high-dimensional approximation, $\tilde{\bm{x}}$.

The current work focuses only on constructing an affine trial manifold that exist in the Steifel manifold, i.e. for a full-column-rank matrix, $\mathcal{A}\in\mathbb{R}^{q \times p}$,  the Steifel manifold is defined by $\mathcal{V}_p(\mathbb{R}^q) \equiv \{\boldsymbol{\mathcal{A}} \in \mathbb{R}^{q\times p} \: |\: \boldsymbol{\mathcal{A}}^{T} \boldsymbol{\mathcal{A}} = \mathbf{I}\}$. This affine manifold is expressed as $\bm{g}: \hat{\boldsymbol{\xi}} \mapsto \boldsymbol{\Phi} \hat{\boldsymbol{\xi}}$, where $\boldsymbol{\Phi}\in\mathcal{V}_{M}({\mathbb{R}^{N_d}})$, where the current work constructs a POD \cite{holmes2012turbulence} basis matrix as the mapping operator, $\boldsymbol{\Phi}$.
Finally, for the current POD basis, the reference state can be set as $\bm{x}_{\textup{ref}}=\bm{x}^{0}(\boldsymbol{\mu})$, and so the full-order model state vector can be approximated as 
\begin{equation}
\tilde {\bm{x}}= \bm{x}^0+\boldsymbol{\Phi}\hat{\bm{x}}.
\label{statePODApproximation}
\end{equation}

 It's important to note that although the current work has been restricted to affine trial manifolds, recent works in \cite{lee2020model} have generalized projection based dimensionality compression via nonlinear trial manifolds. Future work will look into generalizing the PTROM framework by adopting these nonlinear mappings, as they have shown to out-perform POD bases for advection dominated physics. 

\subsection{Constructing the proper-orthogonal decomposition basis} \label{Section_PODConstruction}

As previously mentioned, the projection operator, $\boldsymbol{\Phi}$, is based on constructing the POD basis which is a main tool for building the PTROM. Thus, the POD construction procedure will be discussed here. To build $\boldsymbol{\Phi}$, the method of snapshots is employed, where the singular-value decomposition (SVD) is used to factor the snapshot data matrix  $\boldsymbol{\mathcal{S}} \in \mathbb{R}^{N_d \times N_t}$, where
\begin{equation}
\boldsymbol{\mathcal{S}}=\left[\bm{x}^1,\:\bm{x}^2, \ldots, \: \: \bm{x}^{N_t-1},\: \bm{x}^{N_t}\right],
\end{equation}
and where columns of $\boldsymbol{\mathcal{S}}$ represent the time history of the state vector, $\bm{x}$. By factoring the snapshot matrix via the SVD, we obtain
\begin{equation}
\boldsymbol{\mathcal{S}}=\bm{U}\boldsymbol{\Sigma}\bm{V}^{T},
\end{equation}
where the left-singular matrix $\bm{U} \in \mathcal{V}_{N_d}(\mathbb{R}^{N_d})$, the singular-value matrix $\boldsymbol{\Sigma}\equiv \textup{diag}(\sigma_i) \in \mathbb{R}^{N_d \times N_t}$ has diagonal entries that follow a monotonic decrease such that, $\sigma_1 \ge \ldots \ge \sigma_{N_d} \ge 0$, and the right-singular matrix $\bm{V} \in \mathcal{V}_{N_t}(\mathbb{R}^{N_t})$. The POD basis used to build the low-dimensional subspace is constructed by taking the $M$ left singular vectors of $\bm{U}$, such that $M \ll \min\left({N_d, N_t} \right)$ where $\boldsymbol{\Phi} \equiv \left[U^1, \ldots, U^{M} \right]$. Constructing this POD basis is performed as a training step \emph{a priori}, before any online simulations are performed. 

\subsection{Least-Squares Petrov--Galerkin projection}\label{subsection_LSPGProjection}
The least-squares Petrov--Galerkin (LSPG) method constructs a projection-based and time-discrete residual minimization framework, where the projection-based state approximation, Eq.~\ref{statePODApproximation}, is substituted into the time-discrete residual, Eq.~\ref{linKstep}, and cast into a nonlinear least-squares formulation. The LSPG method provides discrete optimality of the residual, $\bm{r}^n(\tilde{\bm{x}},\boldsymbol{\mu})$, at every time-step \cite{carlberg2013gnat}, such that
\begin{equation}
\hat{\bm{x}}=\argmin_{{\bm{z}} \in  \mathbb{R}^{M}}\lVert  \bm{r}(\bm{x}^0+\boldsymbol{\Phi}{\bm{z}}) \rVert_{2}^2.
\label{LSPGminimization}
\end{equation}
It can be shown that this time-discrete residual minimization is akin to a Petrov-Galerkin projection at the time-discrete level, where the test basis is defined by the residual Jacobian $\bm{J}:=\partial \bm{r}(\hat{\boldsymbol{\xi}},\boldsymbol{\mu})/ \partial \hat{\boldsymbol{\xi}}$ and the trial basis is the familar POD basis, $\boldsymbol{\Phi}$, such that $\bm{J}^T\boldsymbol{\Phi}^T \bm{r}=0$, hence the name ``\emph{least-squares} Petrov--Galerkin projection". 

The solution to Eq.~\ref{LSPGminimization} yields the following iterative linear least-squares formulation via the Gauss-Newton method:
\begin{equation}
\hat{\bm{x}}^{n(k)} = \argmin_{{\bm{z}} \in  \mathbb{R}^{M}} \big\lVert \bm{J}^{n}(\bm{x}^0+\bm{\Phi}\hat{\boldsymbol{z}}^{n(k)}; \boldsymbol{\mu}) \boldsymbol{\Phi} \bm{z} + \bm{r}^{n(k)} (\bm{x}^0+\bm{\Phi}\hat{\boldsymbol{z}}^{n(k)}; \boldsymbol{\mu})\big\rVert_2^2
\label{GNLinearLeastSquares}
\end{equation}
and updates to the iterative solution are given by
\begin{equation}
\tilde{\bm{x}}^{n(k+1)}=\tilde{\bm{x}}^{n(k)}+\alpha^{n(k)}\boldsymbol{\Phi}\Delta\hat{\bm{x}}^{n(k)}
\label{GNUpdate}
\end{equation}
for $k=0, \ldots, K$ and where $\alpha^{n(k)} \in \mathbb{R}$ denotes a step length in the search direction, $\Delta \hat{\bm{x}}^{n(k)}$, that can be computed to ensure global convergence (e.g., satisfy the strong Wolfe conditions \cite{nocedal2006numerical}). Here, the initial guesses $\hat{\bm{x}}^{n(0)}$ for the iterative problem are taken as $\hat{\bm{x}}^{n-1}$. 

\subsection{Hyper-reduction}
Despite the low-dimensional sub-space of the generalized coordinates and trial manifold, for all time steps the LSPG method requires the evaluation of $\mathcal{O}(kN_d)$ residual minimization operations and $\mathcal{O}(N_d^2)$ pair-wise operations of the kernel. Therefore, a layer of reduction is required to reduce the OCC of the residual minimization counts in Eq.~\ref{GNLinearLeastSquares} and \ref{GNUpdate}. The current PTROM employs hyper-reduction to perform sparse residual minimization. Specifically the GNAT hyper-reduction approach \cite{carlberg2013gnat} is employed, which performs LSPG residual minimization on a weighted $l^2$-norm, such that
\begin{equation}
\hat{\bm{x}}^n \approx \argmin_{\tilde{\mathbf{x}} \in  \mathbb{R}^{M}}\lVert \bm{\Theta} \bm{r}(\bm{x}^0+\boldsymbol{\Phi}{\bm{z}}) \rVert_2^2.
\label{weightedLSPGminimization}
\end{equation}
Here, the weighting matrix, $\boldsymbol{\Theta}$, is constructed by a gappy POD approach \cite{everson1995karhunen}, where the time-discrete residual is approximated and minimized over a sparse set of entries. The residual approximation, $\tilde{\bm{r}}\approx \bm{r}$, is constructed by way of a time-discrete residual POD basis employing the offline training procedure discussed in Section \ref{Section_PODConstruction}, where $\tilde{\bm{r}}=\boldsymbol{\Phi}_{\bm{r}}\hat{\bm{r}}$, such that $\boldsymbol{\Phi}_{\bm{r}}\in \mathcal{V}_{M_{\bm{r}}}(\mathbb{R}^{N_d})$, $\hat{\bm{r}}\in \mathbb{R}^{M_{\bm{r}}}$, and $M_{\bm{r}} \ll N$ is the number of retained SVD  singular vectors in $\bm{U}_{\bm{r}}$. Next, the residual minimization over a sparse set of entries is performed by the following linear least-squares problem,
\begin{equation}
\hat{\bm{r}}=\argmin_{\bm{z}_r \in \mathbb{R}^{n_{\bm{r}}}}\left \lVert  \: \bm{P}\boldsymbol{\Phi}_{\mathbf{r}}\bm{z}_r - \bm{P}{\bm{r}(\tilde{\mathbf{x}})}  \: \right \rVert_2^2,
\label{approx_resi}
\end{equation}
where the matrix $\bm{P}\in\{0,1\}^{n_{d} \times N_d}$ is a sampling matrix consisting of sparse $n_{d}$ selected rows of the identity matrix, which also correspond to the same rows in the time-discrete residual vector, where $n_{d} \ll N_d$. Note $\breve{n}$ is the number of sparsely sampled particles, and $n_{d}=d\breve{n}$, which correspond to the number of sparsely sampled degrees-of-freedom. The solution to Eq.\ref{approx_resi} yields
\begin{equation}
\hat{\bm{r}}= \lbrack \bm{P}\boldsymbol{\Phi}_{\bm{r}} \rbrack^{+}\bm{P}\mathbf{r}(\tilde{\bm{x}}).
\label{gappySolution}
\end{equation}
Substituting Eq.~\ref{gappySolution} into $\tilde{\bm{r}}=\boldsymbol{\Phi}_{\bm{r}}\hat{\bm{r}}$ yields the residual approximation
\begin{equation}
\tilde{\bm{r}}=\boldsymbol{\Phi}_{\bm{r}}\lbrack \bm{P}\boldsymbol{\Phi}_{\bm{r}} \rbrack^{+}\bm{P}\bm{r}(\tilde{\bm{x}}),
\label{residual_approximation}
\end{equation}
whereby via the approximation, $\tilde{\bm{r}}\approx \bm{r}$, the substitution of Eq.~\ref{residual_approximation} into the weighted LSPG minimization in Eq.~\ref{weightedLSPGminimization} yields the following residual minimization,
\begin{equation}
\hat{\bm{x}} \approx \argmin_{\tilde{\bm{x}} \in  \mathbb{R}^{M}}\big\lVert \bm{\Theta}\bm{r}(\bm{x}^0+\boldsymbol{\Phi}{\bm{z}}) \big\rVert_{2}^2,
\label{GNATminimization}
\end{equation}
where $\boldsymbol{\Theta}:= \lbrack  \bm{P}\boldsymbol{\Phi}_{\bm{r}}\rbrack^{+} \bm{P}$. A visual representation of the GNAT hyper-reduction technique employed for $N$-body pair-wise interactions is illustrated in Fig.~\ref{HROM_forNbodies} below.

\begin{figure}[t]
	\includegraphics[scale=0.8, trim=1cm 10cm 0 7cm]{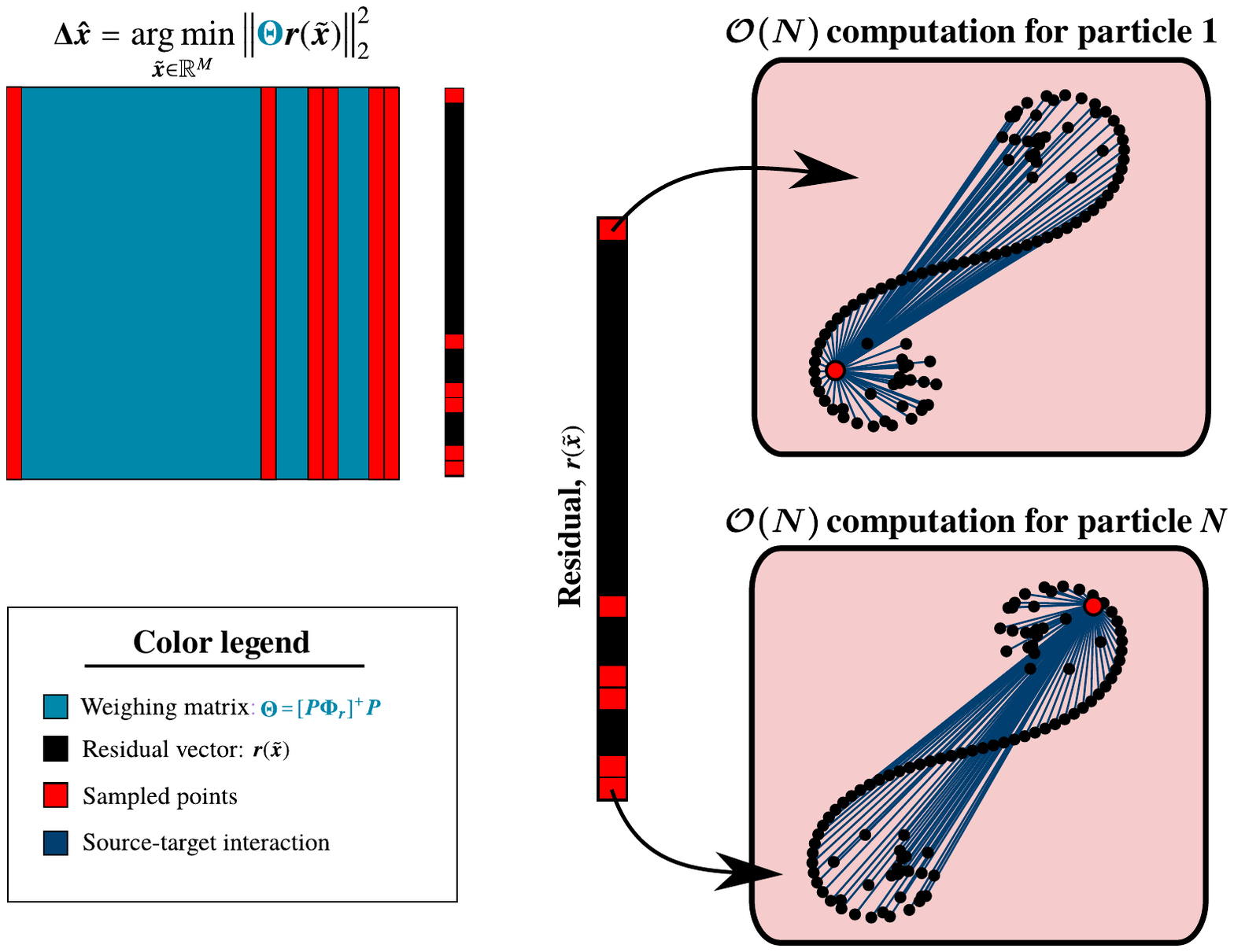}
	\caption{Illustration of the GNAT hyper-reduction approach applied to an $N$-body pairwise interaction computational framework.}
	\label{HROM_forNbodies}
\end{figure}

\subsection{Constructing the sampling matrix}
To enable the sparse residual minimization of Eq.~\ref{GNATminimization}, a sampling matrix $\bm{P}$ is constructed strategically. Work presented in \cite{carlberg2013gnat} constructed a sampling matrix tailored to a computational fluid dynamics (CFD) mesh via a greedy algorithm based on mitigating the maximum error of the POD residual basis and POD residual Jacobian basis. The current sampling approach is similar to that presented in \cite{carlberg2013gnat} but is tailored to a computational domain for $N$-body problems and only attempts to mitigate the error of the POD residual basis.  The greedy algorithm to construct the sampled matrix $\bm{P}$ is presented below in Algorithm \ref{samplingAlgorithm}.

\begin{algorithm}[t]
	\DontPrintSemicolon
	\caption[]{Greedy algorithm for selecting sample \emph{N}-bodies in a domain. Adapted from \cite{carlberg2013gnat}	\label{samplingAlgorithm}}
	\KwIn{ $\boldsymbol{\Phi}_{\bm{r}}$; user-defined target number of sample \emph{N}-bodies, $\breve{n}$; sampled \emph{N}-body set, $\mathcal{N}$ (can have \\ user-defined preseeded \emph{N}-bodies); the minimum number of either retained columns in $\boldsymbol{\Phi}_{\bm{r}}$, or target degrees of freedom, i.e.~$n_c \le \textup{min}\left(n_{\bm{r}}, n_d\right)$} 
	\KwOut{ Sampled set of \emph{N}-bodies, $\mathcal{N}$}
	{Let $\mathbf{I}$ be the identity matrix, $\mathcal{N}(\delta)$ denote the degrees of freedom of the sampled node set $\mathcal{N}$, and $\bm{I}(\mathcal{N}(\delta))$ denote the entries of the identity matrix associated with $\mathcal{N}$.}\\
Compute additional number of bodies left to sample $n_a=\breve{n}-\lvert \mathcal{N} \rvert$ \\
 Initialize counter for the number of working basis vectors used: $n_b \leftarrow 0$ \\
Set the number of greedy iterations to perform: $n_{it}=\textup{min}(n_c, n_a)$ \\
Compute the maximum number of right-hand sides in the least-squares problem:  $n_{\textup{rhs}}=\textup{ceil}(n_a n_{\textup{rhs}}/n_c) ^{\dagger}$\\
Compute the minimum number of working basis vectors per iteration: $n_{ci,\textup{min}}=\textup{floor}(n_c/n_{it})$\\
Compute the minimum number of sample bodies to add per iteration: $n_{ci,\textup{min}}=\textup{floor}(n_c/n_{it})$\\
		\For(\tcp*[h]{begin greedy sampeling loop}){$i$=$1, \ldots, n_{\textup{it}}$}{ $n_{ci} \leftarrow n_{ci, \textup{min}}$ \tcp*[h]{computing the number of working basis vectors for this iteration}\\
		\If{$i \le n_c \: \textup{mod} \: n_{it}$}{$n_{ci} \leftarrow n_{ci}+1$}
$n_{ai} \leftarrow n_{ai,\textup{min}}$ \tcp*[h]{computing the number of sampled N-bodies to add during this iteration}\\
 \If{$n_{\textup{rhs}}=1$ $\textup{and}$ $i \le n_a \: \textup{mod} \:\: n_c$}{$n_{ai} \leftarrow n_{ai}+1$}
\If{i=1}{$\lbrack \bm{R}^1 \: \cdots \: \bm{R}^{n_{ci}}\rbrack \leftarrow \lbrack \boldsymbol{\phi}^1_R \: \cdots \: \boldsymbol{\phi}^{n_{ci}}_R \rbrack$}

\ElseIf{$i\neq 1$}{
		\For(\tcp*[h]{loop for selecting a sparse set of \emph{N}-bodies}){$j=1, \ldots, n_{ai}$:}{
		$\breve{n} \leftarrow \argmax\limits_{l \notin \mathcal{N}} \sum_{q=1}^{n_{ci}}\left(\sum_{i \in \delta(l)} (\bm{R}_i^q)^2 \right)$ \tcp*[h]{choose $N^{\textup{th}}$ body with largest average error}\\
		\tcp*[h]{where $\delta(l)$ denotes the degrees of freedom associated with the $l^{\textup{th}}$ body} \\
		$\mathcal{N}\leftarrow \mathcal{N} \cup \breve{n}$}
}
$n_b \leftarrow n_b + n_{ci} $
}
$\bm{P} \leftarrow \bar{\bm{I}}(\mathcal{N}(\delta))$, where the over-bar denotes minimum-cardinality \\ 
\footnotesize{$\dagger$ $n_{\textup{rhs}}$ ensures the system of equations in the minimization problem remains over-determined}

\end{algorithm} 

\subsection{Remarks on the current projection-based reduced reduced order model}\label{SubsectionHyperReductionNdependent}
Implementing GNAT hyper-reduction into an LSPG method drastically reduces the residual minimization count over the $N_d$ residual vector space. For grid-based methods, this hyper-reduction step is sufficient to achieve $N$-independence. However, employing GNAT hyper-reduction on computational Lagrangian $N$-body methods only reduces the number of target bodies in the $N$-body pairwise interaction, which still requires the knowledge of \emph{all} $N$ sources in the domain that act on targets. Therefore, even though integrating GNAT hyper-reduction into the $N$-body problem reduces overall  OCC, the resulting cost remains $N$-dependent, and requires an additional layer of reduction to reduce the number of $N$ sources.

\section{The Barnes--Hut tree method } \label{Section_QuadTreeDecomp}

\begin{figure}[t]
	\includegraphics[scale=0.4, trim=-6cm 0cm 0cm 0cm]{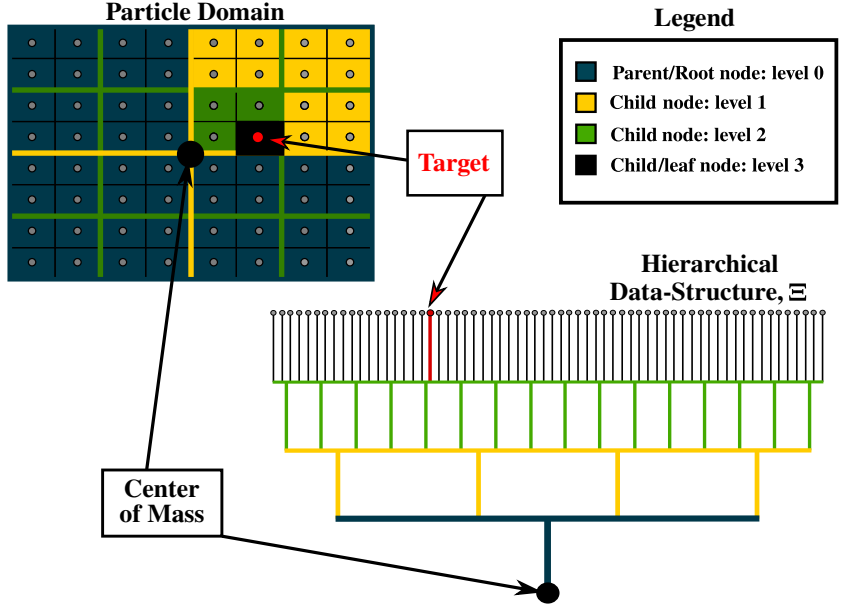}
	\caption{Schematic of the different levels in a hiearchical decomposition generated by the Barnes-Hut method. In the top left, a domain schematic highlighting the recursively partitioned levels are shown. In the bottom center the hierarchical data structure $\boldsymbol{\Xi}$ is illustrated. }
	\label{Tree_flowchart}
\end{figure}

Prior sections focused on drastically decreasing OCC and compute-time by finding a low-dimensional embedding over a hyper-reduced numerical domain. Now, it also necessary to reduce the OCC dependencies associated with $N$ sources over the sparse $\breve{n}$ residual entries in Eq.~\ref{GNATminimization} to realize $N$-independence and further improve the efficiency of the current framework. Here, the PTROM employs hierarchical decomposition and source agglomeration via the Barnes--Hut tree method \cite{barnes1986hierarchical, pfalzner2005many}. The Barnes--Hut tree method builds a hierarchical quad-tree (or in three-dimension, oct-tree) data structure, $\boldsymbol{\Xi}$, that performs recursive partitioning over the entire domain (the root node) that contains all $N$ bodies. Recursive partitioning generates branch nodes until a desired number of bodies are contained per partition, where this final level of partitioning is known as the leaf node. Figure ~\ref{Tree_flowchart} illustrates the hierarchical data-structure generated by the Barnes--Hut tree method. The Barnes--Hut tree method is well-documented in the literature, where pseudo codes and flowcharts to build the hierarchical data structure can be found in \cite{pfalzner2005many}. 

\subsection{Source clustering }  
The Barnes--Hut data structure enables OCC reduction by employing branch node agglomeration that correspond to negligible far-field sources. The PTROM explores two criteria to identify which branch nodes to cluster: 1) A neighbor search criteria, and 2) the classical Barnes--Hut clustering criteria. Before presenting the details of the clustering criteria, useful terminology is introduced:

\begin{definition} \normalfont
Let the Cartesian position vector of a target be expressed by $\bm{s}_i^{\tau} \in \mathbb{R}^2$. Next, let the set of far-field sources be expressed by $\{\zeta \: | \: \zeta \in \mathbb{N}, \zeta \le N\}$, and let $\tilde{\zeta}$ denote the surrogate source representing the clustered set $\zeta$. The Cartesian position vector for a source $j$ in the set $\zeta$, is defined by $\bm{s}^{\zeta}_{j} \in \mathbb{R}^2$, and so the Cartesian position vector for the surrogate source $\tilde{\zeta}$ is defined by some weighted mean, $\bm{s}^{\tilde{\zeta}}=\left\{\sum\limits_{j=1}^{n(\zeta)} s_{\chi,j}^{\zeta}\Gamma_j/\sum\limits_{j=1}^{n(
	\zeta)} \Gamma_j,\sum\limits_{j=1}^{n(\zeta)} s_{\psi,j}^{\zeta}\Gamma_j/\sum\limits_{j=1}^{n(\zeta)} \Gamma_j\right\}$, where the subscripts $\chi$ and $\psi$ denote Cartesian coordinate components of $\bm{s}^{\zeta}_j$ and $n(\zeta)$ is the cardinality of $\zeta$. Note that the weighted mean is computed within the context of the Biot--Savart kernel, such that the circulation of source bodies in $\zeta$ correspond to the weights in the mean computations. 
\end{definition}

\begin{definition} \normalfont
Let $\breve{\mathcal{N}_{\zeta}}$ denote the quad-tree quadrilateral source node containing $\zeta$. The width $w_{\breve{\mathcal{N}_{\zeta}}}\in\mathbb{R}_{+}$ of $\breve{\mathcal{N}_{\zeta}}$ is defined as the maximum length of its sides, and node $\breve{\mathcal{N}_{\zeta}}$ contains all $\zeta$ inside the boundary $ {b}^{\breve{\mathcal{N}_{\zeta}}}_{\chi,\textup{min}} \leq \chi \leq {b}^{\breve{\mathcal{N}_s}}_{\chi,\textup{max}}$ and  $ {b}^{\breve{\mathcal{N}_s}}_{\psi,\textup{min}} \leq \psi \leq {b}^{\breve{\mathcal{N}_{\zeta}}}_{\psi,\textup{max}}$, where  $\bm{b}^{\breve{\mathcal{N}_{\zeta}}}:=\left\{{b}^{\breve{\mathcal{N}_{\zeta}}}_{\chi,\textup{min}}, {b}^{\breve{\mathcal{N}_{\zeta}}}_{\chi,\textup{max}},{b}^{\breve{\mathcal{N}_{\zeta}}}_{\psi,\textup{min}},{b}^{\breve{\mathcal{N}_{\zeta}}}_{\psi,\textup{max}}\right\}$ is the set of boundaries of $\breve{\mathcal{N}_{\zeta}}$ and its components are in $\mathbb{R}$. Similarly, let  $\breve{\mathcal{N}_{\tau}}$ denote the quad-tree quadrilateral leaf node containing a set of targets, $\{\tau \: | \: \tau \in \mathbb{N}, \tau \le N\}$, where the position vector for an $i^{\textup{th}}$ target in $\breve{\mathcal{N}_{\tau}}$ is defined by $\bm{s}_i^{\tau}$. The width $w_{\breve{\mathcal{N}_{\tau}}}\in\mathbb{R}$ of $\breve{\mathcal{N}_{\tau}}$ is defined as the maximum length of its sides, and node $\breve{\mathcal{N}_{\tau}}$ contains all $\tau$ inside the boundary $ {b}^{\breve{\mathcal{N}_{\tau}}}_{\chi,\textup{min}} \leq \chi \leq {b}^{\breve{\mathcal{N}_{\tau}}}_{\chi,\textup{max}}$ and  $ {b}^{\breve{\mathcal{N}_{\tau}}}_{\psi,\textup{min}} \leq \psi \leq {b}^{\breve{\mathcal{N}_{\tau}}}_{\psi,\textup{max}}$, where  $\bm{b}^{\breve{\mathcal{N}_{\tau}}}:=\left\{{b}^{\breve{\mathcal{N}_{\tau}}}_{\chi,\textup{min}}, {b}^{\breve{\mathcal{N}_{\tau}}}_{\chi,\textup{max}},{b}^{\breve{\mathcal{N}_{\tau}}}_{\psi,\textup{min}},{b}^{\breve{\mathcal{N}_{\tau}}}_{\psi,\textup{max}}\right\}$ is the set of boundaries of $\breve{\mathcal{N}_{\tau}}$ and its components are in $\mathbb{R}$. Finally, the neighborhood of the target node, $\breve{\mathcal{N}_{\tau}}$, is defined by the following set of boundaries $\mathcal{H}:=\left\{\bm{b}^{\breve{\mathcal{N}_{\tau}}} \oplus {\bm{w}}\:| \: \mathcal{H}=\{{h}_{\chi,\textup{min}},{h}_{\chi,\textup{max}},{h}_{\psi,\textup{min}},{h}_{\psi,\textup{max}}\}\right\}$, where each component of $\mathcal{H}$ is in $\mathbb{R}$, $\bm{w}:=\left\{-p_cw_{\breve{\mathcal{N}_{\tau}}}, p_cw_{\breve{\mathcal{N}_{\tau}}}, -p_cw_{\breve{\mathcal{N}_{\tau}}}, p_cw_{\breve{\mathcal{N}_{\tau}}}\right\}$, $p_c\in\mathbb{R}_{0,+}$ is a factor that is added to the target node boundaries to extend the neighborhood of the target node, and $\oplus$ denotes the direct sum.
\end{definition}

\begin{figure}[t]
	\includegraphics[scale=0.6, trim=-3cm 7cm 0cm 6cm]{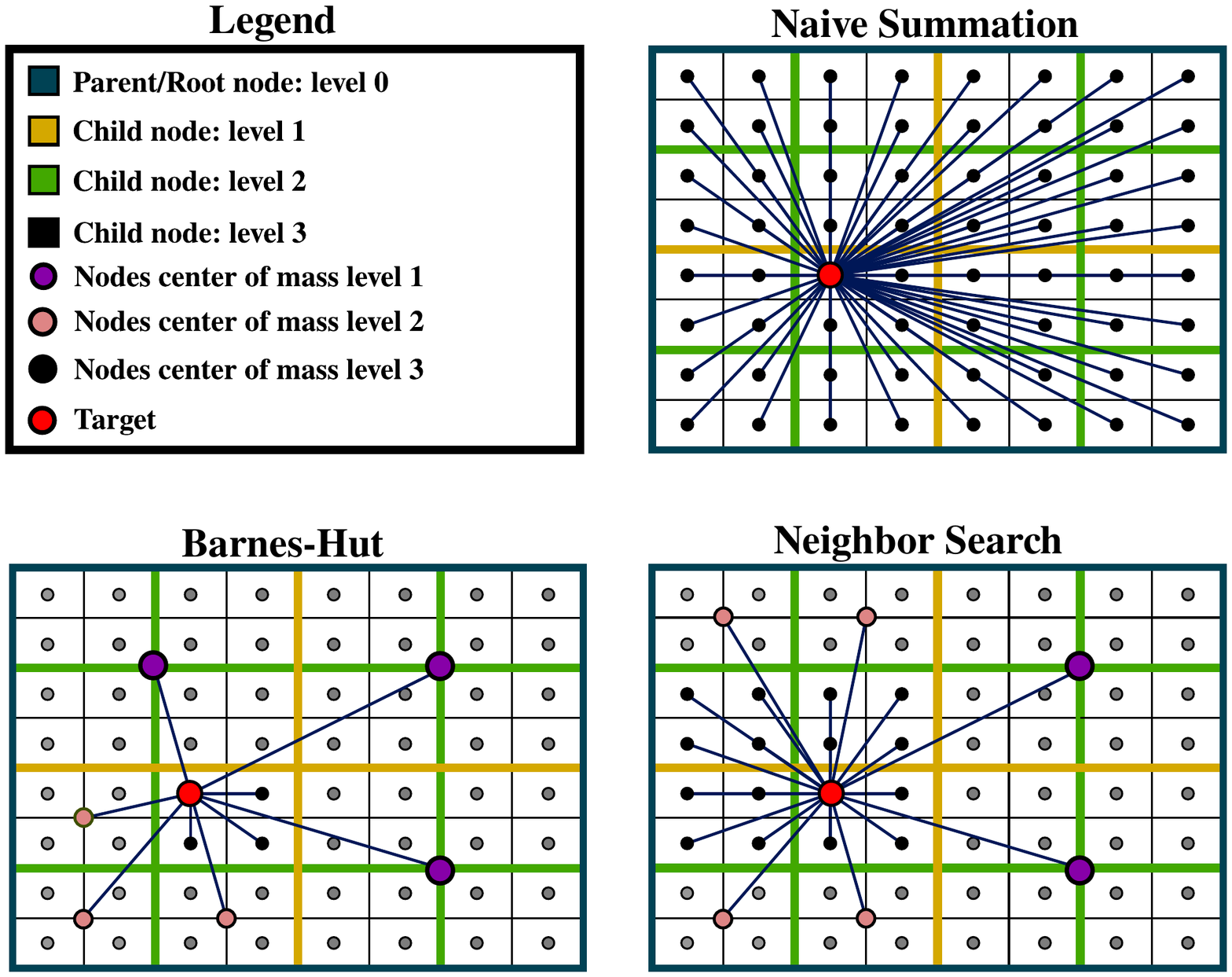}
	\caption{Illustration of the naive summation (top-right) of an $N$-body problem, Barnes--Hut clustering (bottom-left) with $\theta=2$, and the neighbor search clustering (bottom-right), with neighborhood boundary defined by $\frac{1}{2}{w}_{\mathcal{N}_\mathcal{T}}$ .  }
	\label{Clustering}
\end{figure}

The neighbor search criteria employed by the PTROM is based on identifying any overlapping source node corners, within the neighborhood boundaries of a target node. Specifically, clustering occurs if, and only if, a source node corner does not overlap with a target neighborhood, i.e. $\left\{h_{\chi,\textup{max}}> b_{\chi,\textup{min}}^{\breve{\mathcal{N}_{\tau}}}, h_{\chi,\textup{min}}< b_{\chi,\textup{max}}^{\breve{\mathcal{N}_{\tau}}}, h_{\psi,\textup{max}}> b_{\psi,\textup{min}}^{\breve{\mathcal{N}_{\tau}}}, h_{\psi,\textup{min}}< b_{\psi,\textup{max}}^{\breve{\mathcal{N}_{\tau}}}\right\}$ must all be false to cluster. The Barnes--Hut clustering criteria is based on clustering $\zeta$ when the following is met:
If $ w_{\breve{\mathcal{N}_{\zeta}}}/\lVert \bm{s}^{\tilde{\zeta}}_j -  \bm{s}_i^{\tau}\rVert_2 \le \theta $, where $\theta \in \mathbb{R}_{0+}$ and is a user-defined clustering parameter, then clustering of $\zeta$ is performed. An illustration of the two clustering approaches is shown in Fig.~\ref{Clustering}. A pseudo algorithm for both clustering approaches will be provided in the proceeding sub-section.

\subsubsection{Hierarchical decomposition of the projection basis}
\begin{figure}[t]
	\includegraphics[scale=0.55, trim= 0cm 1cm 0cm 0cm]{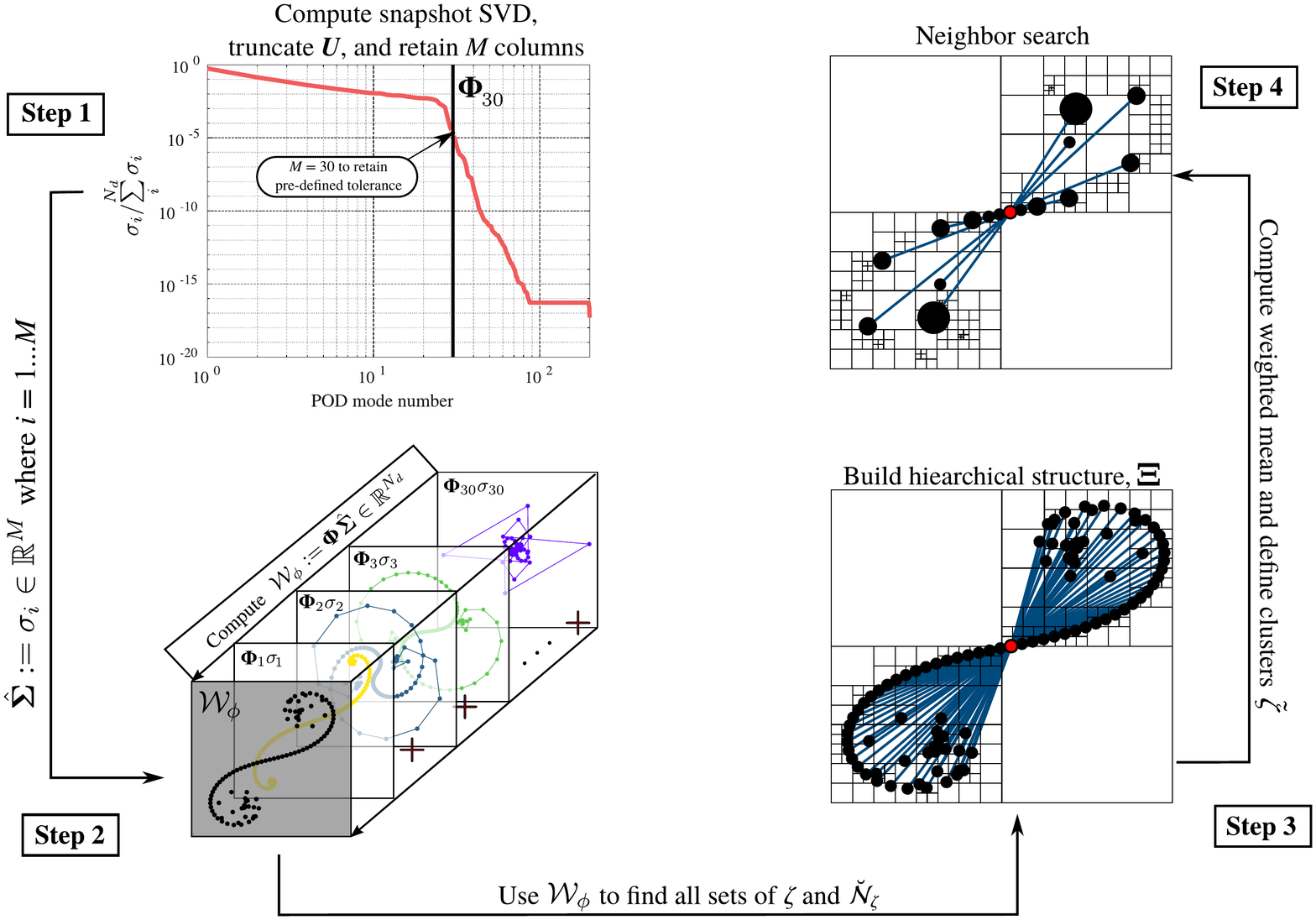}
	\caption{Schematic of the weighted POD space, $\mathcal{W_{\phi}}$, construction. }
	\label{OfflinePODClustering}
\end{figure}

Traditionally, computing the Barnes-Hut tree decomposition and corresponding source clustering is performed over the $N$-body state-space, where both tree decomposition and clusters are updated at incremental time-steps throughout a simulation. However, building the tree data structure and performing source clustering are $N$-dependent operations \cite{pfalzner2005many}, which would not overcome the $N$-dependent OCC barrier in the hyper-reduction step, as discussed in Section \ref{SubsectionHyperReductionNdependent}. To overcome the need to perform multiple online tree construction and clustering of the state space the PTROM constructs the hierarchical data structure and source clustering in a \emph{weighted POD space}, offline and only once. This weighted POD space is denoted by $\mathcal{W}_{\phi}:\hat{\boldsymbol{\Sigma}} \mapsto \boldsymbol{\Phi}\hat{\boldsymbol{\Sigma}}$, where $\hat{\boldsymbol{\Sigma}} \in \mathbb{R}^M$ and its entries correspond to the diagonal entries of $\boldsymbol{\Sigma}$ up to the $M^{\textup{th}}$ retained singular value. In other words,  $\mathcal{W}_{\phi}$ is a linear combination of the POD trial bases where the weights are defined by the corresponding singular values of the retained columns in the singular matrix, $\bm{U}$. Figure \ref{OfflinePODClustering} illustrates the offline procedure to compute the weighted POD space, construct the data structure, $\boldsymbol{\Xi}$, and perform source clustering.

\begin{figure}[h!]
	\includegraphics[scale=0.5, trim=-2cm 2cm 1cm 1cm]{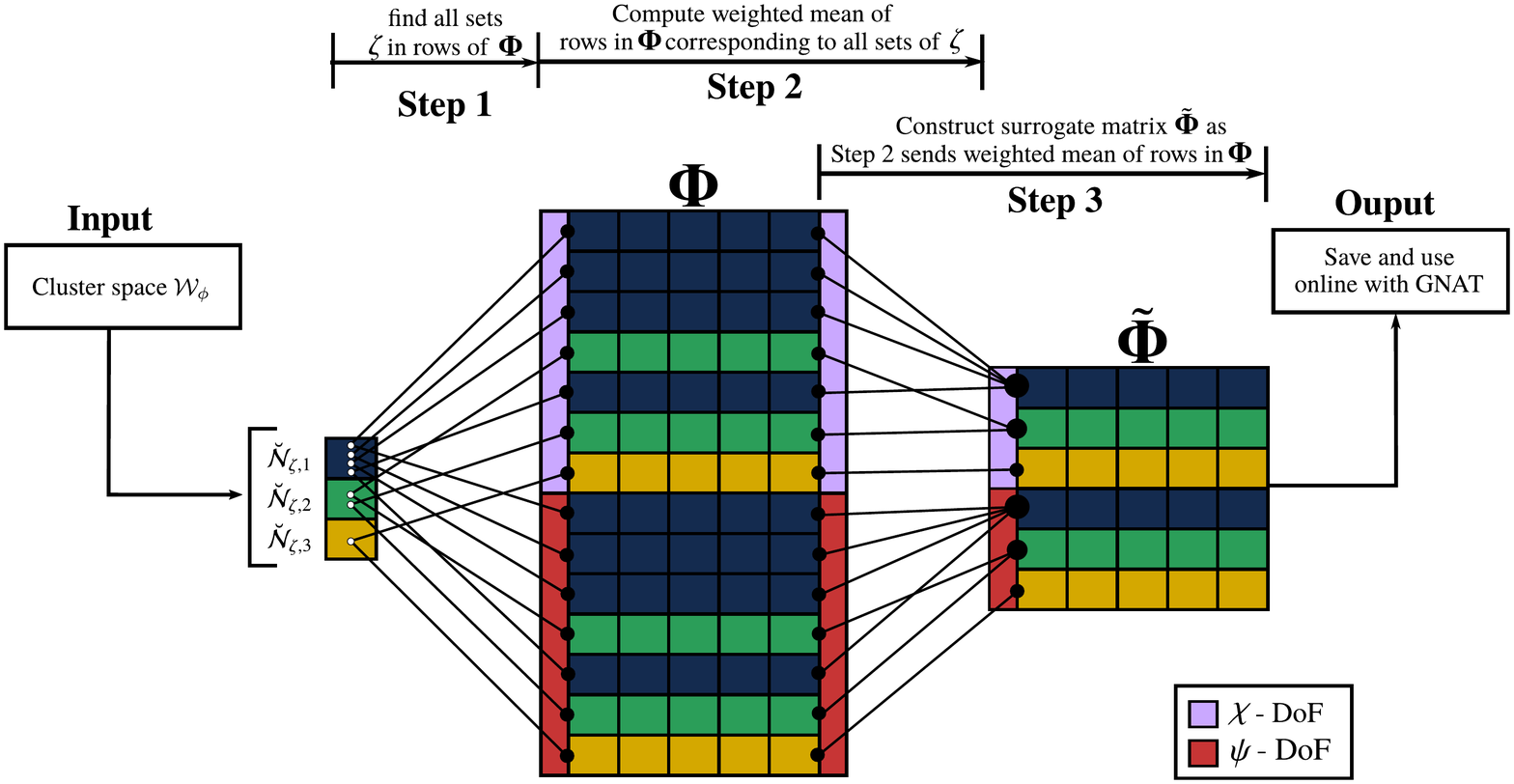}
	\caption{Illustration of the map from the weighted POD space, $\mathcal{W}_{\phi}$, to the clustered surrogate source POD matrix $\tilde{\boldsymbol{\Phi}}$. }
	\label{ClusteringMapping}
\end{figure}

By computing the hierarchical decomposition and clustering in $\mathcal{W}_{\phi}$, all $\mathcal{F}$ clusters and $\breve{\mathcal{N}_s}$ nodes identified over the particle domain can be mapped to associated degrees-of-freedom per column in the POD basis matrix, $\boldsymbol{\Phi}$. This mapping of the degrees-of-freedom from $\mathcal{W}_{\phi}$ to $\boldsymbol{\Phi}$ allows the construction of a ``source surrogate POD basis" data structure that reduces the dimensionality of $\boldsymbol{\Phi}$ by approximating the structure of the source POD mode shapes as observed by the sparsely sampled targets selected by Algorithm \ref{samplingAlgorithm}, which ultimately enables $N$-independence. Mapping from $\mathcal{W}_{\phi}$ to $\boldsymbol{\Phi}$ and the associated construction of the source surrogate POD basis is depicted in Figure \ref{ClusteringMapping}, where the algorithm to perform the construction of the surrogate POD basis data structure is presented in Algorithm \ref{PODsurrogateAlgorithm}.The construction of $\tilde{\boldsymbol{\Phi}}$ occurs during the clustering of $\mathcal{W}_{\phi}$, such that the degrees of freedom clustered in the weighted POD space are mapped to corresponding row entries of $\boldsymbol{\Phi}$, and these associated row entries of $\boldsymbol{\Phi}$ are clustered to generate  $\tilde{\boldsymbol{\Phi}}$.

\begin{figure}[h!]
	\includegraphics[scale=0.675, trim=-1cm 0cm 0cm 0cm]{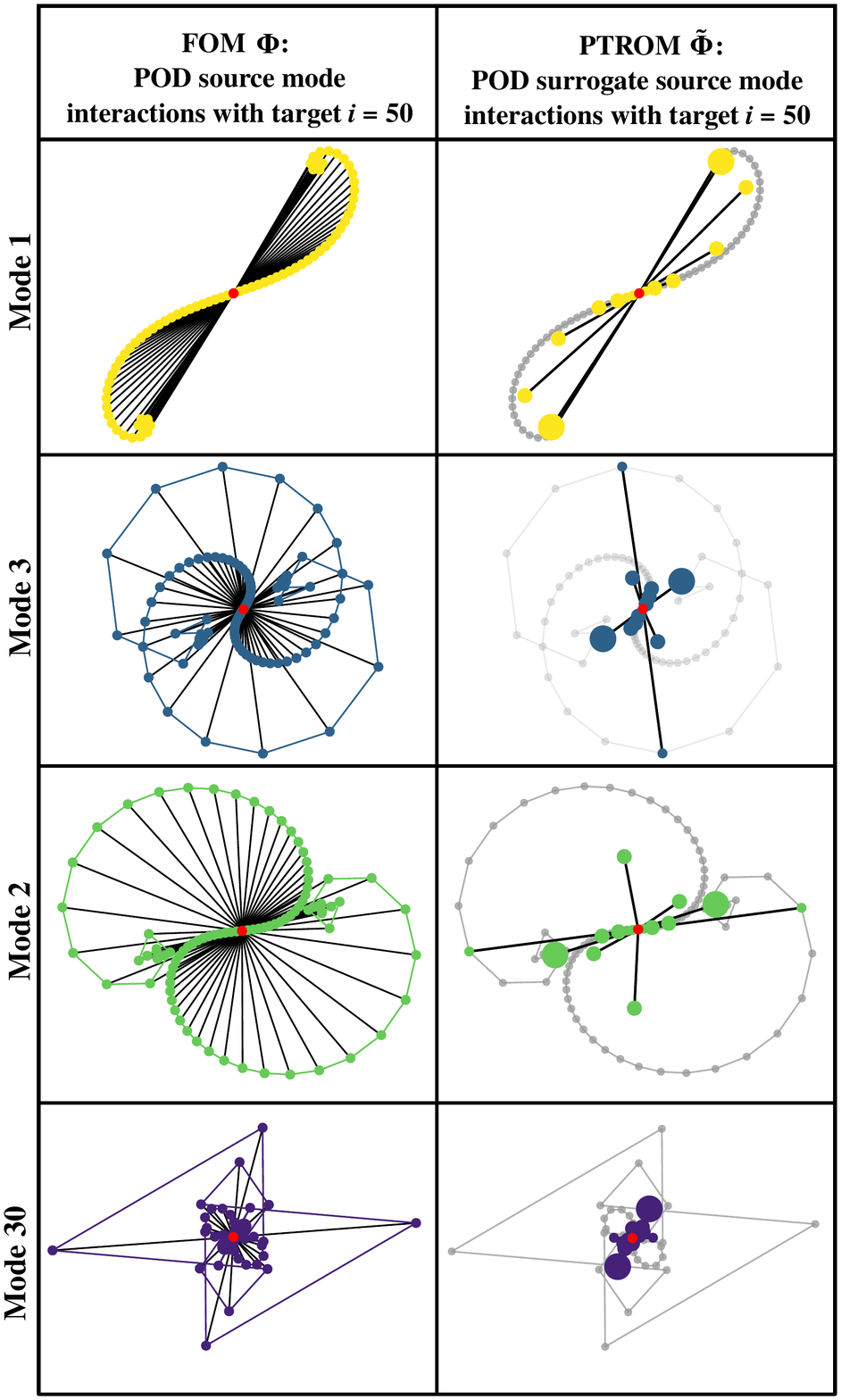}
	\caption{Comparisons of the FOM POD modes and surrogate source POD matrices as observed by target point $i=50$}
	\label{ModeComparisons}
\end{figure}

As mentioned earlier, the data-structure $\tilde{\boldsymbol{\Phi}}$ contains approximations of POD modes as observed by individual sampled targets. In other words,  $\tilde{\boldsymbol{\Phi}}$ contains a library of $\breve{n}$ POD mode source approximations for all $\breve{n}$ sampled targets. Figure \ref{ModeComparisons} compares the interaction between a sampled target, say particle $i=50$, and all source POD modes against the interaction between a sampled target and clustered source POD modes. It is important to take notice that clusters of the POD modes in the weighted space $\mathcal{W}_{\phi}$ are generally not unique for each sampled target. In fact, most targets in the same neighborhood share the same clusters. As a result, there exists a unique number of source POD surrogate clusters, $N_c$ and corresponding degree-of-freedom $N_{c,d}$ for all sampled targets, where $N_{c, d}=dN_c$ and $N_c\ll N$ and $N_{c,d}\ll N_d$. These unique source POD structures can then be structured into the final form of the source POD basis surrogate matrix,  $\tilde{\boldsymbol{\Phi}}\in \mathbb{R}^{N_{c,d}\times M}$. Similarly, the same unique clusters have a unique agglomerated circulation, which must also be structured into the compact vector $\tilde{\boldsymbol{\Gamma}}\in \mathbb{R}^{N{\tilde{\gamma}}}$, where $N_{\tilde{\gamma}}$ is the  number of unique source strengths and $N_{\tilde{\gamma}}\ll N$.  The process of finding unique source POD surrogate clusters is straight-forward and there is no specific search algorithm that must be used. As a result, Algorithm \ref{PODsurrogateAlgorithm} denotes the unique cluster search by the function \textproc{\textbf{UniqueSearch}}.\\

\clearpage
 \begin{algorithm}[H]
 	\caption[]{\textproc{\textbf{ClusterPOD}}; Offline agglomeration of source POD modes. \label{PODsurrogateAlgorithm}}
 	\KwInput{POD basis, $\boldsymbol{\Phi}$; tree data-structure of the weighted POD space, $\boldsymbol{\Xi}(\mathcal{W}_{\phi})$; circulation vector, $\boldsymbol{\Gamma}$. }  
 	\KwOutput{Unique surrogate source POD matrix, $\tilde{\boldsymbol{\Phi}}$; Unique surrogate source circulation vector, $\tilde{\boldsymbol{\Gamma}}$} 
 	{$\boldsymbol{\Xi}_i^l(\mathcal{W}_{\phi})$	denotes the hierarchical data-structure of the POD weighted space at some level $l$ and node $i$. $\mathcal{L}_i=\{\boldsymbol{\Xi}_i^{l_f}(\mathcal{W}_{\phi})\: | \: i \subseteq n_f\}$ denotes a set corresponding to the leaf node $i$, where $l_f$ is the leaf level and $n_f$ denotes the number of leaves. $\mathcal{L}_i(\hat{n})$ is the position vector of particle $\hat{n}$ at leaf node $i$. $\bm{p}_i^l(\hat{n}(\delta)):\boldsymbol{\Xi}_i^l(\mathcal{W}_{\phi}(\hat{n}(\delta)))$ $\mapsto \boldsymbol{\Phi}(\boldsymbol{\Xi}_i^l(\mathcal{W}_{\phi}(\hat{n}(\delta))))$ denotes the mapped particles $\hat{n}$ with degrees-of-freedom $\delta$, in node $i$, at level $l$ from $\boldsymbol{\Xi}(\mathcal{W}_{\phi})$ to $\boldsymbol{\Phi}$, and where $\mathcal{P}_i^l  \equiv \lvert\bm{p}_i^l(\hat{n}(\delta))\rvert$. \textbf{Note}: All operations are performed per POD column.} \\
 	
 	\SetKwFunction{KwFn}
 	\DontPrintSemicolon
 	\SetKwFunction{FMain}{\rm\textbf{{\textproc{WeightedMean}}}}
 	\SetKwProg{Pn}{Function}{:}{\KwRet}
 	\Pn{\FMain{$\boldsymbol{\Xi}(\mathcal{W}_{\phi}),\: \boldsymbol{\Phi},\: \boldsymbol{\Gamma}$}}{
 		{$l=l_f$}\\

 		\For(\tcp*[h]{loop over all leaf nodes} ){ $i=1\dots n_f$}{
 			$\tilde{\boldsymbol{\Gamma}}_i^l=\sum\limits_{k=1}^{\mathcal{P}_i}\boldsymbol{\Gamma}_{\bm{p}_i^l(\hat{n}(\delta))}$ and  
 			$\:\tilde{\boldsymbol{\Phi}}_{i}^{l}=  \sum\limits_{k=1}^{\mathcal{P}_i} \boldsymbol{\Phi}_{\bm{p}_i^l(\hat{n}(\delta))}\:\boldsymbol{\Gamma}_{\bm{p}_i^l(\hat{n}(\delta))}/\sum\limits_{k=1}^{\mathcal{P}_i}\boldsymbol{\Gamma}_{\bm{p}_i^l(\hat{n}(\delta))}$  \tcp*[h]{compute weighted mean of POD modes } }
 		$l\leftarrow l_f - 1$\\
 		 	
 		\While(\tcp*[h]{compute clustering over all children at level $l_{\textup{children}}$ of cell $i$}){$l\ge0$}{
 				\For{i=1\ldots\textup{all nodes in level} $l$}{
 			$\tilde{\boldsymbol{\phi}} = \mathbf{0},\:\tilde{\boldsymbol{\Gamma}}_i^l = \mathbf{0}$\\
 			\For(\tcp*[h]{cluster the children of this node $i$}){k=1 \ldots 4}{ 
 				$\tilde{\boldsymbol{\gamma}}^l_i \leftarrow \tilde{\boldsymbol{\gamma}}^l_i + \tilde{\boldsymbol{\Gamma}}^{l_\textup{children}}_k $ and 
 				$\: \tilde{\boldsymbol{\phi}}_i \leftarrow \tilde{\boldsymbol{\phi}}_i + \tilde{\boldsymbol{\Phi}}_k^{l_\textup{children}} \tilde{\boldsymbol{\Gamma}}^{l_\textup{children}}_k$}
 			$\tilde{\boldsymbol{\phi}}_i^l=\tilde{\boldsymbol{\phi}}/ \tilde{\boldsymbol{\gamma}}$,  \tcp*[h]{compute the source POD mode surrogate  of node $i$ at level $l$}\\

 	}
  	        $l\leftarrow l - 1$	
 }
 	}

 	\vspace{0cm}
 	\SetKwFunction{KwFn}
 	\DontPrintSemicolon
 	\SetKwFunction{FMain}{\rm\textbf{{\textproc{FindClusters}}}}
 	\SetKwProg{Pn}{Function}{:}{\KwRet}
 	\Pn{\FMain{$\boldsymbol{\Xi}(\mathcal{W}_{\phi}),\: \tilde{\boldsymbol{\phi}},\: \tilde{\boldsymbol{\gamma}}$}}{
 		
 		\For{${k}=1\ldots n_f$}{
 			\For{$i=1\ldots \lvert\mathcal{L}_k\rvert$ }{
 		$\tau=\mathcal{L}_k(i)$; $l=0$; $j=1$;\\
 			\SetKwFunction{KwFn}
 		\DontPrintSemicolon
 		\While {$l<l_f$}{
 			\SetKwFunction{FMain}{\rm\textbf{{\textproc{Traverse}}}}
 			\SetKwProg{Pn}{Function}{:}{\KwRet}
 			\Pn{\FMain{$\boldsymbol{\Xi}_j^l(\mathcal{W}_{\phi}),\: \tilde{\boldsymbol{\phi}},\: \tilde{\boldsymbol{\gamma}}$}}{
 				Prune=\textproc{\textbf{checkPrune}}$(\boldsymbol{\Xi}_{j}^l(\mathcal{W}_{\phi}), \bm{s}_i^{\tau}, \bm{s}^{\tilde{\zeta}})$ \tcp*[h]{cluster the current source node?}\\
 				\If (\tcp*[h]{ if false, check children for pruning instead}){\textup{Prune = False}}{
 					 			$l\leftarrow l + 1$ \tcp*[h]{climb up to the next child level}\\
 					\For {j=1 \ldots 4}{
 						${\textproc{\textbf{Traverse}}} (\boldsymbol{\Xi}_j^l(\mathcal{W}_{\phi}),\: \tilde{\boldsymbol{\phi}},\: \tilde{\boldsymbol{\gamma}})$
 				}}
 			  \If{\textup{Prune = True}}{
 			  	$\prescript{}{i}{\tilde{\boldsymbol{\Gamma}}}_{j}^l\leftarrow  \prescript{}{i}{\tilde{\boldsymbol{\gamma}}}_{j}^l$ and
 			  	$\: \prescript{}{i}{\tilde{\boldsymbol{\Phi}}}_{j}^l\leftarrow  \prescript{}{i}{\tilde{\boldsymbol{\phi}}}_{j}^l$  \tcp*[h]{the pre-subscript $i$ denotes "belongs to the $i^{\textup{th}}$ target"}} 
 				
 			}
 		}

} 			

 		}
 		
$\tilde{\boldsymbol{\Phi}}$ $\leftarrow$ 	\textproc{\textbf{UniqueSearch}}($\tilde{\boldsymbol{\Phi}}$) and
$\tilde{\boldsymbol{\Gamma}}$ $\leftarrow$ 	\textproc{\textbf{UniqueSearch}}($\tilde{\boldsymbol{\Gamma}}$) \\

 	}
 	
 \end{algorithm}

\newpage

\begin{algorithm}[h!]
	\caption[]{\textproc{\textbf{checkPrune}}; Perform Barnes--Hut or neighbor search pruning. 	\label{Pruning}}
	\KwInput{ Tree data structure of the weighted POD space, $\boldsymbol{\Xi}(\mathcal{W}_{\phi})$;  $\theta$ =  Barnes--Hut pruning ratio; or $p_c$ \\ neighborhood scaling factor, target position vector, $\bm{s}_i^{\tau}$,  current source cluster position vector, $\bm{s}^{\tilde{\zeta}}$, leaf node $\mathcal{L}_i$ containing $\tau$ }  
	\KwOutput{Cluster decision Boolean, i.e. ``True = prune" or ``False = do not prune" } 
   {Let $\breve{\mathcal{N}_i^l}=\{\boldsymbol{\Xi}_i^{l}(\mathcal{W}_{\phi})\}$ denote a set corresponding to a source node $i$ at level $l$. Next, let $\bm{s}$ be the position vector of the weighted mean of particles in $\breve{\mathcal{N}_i^l}$ computed by the function \textbf{{\textproc{WeightedMean}}}. }\\
	\If {\textup{Prune Technique = ``Barnes--Hut"}}{
		$w_{\breve{\mathcal{N}_{\zeta}}} \leftarrow $ query the source node width $\breve{\mathcal{N}_i^l}$ from $\Xi(\mathcal{W}_{\phi})$ see \textbf{Def. 2}\\
		\If {$w_{\breve{\mathcal{N}_{\zeta}}} /\lVert \bm{s}^{\tilde{\zeta}}-\bm{s}^{\tau}_i \rVert_2 \leq \theta$}{
			Prune = True}
		\ElseIf{$w_{\breve{\mathcal{N}_{\zeta}}} / \lVert \bm{s}^{\tilde{\zeta}}-\bm{s}^{\tau}_i \rVert_2 \geq \theta$}{
			Prune = False}}
	\ElseIf{\textup{Prune Technique = ``neighbor search"}}{
		{$w_{\breve{\mathcal{N}_{\tau}}}\leftarrow$  query the target node width  $\mathcal{L}_i$ from $\Xi(\mathcal{W}_{\phi}),$ see \textbf{Def. 2}}\\
     { $\bm{b}^{\breve{\mathcal{N}_{\tau}}}:=\{{b}^{\breve{\mathcal{N}_{\tau}}}_{\chi,\textup{min}}, {b}^{\breve{\mathcal{N}_{\tau}}}_{\chi,\textup{max}},{b}^{\breve{\mathcal{N}_{\tau}}}_{\psi,\textup{min}},{b}^{\breve{\mathcal{N}_{\tau}}}_{\psi,\textup{max}}\}$ $\leftarrow$ query the target node boundaries $\mathcal{L}_i$ from $\Xi(\mathcal{W}_{\phi}),$ see \textbf{Def. 2}}\\
		{$\bm{b}^{\breve{\mathcal{N}_{\zeta}}}:=\{{b}^{\breve{\mathcal{N}_{\zeta}}}_{\chi,\textup{min}}, {b}^{\breve{\mathcal{N}_{\zeta}}}_{\chi,\textup{max}},{b}^{\breve{\mathcal{N}_{\zeta}}}_{\psi,\textup{min}},{b}^{\breve{\mathcal{N}_{\zeta}}}_{\psi,\textup{max}}\}$ $\leftarrow$ query the source node boundaries $\breve{\mathcal{N}_i^l}$ from $\Xi(\mathcal{W}_{\phi})$, see \textbf{Def. 2}}\\
		{$\mathcal{H}(\bm{b}^{\mathcal{\breve{N}_{\tau}}},\bm{w}(p_c, w_{\breve{\mathcal{N}_{\tau}}}))=\{{h}_{\chi,\textup{min}},{h}_{\chi,\textup{max}},{h}_{\psi,\textup{min}},{h}_{\psi,\textup{max}}\}$	$\leftarrow$ define the neighborhood of the target, see \textbf{Def. 2}. }\\ 
		\tcp*[h]{Check overlap between target and source cells:}\\
		overlap$_{\chi}=h_{\chi,\textup{max}}> b_{\chi,\textup{min}}^{\breve{\mathcal{N_{\zeta}}}}$ and $h_{\chi,\textup{min}}< b_{\chi,\textup{max}}^{\breve{\mathcal{N_{\zeta}}}}$ \\
		overlap$_{\psi}= h_{\psi,\textup{max}}> b_{\psi,\textup{min}}^{\breve{\mathcal{N_{\zeta}}}}$ and $h_{\psi,\textup{min}}< b_{\psi,\textup{max}}^{\breve{\mathcal{N_{\zeta}}}}$ \\
	}
	\If{\textup{(overlap}$_{\chi}$ \textup{and overlap}$_{\psi}$\textup{)} $=$ \textup{False}}{
		Prune = True}
	\ElseIf{\textup{(overlap}$_{\chi}$ \textup{and overlap}$_{\psi}$\textup{)} $=$ \textup{True}}{
		Prune = False
	}
\end{algorithm}

\section{Projection-tree reduced order modeling} \label{Section_PTROM}

\begin{figure}[h!]
	\includegraphics[scale=0.8, trim=1cm 10cm 0 7cm]{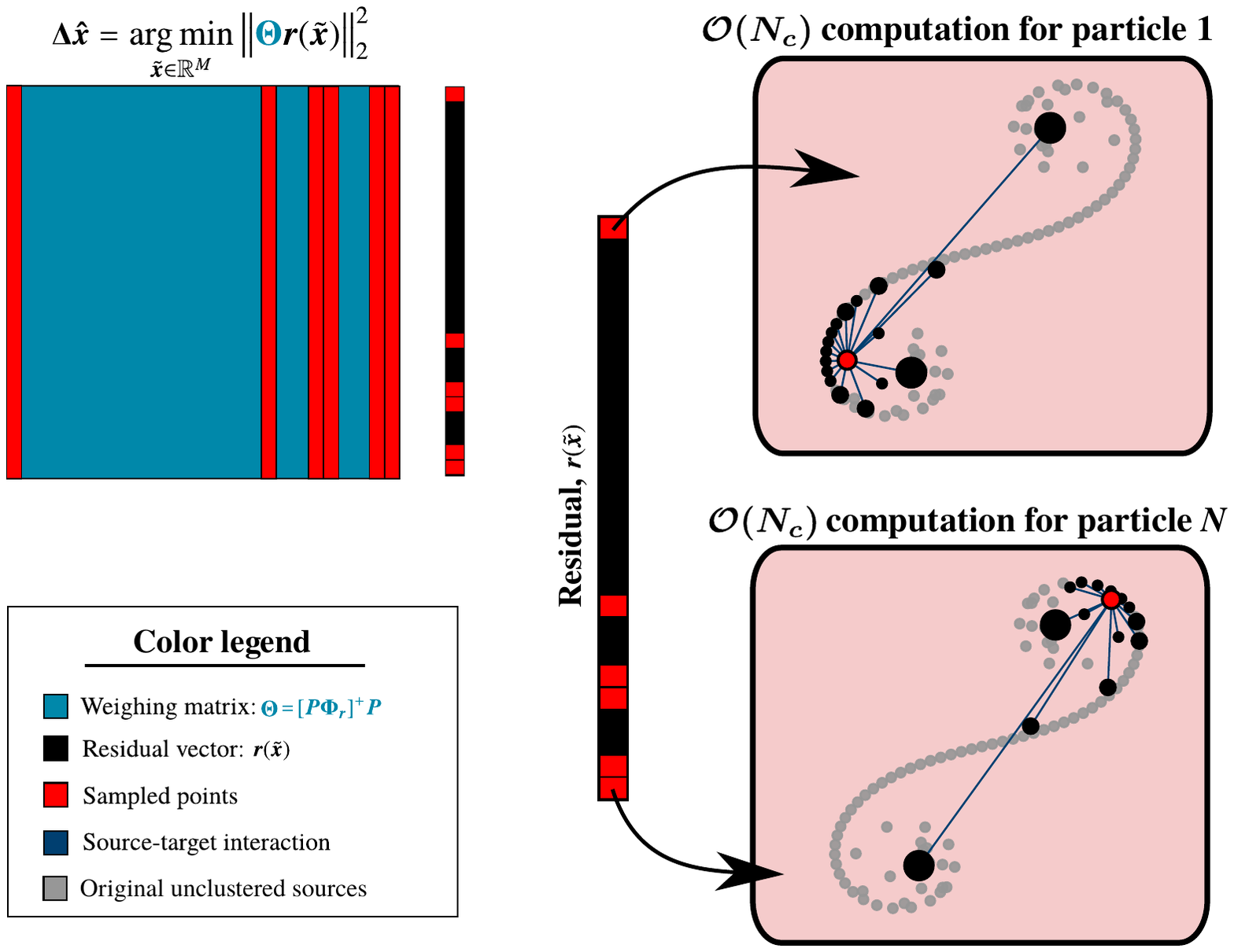}
	\caption{Illustration of the PTROM hyper-reduction and source clustering approach applied to an $N$-body pairwise interaction computational framework.}
	\label{PTROM_schematic}
\end{figure}

The combination of projection-based dimensionality reduction, hyper-reduction, and tree-based hierarchical decomposition constitutes the presented ``projection-tree reduced order model" to rapidly compute $N$-body problems. Figure \ref{PTROM_schematic} illustrates the underlying concept of the PTROM framework, which is a sparse residual minimization problem in a low-dimesional embedding over a clustered set of sources. In this section, a discussion on the training hierarchy is presented, along with a presentation of the deployed online computations, discussions on the $N-$independent OCC, and error bounds of the resulting framework.

\subsection{Training hierarchy}
To execute the presented PTROM, an offline training stage must be executed to collect the low-dimensional bases of the state vectors $\boldsymbol{\Phi}$ and residuals $\boldsymbol{\Phi}_{\bm{r}}$, perform sparse greedy sampling of the particles, and to compute the source POD basis surrogate. The offline training is comprised of the following four-stages:
\begin{itemize}[align=parleft, labelwidth=4em,leftmargin =\dimexpr\labelwidth+\labelsep\relax]
	\item [Stage 1:]{Perform the full-order pair-wise interaction model over sampled points in parametric space $\mathcal{D}$ and collect the state-vector time-history}
	\item [Stage 2:]{Compute the POD basis, $\boldsymbol{\Phi}$, of the parametric state-vector snapshot matrix $\boldsymbol{\mathcal{S}}$ and perform hierchical decomposition to collect the source POD basis surrogate $\tilde{\boldsymbol{\Phi}}$ via Algorithm \ref{PODsurrogateAlgorithm}.}
	\item [Stage 3:]{Perform a least-squares Petrov--Galerkin simulation over the parametric points in $\mathcal{D}$ and employ the reduced bases to compute the target state-vector approximation, $\tilde{\bm{x}}(\bm{x}^0+ {\boldsymbol{\Phi}} \hat{\bm{x}})$, and the source surrogate state-vector, $\tilde{\bm{x}}^{\textup{QT}}(\bm{x}^0+ {\tilde{\boldsymbol{\Phi}}} \hat{\bm{x}})$. Then build a snapshot matrix of the residual vector for each iteration over all time steps, and construct the resulting residual vector POD basis, $\boldsymbol{\Phi}_{\bm{r}}$.}
	\item[Stage 4:]{Perform Algorithm \ref{samplingAlgorithm} to sample particles that enable online hyper-reduction.}
\end{itemize}

\begin{figure}[t]
	\includegraphics[scale=0.5, trim=-2cm 6cm 0cm 5cm]{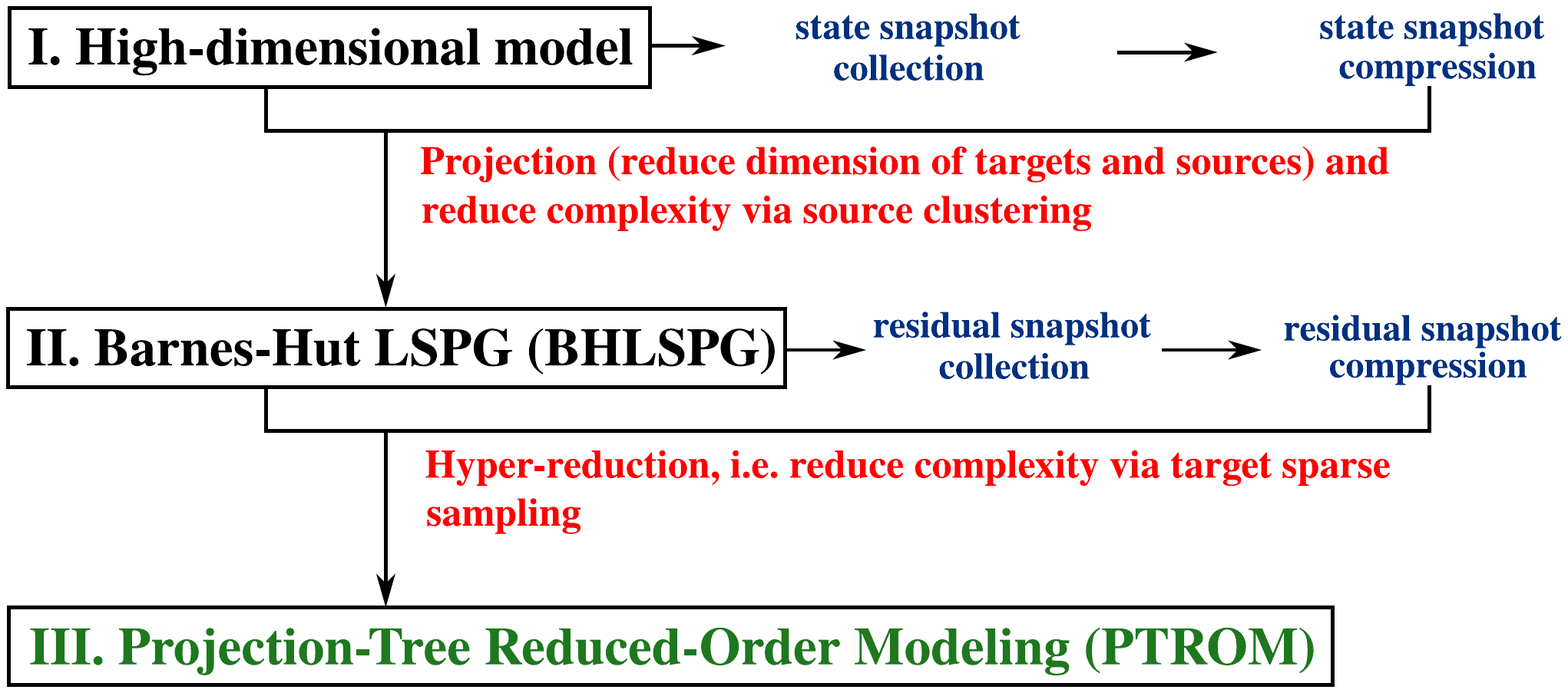}
	\caption{Model hierarchy, where data post-processing procedures are shown in blue text, approximations are shown in red, and the PTROM online procedure is shown in green. Adapted from \cite{carlberg2013gnat} }
	\label{ModelingHierarchy}
\end{figure}

The resulting offline training stages generate a modeling hierarchy, shown in Fig.~\ref{ModelingHierarchy} similar to that presented in \cite{carlberg2013gnat}. However, the modeling hierarchy for the PTROM includes additional complexity reduction stages required to achieve $N$-independence for the Lagrangian $N$-body framework as opposed to the original GNAT method developed for grid-based methods presented in \cite{carlberg2013gnat}. 

It is important to note that the current training procedure employed for the PTROM requires data from the Tier I and II models to gather underlying information about the residual vector POD basis under the approximation of the clustered sources. As a result, it is not an option in the presented PTROM to employ tier I modeling as the only training run as it is in \cite{carlberg2013gnat} (see Section 3.4.2).

\subsection{Online computations}
The online deployment of the PTROM is described by Algorithms \ref{GNATloop} and \ref{hyperPair}. In this study, the kernel under consideration employs the circulation of individual particles as the parametric variable. As a result, it is necessary to update the source POD clusters with the associated parametric variations of circulation, $\boldsymbol{\Gamma}_{\mathbf{\mu}}$. However, no additional training is required to update the clustered sources, as the hierarchical data-structure contains particle identification inside of clusters to reassign cluster circulation variations. Other parametric studies, such as performing a parametric sweep of super-imposed inflow conditions, would not require this step.
\begin{algorithm}[t]
	
	\KwInput{ POD surrogate source basis, $\tilde{\boldsymbol{\Phi}}$; POD basis $\boldsymbol{\Phi}$; Offline-computed hyper-reduction matrix, ${\bm{A}}=$\\ $\left[{\bm{P}}\bar{\boldsymbol{\Phi}}_{\bm{r}}\right]^{+}$; Initial conditions $\tilde{\bm{x}}^0$ and $\tilde{\bm{f}}^0$. Newly loaded circulation clustered vector $\tilde{\boldsymbol{\Gamma}}_{\mathbf{\mu}}$. \textbf{Note:} an over-bar denotes the minimum cardinality of a vector.}
	\KwOutput{Time histories of approximate state vectors $\bar{\tilde{\bm{x}}}$, approximate velocity kernel $\bar{\tilde{\bm{f}}}$ at sampled entries, and \\ generalized coordinates, $\hat{\bm{x}}$}
	$k=1$ \tcp*[h]{initialize Gauss-Newton loop iteration counter}\\
	$\textup{tol}=$ user-defined tolerance\\
   \For(\tcp*[h]{loop over all time steps}){$n=1\ldots N_t$}{
	\While{ $\epsilon< \textup{tol}$}{
		$[\bar{\bm{r}}, \bar{\bm{J}}]=\textproc{\textbf{HyperPair}}({\bar{\boldsymbol{\Phi}}}, \tilde{\boldsymbol{\Phi}},\tilde{\boldsymbol{\Gamma}}_{\mathbf{\mu}}, \hat{\bm{x}}, \bar{\tilde{\bm{x}}},\bar{\tilde{\bm{f}}}) $ \tcp*[h]{compute the hyper-reduced pairwise interaction}\\  \vspace{0.175cm}
		${\bm{C}}={\bm{P}}\bar{\bm{J}}(\bar{\tilde{\bm{x}}}_k^n)\bar{\boldsymbol{\Phi}}$ and
		${\bm{D}}=\bar{\bm{P}}\bar{\bm{r}}(\bar{\tilde{\bm{x}}}_k^t)$\\  \vspace{0.175cm}
${\Delta\hat{\bm{x}}}= \argminl_{\nu\in \mathbb{R}^{M}} \lVert {\bm{A}}{\bm{C}}\nu+{\bm{A}}{\bm{D}}\rVert_2$ \tcp*[h]{compute the linear least-squares problem equivalent to Eq.~\ref{GNATminimization}/\ref{GNATminimization2}}\\ \vspace{0.175cm}
		$\hat{\bm{x}}_{k+1}^n=\hat{\bm{x}}_{k}^n+{\alpha\Delta\hat{\bm{x}}}_k$ and $\bar{\tilde{\bm{x}}}_{k+1}^{n}=\bar{\tilde{\bm{x}}}_{k}^{n}+\alpha\bar{\boldsymbol{\Phi}}{\Delta\hat{\bm{x}}}$ \tcp*[h]{$\alpha$ is computed via line-search or set to 1}\\ \vspace{0.175cm}
		$\epsilon = \lvert\lvert \bar{\boldsymbol{\Phi}}^{\textup{T}}\bar{\bm{J}}(\bar{\tilde{\bm{x}}}_k^n)^{\textup{T}}\bar{\bm{r}}(\bar{\tilde{\bm{x}}}_k^n) \rvert\rvert_2/  \lvert\lvert \bar{\boldsymbol{\Phi}}^{\textup{T}}\bar{\bm{J}}(\breve{\bar{\bm{x}}})^{\textup{T}}\bar{\bm{r}}(\breve{\bar{\bm{x}}})\rvert\rvert_2$, where $\breve{\tilde{\bm{x}}}=\bar{\tilde{\bm{x}}}_1^n$ \tcp*[h]{check relative reduced residual error}\\ \vspace{0.175cm}
		$k\leftarrow k + 1$
	}
	$k=1$
}
	\caption{\textproc{\textbf{Gauss-Newton}}: Online Gauss-Newton residual minimization loop 	\label{GNATloop}}
\end{algorithm}

\begin{algorithm}[h!]
	\KwInput{ POD surrogate source basis, $\tilde{\boldsymbol{\Phi}}$; POD basis, $\boldsymbol{\Phi}$; generalized coordinates, $\Delta \bm{x}$; sampled state vector \\ entry, $\bar{\tilde{\bm{x}}}$; approximate velocity at sampled entries from prior iterations and time steps, $\bar{\tilde{\bm{f}}}$; clustered source circulation $\tilde{\boldsymbol{\Gamma}}$}
	\KwOutput{Approximate velocity, $\bar{\tilde{\bm{f}}}$, residual $\bar{\bm{r}}$, and residual Jacobian $\bar{\bm{J}}$ all at sampled entries.}
	{Here, $\bm{x}^{\textup{QT}}:=\tilde{\mathbf{\Phi}}\Delta\hat{\bm{x}}_k$, where  $\{{x}_i^{\textup{QT}}, {x}_{i+N}^{\textup{QT}}\}^T \mapsto \{\chi_i^{\textup{QT}},\psi_i^{\textup{QT}}\}$. Next, $\bm{x}^{\textup{GP}}:={\bar{{\mathbf{\Phi}}}}\Delta\hat{\bm{x}}_k$, where $\{{x}_i^{\textup{GP}}, {x}_{i+N}^{\textup{GP}}\}^T \mapsto \{\chi_i^{\textup{GP}},\psi_i^{\textup{GP}}\}$. \textbf{Note}: A prescript $i$ denotes ``belonging to the $i^{\textup{th}}$ particle", e.g.~$\prescript{}{i}{\bm{x}^{\textup{QT}}}$ denotes the source cluster locations as observed by the $i^{\textup{th}}$ particle}\\
	{$\tilde{\boldsymbol{\Gamma}}\leftarrow\tilde{\boldsymbol{\Gamma}}_{\mathbf{\mu}}$ \tcp*[h]{Load parametric clustered circulation}}\\
	\For{$i=1 \ldots n$}{
		$N_c=\lvert\prescript{}{i}{\bm{x}^{\textup{QT}} }\rvert/d$ \tcp*[h]{this refers to the number of source clusters $N_c$ observed by particle $i$}\\
		\For{$j=1\ldots N_c$} {
			$r_{\chi}={\chi}^{GP}_i-\prescript{}{i}{{\chi}^{QT}_j}$, \tcp*[h]{compute the $\chi$-component distance between target and source} \\
			$r_{\psi}={\psi}^{GP}_{i}-\prescript{}{i}{{\psi}^{QT}_{j}}$, \tcp*[h]{compute the $\psi$-component distance between target and source}\\
			$\boldsymbol{\tau}_\textup{num}=\tilde{\Gamma}_j\:\left( \hat{\mathbf{e}}_3  \times \bm{r}\right),$ where $\bm{r}=\{r_{\chi},r_{\psi},0\}\:$ and  $\hat{\mathbf{e}}_3=\{0,0,1\}$\\
			$\tau_\textup{den}=2\pi d^2 +\delta_K,$ where $d=\sqrt{r_{\chi}^2+r_{\psi}^2}$\\
			$\bm{k}_i=\sum_{j\neq j}
			\boldsymbol{\tau}_{\textup{num}}/\tau_{\textup{den}}$, \tcp*[h]{ compute the velocity kernel of the target, $i$, and sum over all sources}\\
			$\{k_{i,x},k_{i,y}\} \mapsto \{f_{i},f_{i+N}\}$ \tcp*[h]{map kernel output components to velocity vector}\\
		}
	}
	$\bar{\tilde{\bm{f}}}_k^{t}\leftarrow \bar{\tilde{\bm{f}}}$  \tcp*[h]{Update velocity kernel}\\
    $\bar{\bm{J}}_{\textup{vel}} = \partial \tilde{\bm{f}}/ \partial \bm{r}$, \tcp*[h]{compute the inexact Jacobian, as discussed in Section \ref{Section_ProblemFormulation}} \\
	$\bar{\bm{J}}_k=\bar{\bm{I}}-\frac{\Delta t}{2}\bar{\bm{J}}_{\textup{vel}} $, \tcp*[h]{Compute the residual Jacobian. Note $\bar{\bm{I}}$ is the identity matrix at sampled entries}\\ 
	$\bar{\bm{r}}_k^{t}=\bar{\tilde{\bm{x}}}^{t}_k-\bar{\tilde{\bm{x}}}^{t-1}-\frac{\Delta t}{2}\left(\bar{\tilde{\bm{f}}}^{t}_k+\bar{\tilde{\bm{f}}}^{t-1}\right)$.  \tcp*[h]{Compute the residual Jacobian}
	\caption{\textproc{\textbf{HyperPair}}; Computation of the hyper-reduced pair-wise interaction of the Biot-Savart kernel}
	\label{hyperPair}
\end{algorithm}

\subsection{Computing outputs}
Recall that the PTROM computes state-vector approximations over a sparsely sampled set of particles in the domain. To access all state-vector time-history data, it would be required to compute the POD basis back-projection of the generalized coordinate over all degrees-of-freedom, which incurs an $N$-dependent operational count complexity that scales like $\mathcal{O}(N_dM)$.

However, it's important to note that in many engineering applications of Lagragian $N$-body methods that not all particle time-histories contribute to a quantity of interest. For instance, application of the free-vortex wake method \cite{rodriguez2019strongly, rodriguez2020strongly} traditionally only requires the velocity time-history of a sparse set of particles in the domain, which correspond to lifting surface particles, to compute quantities of interest (QoIs), such as lift and drag coefficients. As a result, the computations of the QoIs can be efficiently executed by projecting the generalized coordinates onto the degrees-of-freedom needed to compute the desired output quantity. Here, let $n_q$ denote the sparse degrees-of-freedom required to compute the QoI. Thus, to access the required subset of degrees-of-freedom 
to compute a QoI will incur $\mathcal{O}(n_qM)$ operations, which is small when $n_q\ll N$. For more information on efficient post-processing and output computations, the reader is referred to \cite{carlberg2013gnat}.

\subsection{Error bounds}
The current PTROM is considered a variant of LSPG projection, and as a result \emph{a posterior} error bounds have been derived in \cite{carlberg2013gnat, lee2020model, parish2020time}, that are directly applicable to the presented framework. In this work, the generalized error bounds presented in \cite{parish2020time} are applicable since they account for \emph{any} arbitrary sequence of approximated solutions $\tilde{\bm{x}}$ and quantity of interest (QoI) functionals (e.g.~functions that output the Hamiltonian, lift coefficient, drag coefficient, etc.). We now briefly present the error bounds presented in \cite{parish2020time}, within the context of the formalisms presented in this work for Lagrangian $N$-body dynamical systems:\\

\noindent Let $q^n$ be some QoI computed by the PTROM where 
\begin{equation}
{q}^n: \boldsymbol{\mu}\mapsto \mathcal{G}({\bm{x}}(\boldsymbol{\mu}),t;\boldsymbol{\mu}), \textup{ and}
\end{equation}
\begin{equation}
\tilde{q}^n: \boldsymbol{\mu}\mapsto \mathcal{G}(\tilde{\bm{x}}(\boldsymbol{\mu}),t;\boldsymbol{\mu})
\end{equation}
where ${q}^n:\mathcal{D}\rightarrow \mathbb{R}$  , $\tilde{q}^n:\mathcal{D}\rightarrow \mathbb{R}$ and $\mathcal{G}: \mathbb{R}^{N_d} \times [0, T] \times \mathcal{D}\rightarrow \mathbb{R}$ denotes the QoI functional. Next, let the normed state error and the quantity of interest errors by defined as:
\begin{align}
&\delta_{\bm{x}}^n(\boldsymbol{\mu}):= \big \lVert \bm{x}^n(\boldsymbol{\mu}) - \tilde{\bm{x}}^n(\boldsymbol{\mu}) \big \rVert_{2}, \: n=1 \ldots N_t, \textup{ and}\\
&\delta_{\bm{q}}^n(\boldsymbol{\mu}):= q^n(\boldsymbol{\mu}) - \tilde{q}^n(\boldsymbol{\mu}),\: n=1 \ldots N_t,
\end{align}
where $\delta_{\bm{x}}^n(\boldsymbol{\mu})$ and $\delta_{\bm{q}}^0(\boldsymbol{\mu})$ are computed explicitly from the initial conditions $\tilde{\bm{x}}^0(\boldsymbol{\mu})$ and ${\bm{x}}^0(\boldsymbol{\mu})$. Then, the error bounds for the state approximation and some QoI functional are expressed by the following:
\begin{erb}
\normalfont For a given parameter instance $\boldsymbol{\mu}\in \mathcal{D}$, if the kernel, $\bm{f}$ (in this case the Biot-Savart kernel and the associated induced velocity), is Lipschitz continuous, i.e.~there is exists a constant  $\kappa > 0$, such that $\lVert\bm{f}(\bm{x},t;\boldsymbol{\mu})-\bm{f}(\bm{y},t;\boldsymbol{\mu})\rVert_{2} \leq \kappa \lVert\bm{x}-\bm{y} \rVert_{2}$ for all $\bm{x}$, $\bm{y} \in \mathbb{R}^{N_d}$ and $t\in\{t^n\}_{i=1}^{N_t}$, and the time step is sufficiently small such that $\Delta t < \lvert \alpha_0 \rvert / \lvert \beta_0 \rvert \kappa$, then the state error bound is defined by\\
\begin{equation}
\delta_{\bm{x}}(\boldsymbol{\mu})\le \frac{1}{h}\big\lVert\bm{r}^n(\tilde{\bm{x}}^n(\boldsymbol{\mu});\tilde{\bm{x}}^{n-1}(\boldsymbol{\mu})), \ldots \tilde{\bm{x}}^{n-\breve{k}}(\boldsymbol{\mu}),\boldsymbol{\mu}) \big\rVert_2 + \sum_{j=1}^{\breve{k}^n}\eta_j\delta^{n-j}_{\bm{x}}(\boldsymbol{\mu})
\end{equation}
for all $n=1,\ldots, N_t$. Here, $h:=\lvert \alpha_0 \rvert - \lvert \beta_0 \rvert \kappa \Delta t$ and $\eta_j:=(\lvert \alpha_j \rvert - \lvert \beta_j \kappa \Delta t\rvert)/h$. Next, regarding the QoI error bound: If the QoI functional $\mathcal{G}$ is Lipschitz continuous, i.e., there is exists a constant $\kappa_{\mathcal{G}} > 0$ such that $\big\lvert \mathcal{G}(\bm{x};t,\boldsymbol{\mu}) - \mathcal{G}(\bm{y};t,\boldsymbol{\mu}) \big\rvert \leq \kappa_{\mathcal{G}} \lVert\bm{x}-\bm{y} \rVert_{2}$ for all $\bm{x},\:\bm{y}\in\mathbb{R}^N$ and $t\in\{t^n\}^{N_t}_{i=1}$, then the QoI error bound is defined by\\
\begin{equation}
\big\lvert \delta_{{q}}(\boldsymbol{\mu}) \big \rvert \le \frac{\kappa_{\mathcal{G}} }{h}\big\lVert\bm{r}^n(\tilde{\bm{x}}^n(\boldsymbol{\mu});\tilde{\bm{x}}^{n-1}(\boldsymbol{\mu})), \ldots \tilde{\bm{x}}^{n-\breve{k}}(\boldsymbol{\mu}),\boldsymbol{\mu}) \big\rVert_2 + \kappa_{\mathcal{G}} \sum_{j=1}^{\breve{k}^n}\eta_j\delta_{q}^{n-j}(\boldsymbol{\mu})
\end{equation}
for all $n=1,\ldots, N_t$. For proof and derivations of these error bounds the reader is referred to \cite{parish2020time,lee2020model, carlberg2017galerkin}.
\end{erb}

\subsection{Operational count complexity}
As mentioned before, the operational count complexity of the online PTROM, i.e.~Algorithms \ref{GNATloop} and \ref{hyperPair} is $N$-independent, which is enabled by both hyper-reduction of target residuals and clustering of the sources. First, it was found that Algorithm \ref{GNATloop} is limited by the OCC associated with solving the linear least-square problem in Algorithm \ref{GNATloop}, such that the OCC scaling goes like $\mathcal{O}(\bar{k}M(M_{\bm{r}}M+M^2))$, where $\bar{k}$ is the mean iteration count of the Gauss-Newton loop, and $\bar{k}, M_{\bm{r}}, M \ll N$.  Next, it was found that in Algorithm \ref{hyperPair} the pair-wise interaction problem between hyper-reduced targets and clustered sources scaled like $\mathcal{O}(\breve{n}\bar{N}_c)$, where $\bar{N}_c$ denotes the mean of source clusters that influence the hyper-reduced targets. Recall that $\breve{n}, N_c \ll N$. However, it was also found that the projection of the generalized coordinates on to the source POD surrogate matrix or onto the POD matrix at the sampled target degrees-of-freedom scales like $\mathcal{O}(\bar{N}_cM)$ or $\mathcal{O}({n}_dM)$. As a result the leading complexity in Algorithm \ref{hyperPair} is associated with user-defined hyper-reduction, dimensional compression, or hierarchical clustering. In other words, depending on the reduction parameters, the OCC of Algorithm \ref{hyperPair} is defined by either $\mathcal{O}(\breve{n}\bar{N}_c)$, $\mathcal{O}(\bar{N}_cM)$, or $\mathcal{O}({n}_dM)$. Ultimately, the PTROM method has a leading OCC of $\mathcal{O}(\bar{k}M(M_{\bm{r}}M+M^2))$ associated with solving the linear least-squares problem, and where $M_{\mathbf{r}}$ and $M$ are user-defined in the residual and state-vector POD basis construction, and $k$ is a tolerance user-defined parameter.

The PTROM $N$-independent OCC  has been an important feature not traditionally available with hierarchical decomposition methods. Traditional acceleration methods like the Barnes--Hut method \cite{barnes1986hierarchical} or the FMM \cite{greengard1987fast, martinsson2007accelerated} have at best scaled with linear dependence or $\mathcal{O}(N \log^{\breve{d}-1}(1/\epsilon))$ \cite{Martinsson2015} (but more precisely scaled with the degrees-of-freedom $\mathcal{O}(N_d \log^{\breve{d}-1}(1/\epsilon))$, where $\breve{d}$ is the dimensionality considered, i.e.~2-$d$ or 3-$d$ kernel, and $\epsilon$ is the error tolerance of the FMM acceleration algorithm . A closer investigation into the PTROM OCC highlights that the overall operational count $\mathcal{O}(\bar{k}M(M_{\bm{r}}M+M^2))$, though independent of $N$, can at times exceed linear scaling operation, $\mathcal{O}(N)$, operations by some factor when user-defined defined tolerances are stringent and the POD bases are of high rank. Nevertheless, the PTROM provides the user the capability of controlling the OCC which scales independent of $N$ and has the capabilities of being lower than a linear OCC scaling, with respect to the number of particles in the domain.

\section{Implementation of the projection-tree reduced order model}
\label{Section_Results}

All of the presented work was computed on MATLAB on an Intel NUC equipped with eight Intel Core i7-
8559U @ 2.70 GHz. However, all FOM, Barnes--Hut, PROM, and PTROM computations were instructed to perform on a single core to perform serial computations via the \code{maxNumCompThreads} command on MATLAB. Future work will incorporate lower-level languages, such as C++, and parallelization to present more applicable savings in CPU hours and for larger scale numerical experiments. The codes employed to generate the performance analysis and parametric investigations were optimized with the built-in code profiler toolbox available in MATLAB. It is also, important to note that all pair-wise interaction algorithms inside of modeling frameworks, i.e.~FOM, Barnes--Hut, LSPH, BHLSPG, GNAT, PTROM employed the same for-loop architecture to generate a consistent comparison between computational methods and their wall-time savings. All inexact Newton loops in the FOM computations and Gauss--Newton loops for the GNAT and PTROM computations were set to converge within $k\le100$ iterations. The GNAT and PTROM computations set $\alpha=1$ as the Gauss--Newton line-search step length. Finally, it was found that building the PTROM surrogate basis, $\tilde{\bm{\Phi}}$ via the Barnes-Hut clustering approach in MATLAB required high levels of random-access memory beyond the resources on the machine used for this investigation. As a result, only the neighbor clustering approach is performed for the PTROM in the proceeding numerical experiments. Future work will entail investigating best clustering approaches on less memory intensive platforms and lower-level languages.

\subsection{Performance metrics and preliminaries}
To assess the application and performance of the PTROM, two parametric numerical experiments and one reproductive experiment were executed. The PTROM's ability to predict the Hamiltonian of the Biot--Savart dynamical system and FOM individual particle path trajectory are the quantities used to measure the method's efficacy in both parametric and reproductive settings. 

The Hamiltonian of the Biot-Savart dynamical system is defined as
\begin{equation}
H^n=\frac{1}{4\pi}\sum_{j}^{N} \sum_{i\neq j}^{N} \Gamma_j \Gamma_i \log \left[ \sqrt{(\chi_j^n-\chi_i^n)^2+(\psi_{j}^n-\psi_{i}^n)^2} \: \right].
\end{equation}
The PTROM Hamiltonian error is measured by the absolute relative error with respect to the FOM Hamiltonian such that

\begin{equation}
\textup{AE}_{H}= \left \vert \frac{H_{\textup{III}}^n-H_{\textup{I}}^n}{H_{\textup{I}}^n}\right \vert,
\label{AE_Hamiltonian}
\end{equation}
where the I and III subscripts denote the tier I and tier III models from Fig.~\ref{ModelingHierarchy}.
 
To measure the accuracy of the PTROM predicted path with respect to the FOM the mean absolute difference of the $l_2$ norm between PTROM particle positions and FOM particle positions is used for a given instant in time normalized by a spatial factor, $l$, as shown below.

\begin{equation}
\textup{MAE}_{D}=\frac{1}{l N}\sum_{i=1}^N \left(\sqrt{ (\chi_{i,\textup{I}}-\chi_{i,\textup{III}})^2+ (\psi_{i,\textup{I}}-\psi_{i,\textup{III}})^2}\right), 
\label{MAE_trajectory}
\end{equation}
where $\chi_{i,\textup{I}}$ and $\psi_{i,\textup{I}}$ denote the tier I model (FOM) position components for the $i^{\textup{th}}$ particle, and $\chi_{i,\textup{III}}$ and $\psi_{i,\textup{III}}$ denote the tier III model (PTROM) position components. The spatial factor, $l$, is used as a relative length scale. In the parametric and reproductive experiments, this relative length scale is defined as the $l_2$ distance between end particles at the initial positions, i.e. $l=\sqrt{(\chi^0_N-\chi^0_1)^2 + (\psi^0_{N}-\psi^0_{1})^2}$.

Computational savings are quantified by the speed-up factor,
\begin{equation}
\textup{SF}=\frac{T_\textup{I}}{T_\textup{III}},
\label{SF_equation}
\end{equation}
where $T_\textup{I}$ is the total wall-time spent on the FOM pair-wise interaction loop and $T_\textup{III}$ is the total wall-time spent on the PTROM Algorithm \ref{GNATloop}, which is the residual minimization loop which also includes the hyperpair-wise interaction loop in Algorithm \ref{hyperPair}.

Finally, the velocity field generated by the FOM particle dynamics is visualized for each experiment by computing the Biot-Savart law on a grid, where each vertex of the grid is a target and each particle is a source. The computation of the velocity field is intended to be an aid and understand the dynamical system of each simulation, and is not part of the meshless domain or method. In addition, to improve the scaling of the visualization, the non-dimensionalized velocity field, $f_g$ is used as the metric mapped on the grid where, 
\begin{equation}
f_{g}= \frac{\lVert \bm{f} \rVert_2 l_g }{\bar{\Gamma}},
\end{equation}
and where $l_g=c_g l$ is the characteristic length scaled up by a factor $c_g$. 

\subsection{Vortex pair parametric experiment}
The PTROM is tested against a parametric numerical experiment where a vortex pair is generated with the Biot--Savart law by assigning high variable circulations to end particles and fixed weak circulations to the remaining particles, i.e. $\Gamma_i \ll \Gamma_1, \Gamma_N$, where $i = 2\ldots N-1$ and in this experiment $N=500$ particles. In the context of the parametric Biot--Savart law, $\bm{\mu} = (\Gamma_1,\Gamma_N)$ are the input parameters and are part of the parametric space $\mathcal{D}=[0.25\Gamma_{\textup{max}}, \Gamma_{\textup{max}}] \times [0.25\Gamma_{\textup{max}}, \Gamma_{\textup{max}}] $, where $\Gamma_{\textup{max}}=255$. All other particles are assigned a circulation strength of $\Gamma_i=0.01$, where $i = 2\ldots N-1$. The initial positions of the particles follow a linear distribution for both $\chi$ and $\psi$ components, starting with $\{\chi_1,\psi_1\}=\{-52.93, -52.93\}$ and ending $\{\chi_N,\psi_N\}=\{52.93, 52.93\}$. Particles have an initial $l_2$ spacing of 0.2121 between each other and the Biot--Savart kernel is assigned a de-singularization cut-off radius $\delta_K$ equal to the initial spacing distance of 0.2121. All simulations were performed in the time interval $t \in [0, 5]$ with $\Delta t = 0.01$ leading to $n_t=500$ time-steps.

\begin{figure}[t]
		\centering
		\includegraphics[scale=0.65, trim=2cm 8.5cm 2cm 8cm]{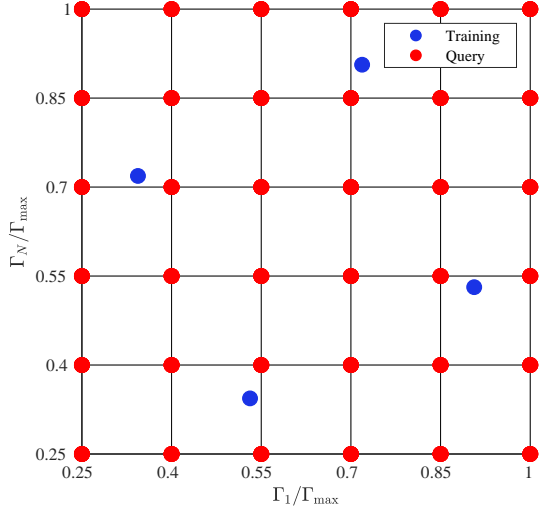}
		\caption[]
		{Parametric space for the vortex pair numerical experiment. Online query points are marked in red and training points are marked in blue. Parametric values are normalized by $\Gamma_{\textup{max}}$ for convenience.}    
		\label{ParametricGrid_Experiment1}
\end{figure}

The PTROM models were trained at four points in the parametric space denoted by the blue points shown in the grid in Fig. \ref{ParametricGrid_Experiment1}. The training points were chosen via the latin hyper-cube sampling method and thirty-six online points were queried online with the PTROM over a $6 \times 6$ grid, where the query points are denoted by the red points in Fig. \ref{ParametricGrid_Experiment1}. All PTROM models in the current experiment were constructed with the following hyper-parameters: $\breve{n} =60, M=85, M_r=110$ and $p_c=1$, which results in $N_c=233$. Here, the Gauss-Newton relative residual error tolerance was set to tol $ =10^{-4}$.

\begin{figure*}[t]
	\centering
   	\begin{subfigure}[b]{0.475\textwidth}  
   	\centering 
   	\includegraphics[scale=0.5, trim=2cm 9cm 25 7cm]{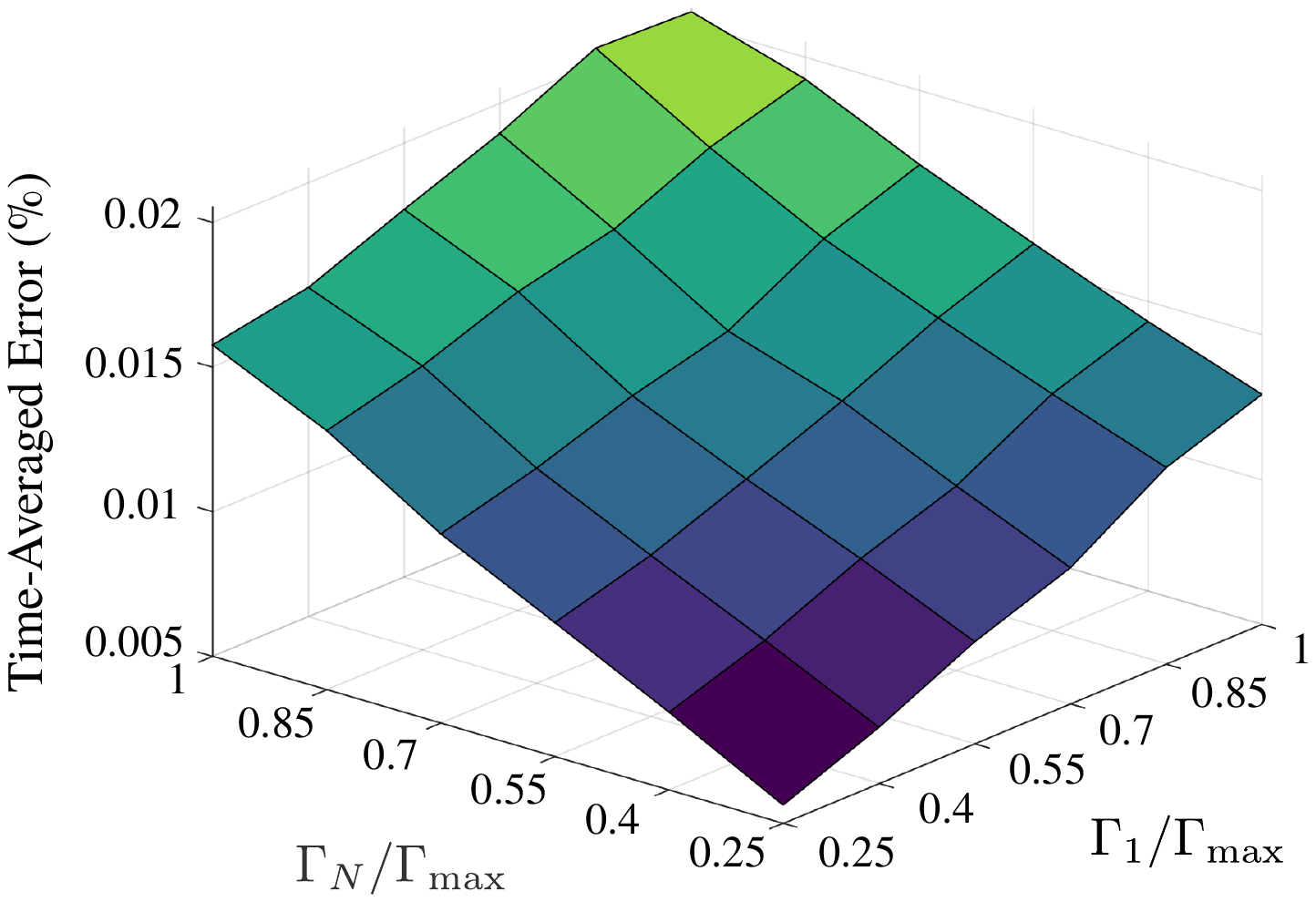}
   	\caption[]%
   	{Average error = 0.0136 \%}
   	\label{MAEP_experiment1}
   \end{subfigure}
   \hfill
   	\begin{subfigure}[b]{0.475\textwidth}  
   	\centering 
   	\includegraphics[scale=0.5, trim=2cm 9cm 25 7cm]{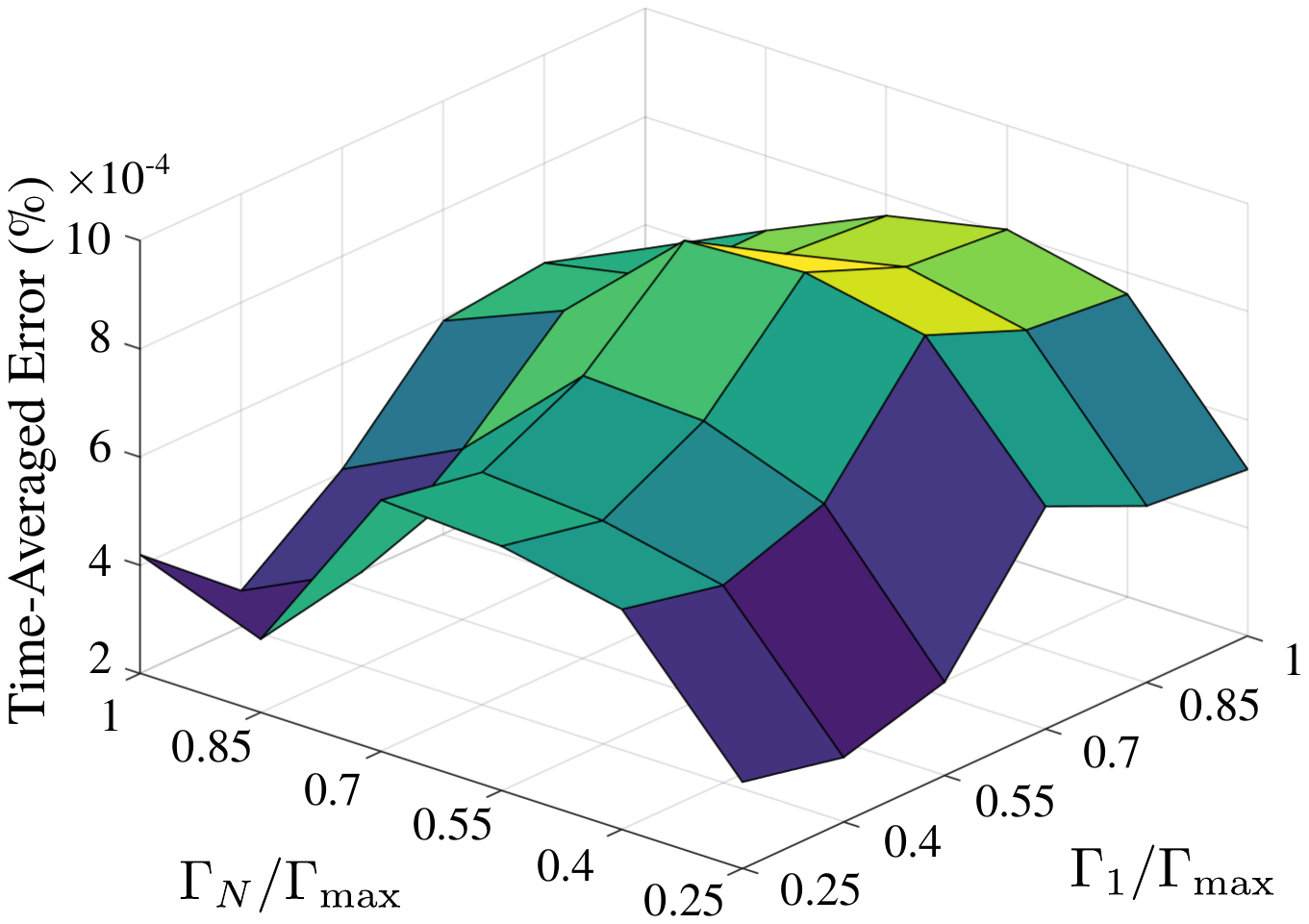}
   	\caption[]%
   	{Average error = $6.09 \times 10^{-4}$ \%}
   	\label{AE_experiment1}
   \end{subfigure}
	\caption[]
	{Left: Surface generated by the time-averaged mean absolute trajectory errors of the online queried points. Right: Surface generated by the time-averaged Hamiltonian absolute error of the online queried points.}  
	\label{Experiment1_ErrorResults}
\end{figure*}

Figure \ref{Experiment1_ErrorResults} show the parametric experimental results. The time-averaged mean trajectory absolute error, normalized by the characteristic length, $l$, has an average value of 0.0136\%. The time-averaged Hamiltonian absolute relative error normalized has an average value of $6.09 \times 10^{-4}$ \%. Results show that the PTROM is capable of predicting FOM QoIs with sub-0.1\% error on average. 

\begin{figure*}[h!]
	\centering
	\includegraphics[scale=0.55, trim=3cm 3cm 50 4cm]{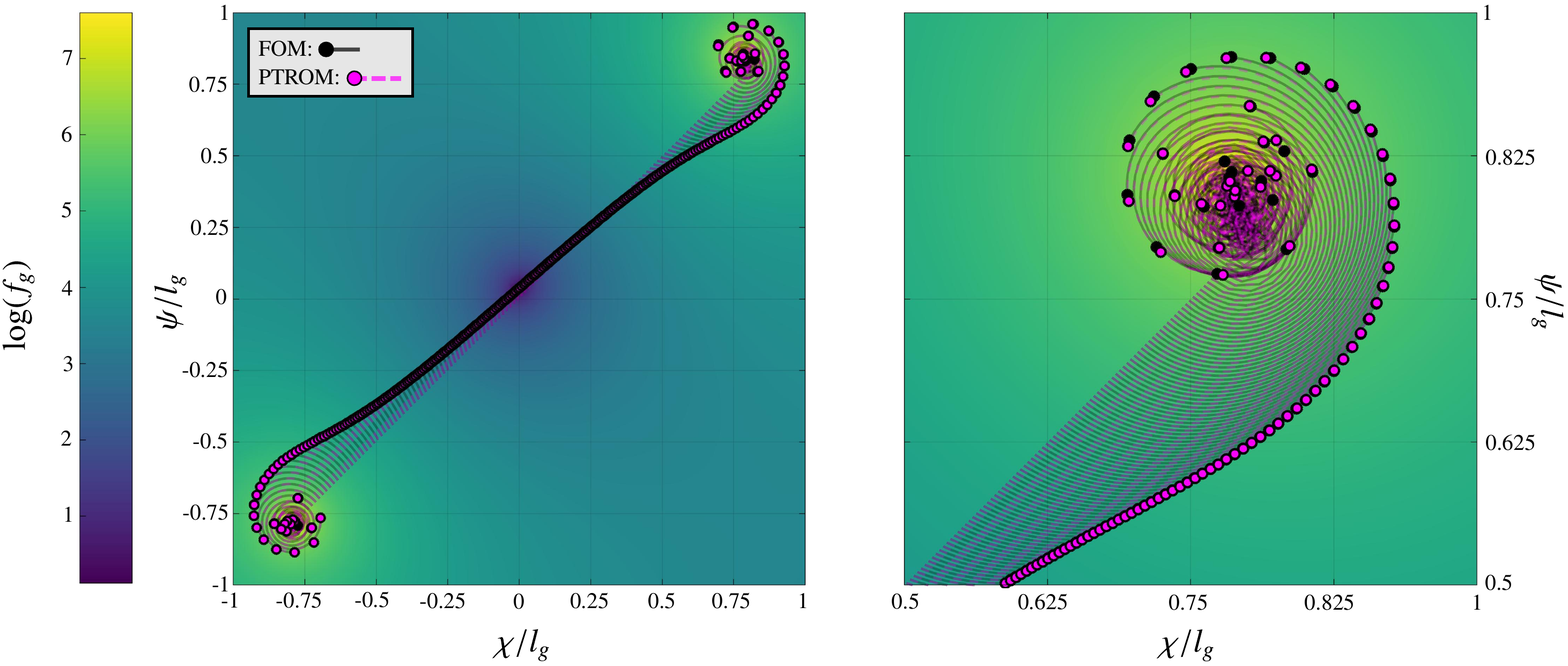}
		\caption[]
	{Vortex particle simulation with its corresponding velocity field at $t=5$. Left: full view of the vortex simulation. Right: zoomed in view of the top right vortex. }
	\label{Experiment1_Simulation}
\end{figure*}

Results presented by the QoI error surfaces in Fig.~\ref{Experiment1_ErrorResults}, are further highlighted by the queried simulation corresponding to parametric values of $\bm{\mu}=(\Gamma_{\textup{max}} ,\Gamma_{\textup{max}})$ in Fig.~\ref{Experiment1_Simulation}. The simulation in Fig.~\ref{Experiment1_Simulation} illustrates the PTROM and FOM (not included in training) particle paths and positions, in addition to the FOM velocity field over a grid with a normalization width of $l_g=1.25l$. Figure \ref{Experiment1_Simulation} shows the multi-scale nature of the vortex pair simulation, where particles near the particle with strong circulation, i.e. particle $\{\chi_1,\psi_1\}$ and $\{\chi_N,\psi_N\}$, orbit the strong particles faster than those further away. The PTROM is capable of following the FOM trajectory with high precision and accuracy for particles not in the neighborhood of the particles with strong circulation, i.e. about $10 \le $ particles away from $\{\chi_1,\psi_1\}$ or $\{\chi_N,\psi_N\}$. Particles in the neighborhood of either $\{\chi_1,\psi_1\}$ or $\{\chi_N,\psi_N\}$ seem to deviate from instantaneous position of the FOM as time passes but are still able to generally follow the path outlined by the FOM.

\begin{figure}[h!]
		\centering
		\includegraphics[scale=0.58, trim=3cm 9cm 2cm 8cm]{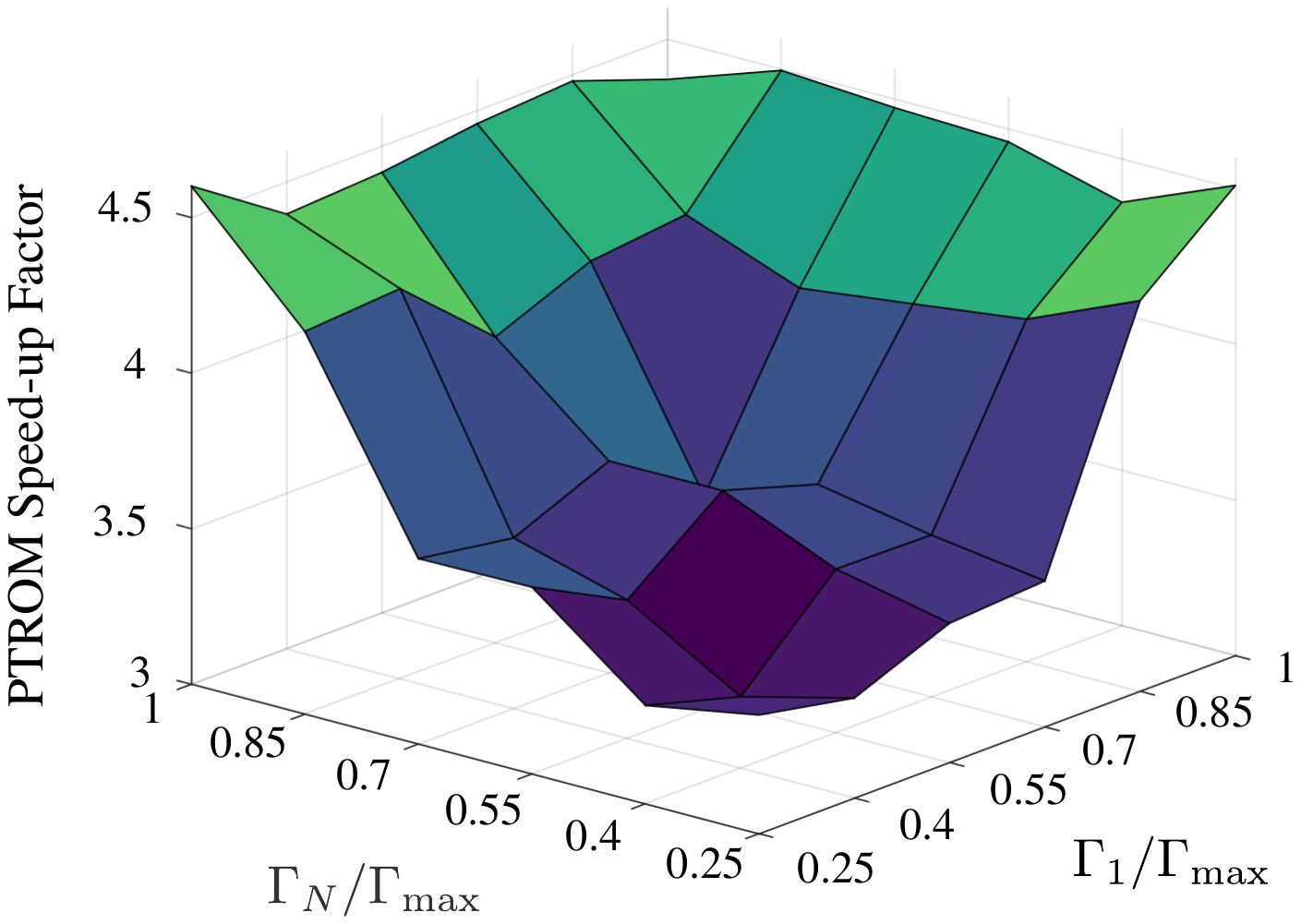}
		\caption[]%
		{Speed-up factor for the vortex pair experiment, where the average speed-up factor = 3.94.}    
		\label{Experiment1SpeedUpFactors}
\end{figure}

Finally, Fig.~\ref{Experiment1SpeedUpFactors} illustrates the surface generated by the query grid speed-up factors. It is seen that the speed-up factor is the highest when the PTROM is running at the edge of the query grid corresponding to either one or both vortex pairs having high circulation. The higher speed-up factor at the edge of the parametric grid corresponds to the increase in wall-time of the FOM needing more Newton iterations to converge as the circulations of $\{\chi_1,\psi_1\}$ and $\{\chi_N,\psi_N\}$ increase. As a result, the increase in Newton iterations also corresponds to more pair-wise interactions incurred that scale like $O(N^2)$. However, because the PTROM performs a hyper-reduced pair-wise interaction, additional iterations due to the stronger circulations do not incur a significant increase in wall-time. Therefore, the significant increase in FOM wall-time near the edge of the parametric grid and relatively low increase in PTROM wall-time incurs a higher speed-up factor than the interior parametric grid. Overall, it is seen that the PTROM is capable of delivering improved wall-time performance when compared to the FOM.

\subsection{Mushroom cloud parametric experiment}
The PTROM is now tested against a parametric numerical experiment where two vortices, each with variable circulation and opposing circulation direction, form a mushroom cloud generated with the Biot--Savart law by assigning high variable circulations to end particles and fixed weak circulations to the remaining particles, as was done in the previous experiment. In this experiment, $\bm{\mu} = (\Gamma_1,\Gamma_N)$ are the input parameters and are part of the parametric space $\mathcal{D}=[-\Gamma_{\textup{max}}, -0.5 \Gamma_{\textup{max}}] \times [0.5\Gamma_{\textup{max}}, \Gamma_{\textup{max}}] $, where $\Gamma_{\textup{max}}=220$. All other particles are assigned a circulation strength of $\Gamma_i=0.01$, where $i = 2\ldots N-1$ and again $N=500$ particles in this experiment. In addition, an inflow condition with a semi-circle profile is added to each particle according to its position on the inflow profile. The inflow profile, $\bm{p}_{\infty}=\{p_{\infty ,\chi }, \: p_{\infty ,\psi }\}$, where  $p_{\infty ,\chi }=0$ and $p_{\infty ,\psi } = 5\sqrt{\left(1.125^2-\chi_{\infty}^2\right)}+0.5$, where $\chi_{\infty}$ is a linear distribution from -1 to 1 with $N$ increments that are assigned to each particle. The initial positions of the particles follow a linear distribution in the $\chi$ component but have a fixed position in $\psi$ component. The starting positions of the end particles are then $\{\chi_1,\psi_1\}=\{-37.43, -10\}$ and ending $\{\psi_N,\psi_N\}=\{37.43, -10\}$. Particles have an initial $l_2$ spacing of 0.15 between each other and the Biot--Savart kernel is assigned a de-singularization cut-off radius $\delta_K$ equal to the initial spacing distance of 0.15. All simulations were performed in the time interval $t \in [0, 5]$ with $\Delta t = 0.005$ leading to $n_t=1000$ time-steps.

\begin{figure}[t]
	\centering
	\includegraphics[scale=0.65, trim=3cm 8cm 50 7cm]{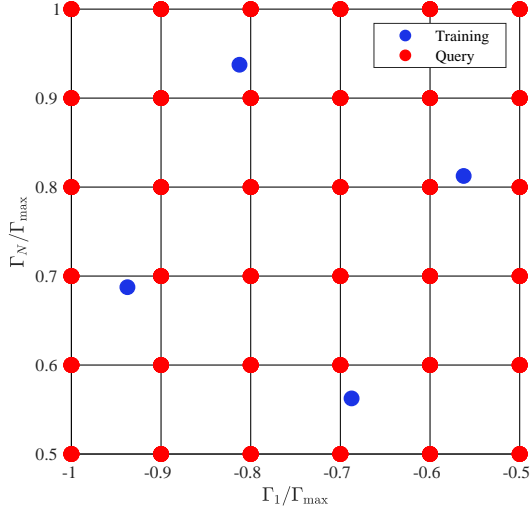}
	\caption[]%
    {Parametric space for the mushroom cloud numerical experiment. Online query points are marked in red and training points are marked in blue. Parametric values are normalized by $\Gamma_{\textup{max}}$ for convenience.}    
	\label{ParametricGrid_Experiment2}
\end{figure}

The PTROM models were trained at four points in the parametric space denoted by the blue points shown in the grid in Fig. \ref{ParametricGrid_Experiment2}. The training points were chosen via the latin hyper-cube sampling method and thirty-six online points were queried online with the PTROM over a $6 \times 6$ grid, where the queried points are denoted by the red points in Fig. \ref{ParametricGrid_Experiment2}. All PTROM models in the current experiment were constructed with the following hyper-parameters: $\breve{n} =75, M=110, M_r=185,$ and $p_c=1$, which results in $N_c=294$. Here, the Gauss-Newton relative residual error tolerance was set to tol $ =10^{-4}$.

\begin{figure*}[t]
	\centering
	\begin{subfigure}[b]{0.475\textwidth}  
		\centering 
		\includegraphics[scale=0.5, trim=2cm 9cm 25 7cm]{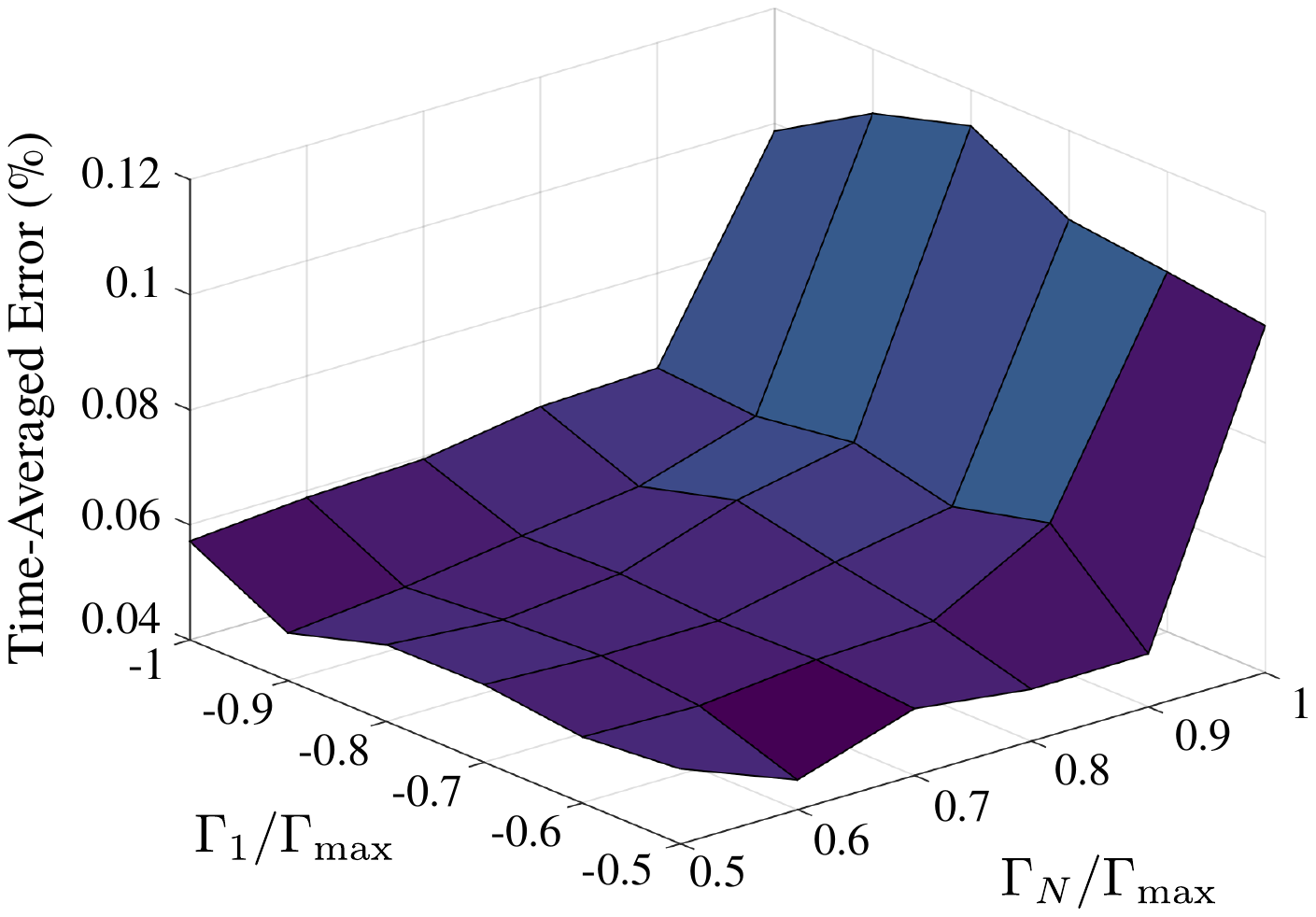}
		\caption[]%
		{Average error = 0.0634 \%}
		\label{MAEP_experiment1}
	\end{subfigure}
	\hfill
	\begin{subfigure}[b]{0.475\textwidth}  
		\centering 
		\includegraphics[scale=0.5, trim=2cm 9cm 25 7cm]{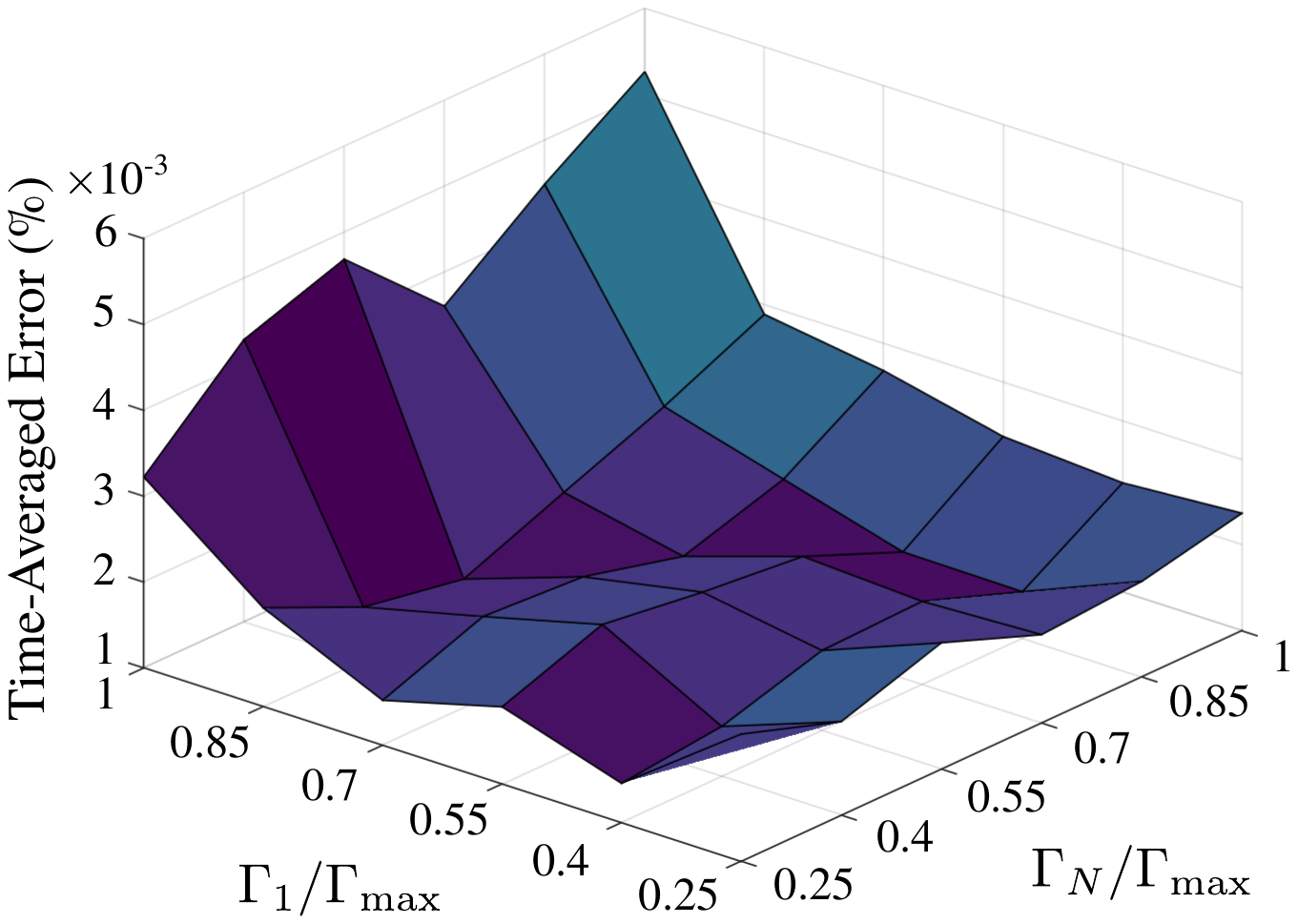}
		\caption[]%
		{Average error = $2.39 \times 10^{-3}$ \%}
		\label{AE_experiment1}
	\end{subfigure}
	\caption[]
	{Left: Surface generated by the time-averaged mean absolute trajectory errors of the online queried points. Right: Surface generated by the time-averaged Hamiltonian absolute error of the online queried points.}  
	\label{Experiment2_ErrorResults}
\end{figure*}

Figure \ref{Experiment2_ErrorResults} show the parametric experimental results. The time-averaged mean trajectory absolute error, normalized by the characteristic length, $l$, has an average value of 0.0634\%. The time-averaged Hamiltonian absolute relative error normalized has an average value of $2.39 \times 10^{-3}$ \%. Results show that the PTROM is capable of predicting FOM QoIs with sub-0.1\% error on average. 

\begin{figure*}[h!]
	\centering
	\includegraphics[scale=0.575, trim=4cm 4cm 50 4cm]{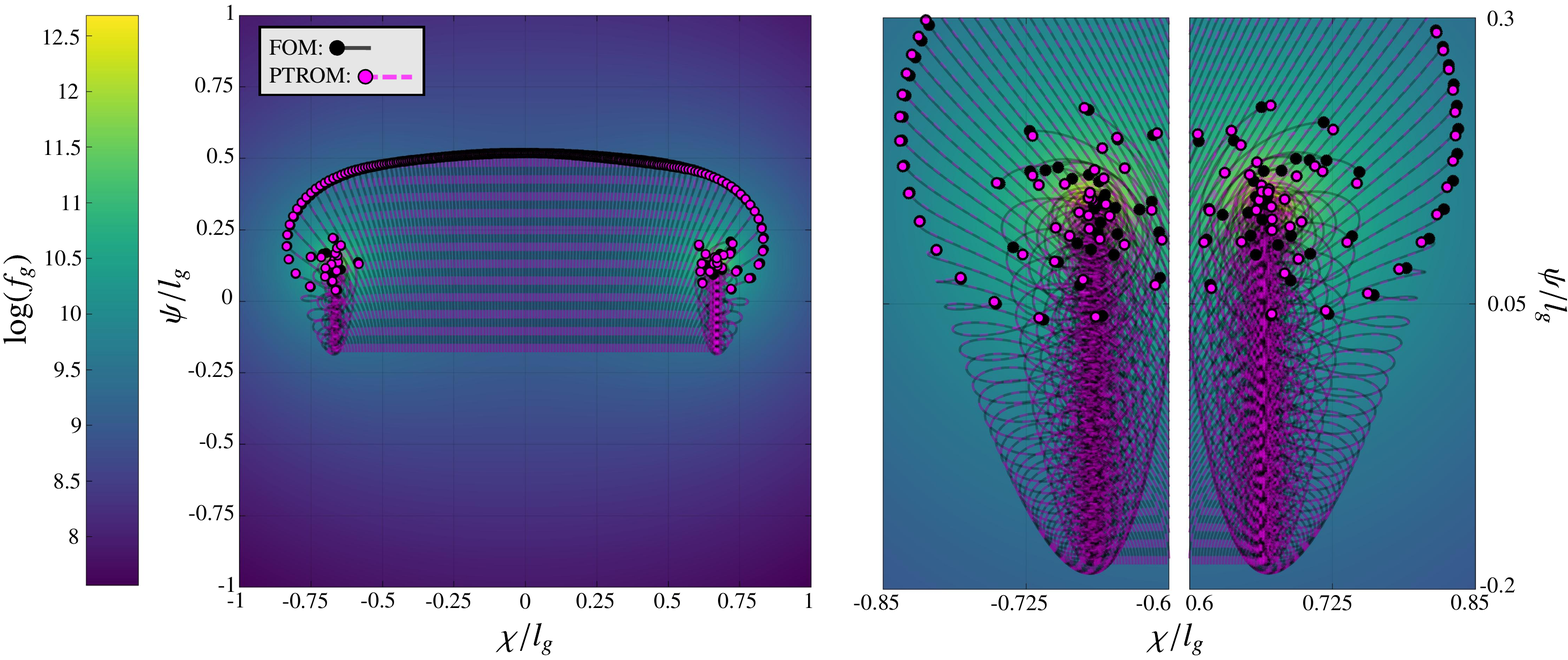}
	\caption[]
	{Vortex particle simulation with its corresponding velocity field at $t=5$. Left: full view of the vortex simulation. Right: Zoomed in view of the left and right ends of the mushroom cloud as they are advected with the inflow condition, $\bm{p}_{\infty}$. }  
	\label{Experiment2_Simulation}
\end{figure*}

Results presented by the QoI error surfaces in Fig.~\ref{Experiment2_ErrorResults}, are further highlighted by the queried simulation corresponding to parametric values of $\bm{\mu}=(\Gamma_{\textup{max}} ,\Gamma_{\textup{max}})$ in Fig.~\ref{Experiment2_Simulation}.  The simulation in Fig.~\ref{Experiment2_Simulation} illustrates the PTROM and FOM (not included in training) particle paths and positions, in addition to the FOM velocity field over a grid with a normalization width of $l_g=1.25l$. Figure \ref{Experiment2_Simulation} shows the multi-scale nature of the vortex pair simulation, where particles near the particle with strong circulation, i.e. particle $\{\chi_1, \psi_1 \}$ and $\{\chi_N, \psi_N \}$, orbit the strong particles faster than those further away. Overall the PTROM exhibits similar performance as in the previous experiment with the vortex pair: the PTROM is capable of following the FOM trajectory with high precision and accuracy for particles not in the neighborhood of the particles with strong circulation, i.e. about $10 \le $ particles away from $\{\chi_1, \psi_1 \}$ or $\{\chi_N, \psi_N \}$. Particles in the neighborhood of either $\{\chi_1, \psi_1 \}$ or $\{\chi_N, \psi_N \}$ seem to deviate from instantaneous position of the FOM as time passes but are still able to generally follow the path outlined by the FOM.

\begin{figure}[h!]
	\centering
	\includegraphics[scale=0.58, trim=3cm 9cm 2cm 8cm]{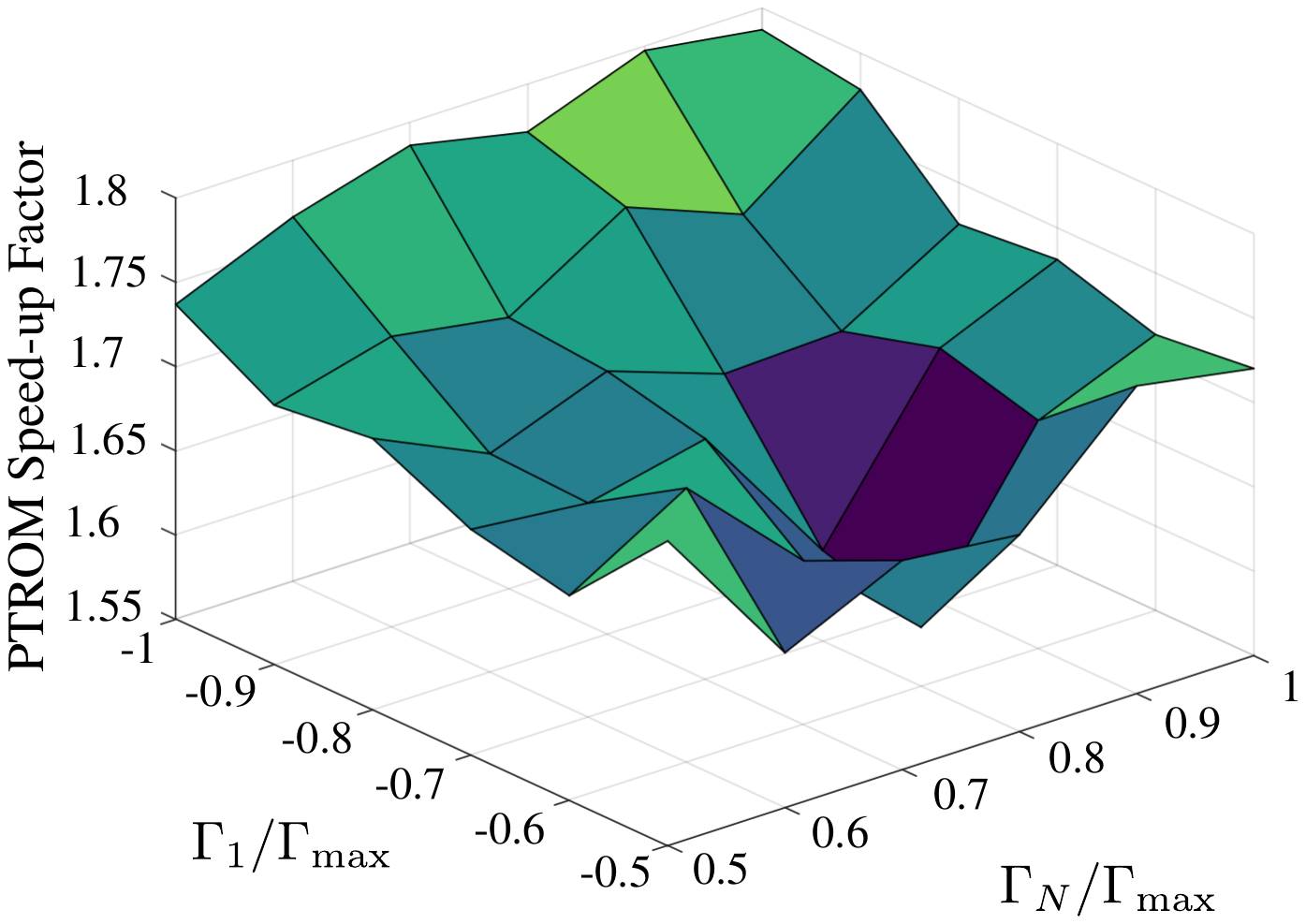}
	\caption[]%
	{Speed-up factor for the vortex pair experiment, where the average speed-up factor = 1.71.}    
	\label{Experiment2SpeedUpFactors}
\end{figure}

Finally, Fig.~\ref{Experiment2SpeedUpFactors} illustrates the surface generated by the query grid speed-up factors. As in the prior experiment with the vortex pair, it is seen that the speed-up factor is the highest when the PTROM is running at the edge of the query grid, which corresponds to either one or both vortex pairs having high circulation. The higher speed-up factor at the edge of the parametric grid corresponds to the same reasons as in the previous vortex pair experiment: the increase in wall-time of the FOM is a result of an increase in Newton iterations and pair-wise computations required to converge, which is due to an increase in end-particle circulations. However, because the PTROM performs a hyper-reduced pair-wise interaction, where additional Gauss-Newton iterations do not incur a significant increase in wall-time, additional iterations due to the stronger circulations do not incur a significant increase in wall-time. Therefore, the significant increase in FOM wall-time near the edge of the parametric grid and relatively low increase in PTROM wall-time incurs a higher speed-up factor than the interior parametric grid. Overall, it is seen that the PTROM is capable of delivering improved wall-time performance when compared to the FOM.

\subsection{Single vortex reproductive experiment}

A reproductive experiment was conducted to assess the performance of the PTROM as the numerical domain increases, i.e.~as the number of particles increase, and compared its performance against explicit time integration, hierarchical decomposition, and the GNAT method. Specifically, the PTROM run-time performance was compared against the results from the implicit trapezoidal rule equipped with the Barnes-Hut method (both neighbor clustering and Barnes--Hut clustering were tested) and the GNAT method. In addition, the incurred wall-times of the explicit modified Euler integration (Heun's integration) and modified Euler integration equipped with the Barnes-Hut method (employing neighbor clustering and Barnes-Hut clustering) were compared to the PTROM.  Comparing the PTROM run-time performance with the modified Euler's time integration helps gauge its performance against traditional rapid time-integrators, such as explicit schemes. Next, QoI reproduction errors as functions of hyper-parameter variation are assessed against incurred wall-times. In other words, Perato fronts for the reproductive experiment are constructed based on QoI error versus wall-time over a range of hyper-parameters. QoIs generated by the explicit integration with and without Barnes--Hut hierarchical decomposition were not compared against the implicit FOM, as it would be inconsistent to measure QoI results of an explicit integration scheme against the implicit FOM QoIs and corresponding GNAT and PTROM results trained by the implicit FOM. 

The current reproductive experiment was performed on a range of particle numbers with varying conditions of single vortex simulation that are show in Table \ref{ReproductiveExperimentConditions}. The initial positions of particles for all cases listed in Table \ref{ReproductiveExperimentConditions} were defined by a linear distribution in space, i.e. \code{linspace(-N,N,N)} in MATLAB syntax, such that the end particles are located at $\{\chi_1,\psi_1\} =\{-N,-N\}$ and $\{\chi_N,\psi_N\} =\{N,N\}$. The initial distance between each particle was defined by the aforementioned linear distribution of the particle initial positions, and the de-singularization cutoff was set to $\delta_K=0$. \\

\noindent \textbf{Note:} A reproductive experiment and not a parametric experiment was performed to avoid expensive singular value decomposition computations and hierarchical data structures on MATLAB that would completely allocate the random access memory on the machine employed in this investigation. Future work will perform larger scale parametric experiments on a more efficient lower-level language.

{\renewcommand{\arraystretch}{1.5}
	\begin{table}[t]
		\centering
		\caption[]{Reproductive experiment conditions for the implicit FOM. Note all particles except the center particle were assigned $\Gamma=0.01$.}
		\begin{tabular}{l  l  l  l l l l l  } 
			\hline\hline
			$N$ & 100 & 500 & 1000 & 2000 & 3000 & 4000 & 5000\\
			\hline
			$\Delta t$ & 0.01 & $2.5\times 10^{-3}$  & $2.5\times 10^{-4}$ & $1.25\times 10^{-4}$ & $10^{-4}$ & $7.5\times 10^{-5}$ & $ 5\times 10^{-5}$  \\
          		$\Gamma_{\textup{center}}$ & 500 & $10^4$ & $10^5$ & $2 \times 10^5$ & $3\times 10^5$ & $4 \times 10^5$ & $6.75\times 10^{5}$ \\
		        $t \in [t_0, t_f]$  & [0, 20] & [0, 5] & [0, 0.5] & [0, 0.25] & [0, 0.2] & [0, 0.15] & [0, 0.1]\\
			\hline\hline
		\end{tabular}
		\label{ReproductiveExperimentConditions}
		\vspace{0.5cm}
\end{table}}

\subsubsection{PTROM reproductive results}

Hyper-parameter selection for the PTROM experiments was based on varying the rank of the state POD basis $M$ and neighborhood width scaling factor $p_c$. The residual POD basis rank and sampled particle hyper-parameters followed the change of the state POD basis such that $M_r=2M$ and $\breve{n}=M_r$. Variation of the state POD basis was selected based on loosening and tightening the relative residual tolerances and adjusting the rank of the POD basis enough to satisfy the max iteration criteria of $k\le100$. A discussion is now presented for each hyper-parametric experiment followed by a discussion on the corresponding reproduction errors of QoIs versus wall-times.

Table \ref{PTROM_NarrowWidthHyperParameters} lists the bases rank variations, $M$, width parameters, $p_c$, and resulting number of POD source clusters, $N$, for narrow neighborhood widths, i.e.~only few neighboring particles have not been clustered. It was found that neighborhood width impacted the rate of convergence of the PTROM and as a result the neighborhood width was widened for $N=3000, 4000,$ and $5000$ at Bases Case 4 to satisfy the max iteration count criteria for corresponding cases. Future work will look into the underlying reasons of how clustering impacts convergence in the PTROM formulation. It is seen in Table \ref{PTROM_NarrowWidthHyperParameters} that as the tolerances tighten, additional bases must be added to satisfy the max iteration criteria. 

\begin{table}[t]
  \caption{PTROM hyper-parameter settings for narrow-width neighborhood clustering.}
\makebox[\textwidth][c]{
    \begin{tabular}{cccccccccc}
\toprule
      &       &  & \multicolumn{7}{c}{Number of particles, $N$, in the domain} \\
\cmidrule(lr){4-10}
& tol & Hyper-parameter & 100  & 500 & 1000  & 2000  & 3000 & 4000 & 5000 \\ \midrule \midrule
\multirow{3}{*} {Bases Case 1 } & \multirow{3}{*} {$10^{-1}$} & $M$ & 13 & 21 & 21  & 21  & 23  & 23  & 23  \\
\hhline{~~~~~}  &   & $ p_c$ & 0 & 0 & 0  & 0  & 0  & 0  & 0  \\
\hhline{~~~~~}  &   & $ N_c$ & 68 & 113 & 121  & 122  & 134  & 131 & 142  \vspace{0.25cm}\\
\multirow{3}{*} {Bases Case 2 } & \multirow{3}{*} {$10^{-2}$} & $M$ & 14 & 22 & 22  & 22  & 23  & 23  & 24  \\
\hhline{~~~~~}  &   & $ p_c$ & 0 & 0 & 0  & 0  & 0  & 0  & 0 \\
\hhline{~~~~~}  &   & $ N_c$ & 69 & 114 & 129  & 127  & 142  & 130  & 140  \vspace{0.25cm}\\
\multirow{3}{*} {Bases Case 3 } & \multirow{3}{*} {$10^{-3}$} & $M$ & 14 & 23 & 23  & 23  & 25  & 25  & 26  \\
\hhline{~~~~~}  &   & $p_c$ & 0 & 0 & 0  & 0  & 0  & 0  & 0\\
\hhline{~~~~~}  &   & $ N_c$ & 70 & 120 & 132  & 134  & 136  & 143  & 156   \vspace{0.25cm} \\
\multirow{3}{*} {Bases Case 4 } & \multirow{3}{*} {$10^{-4}$} & $M$ & 16 & 24 & 24  & 26  & 26  & 26  & 26  \\
\hhline{~~~~~}  &   & $p_c$ & 0 & 0 & 0  & 0  & 0.5  & 0.5  & 0.5 \\
\hhline{~~~~~}  &   & $ N_c$ & 80 & 128 & 130  & 134  & 151  & 159  & 144  \vspace{0.25cm}\\
\bottomrule
\end{tabular}%
}
\label{PTROM_NarrowWidthHyperParameters}%
\end{table}%

\begin{figure*}[h!]
	\centering
	\begin{subfigure}[b]{0.475\textwidth}
		\centering
		\includegraphics[scale=0.575, trim=4cm 8cm 50 8cm]{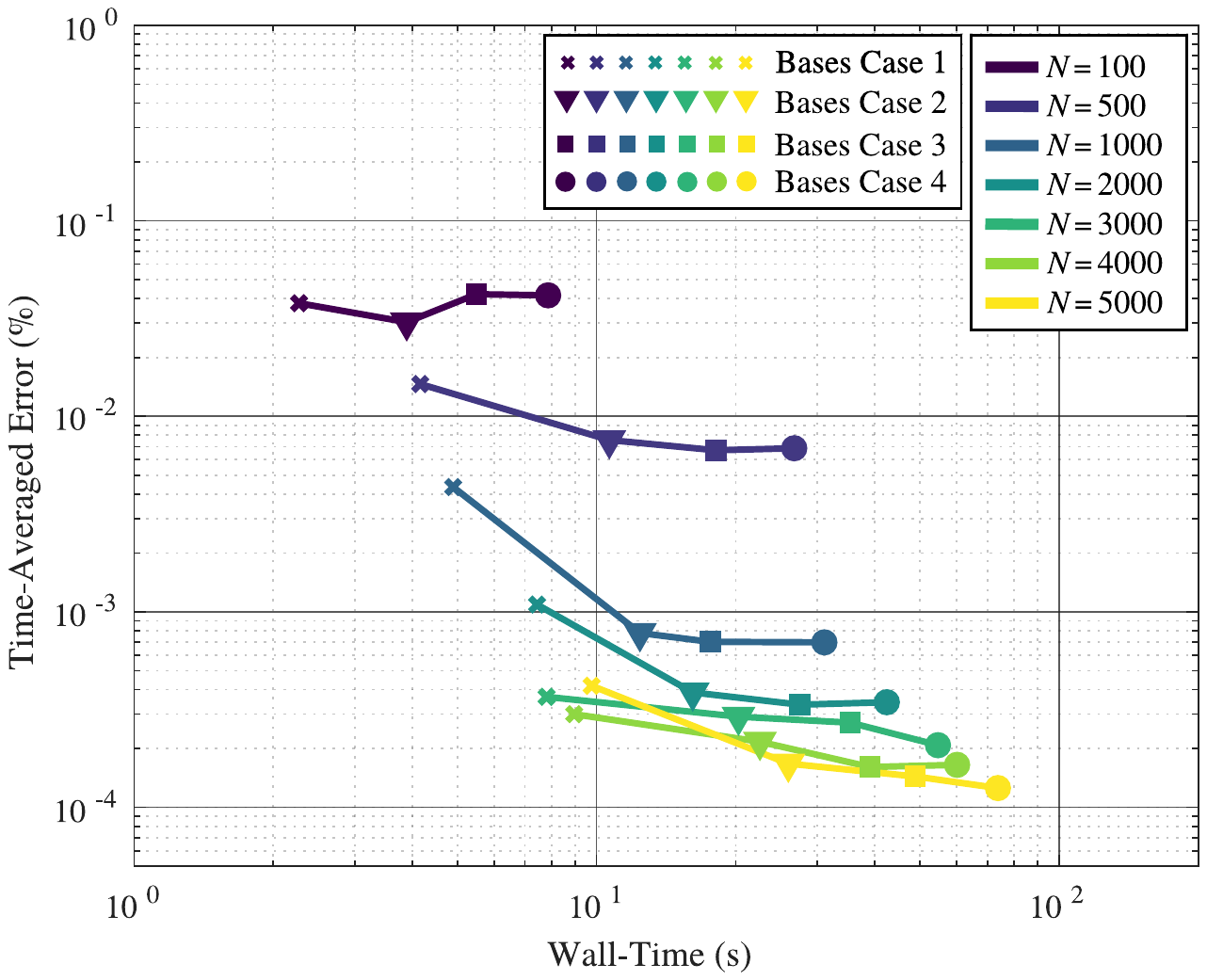}
		\caption[]%
		{Time-averaged $\textup{MAE}_{D}$}    
		\label{MAED_narrow}
	\end{subfigure}
	\hfill
	\begin{subfigure}[b]{0.475\textwidth}  
		\centering 
		\includegraphics[scale=0.575, trim=5cm 8cm 25 8cm]{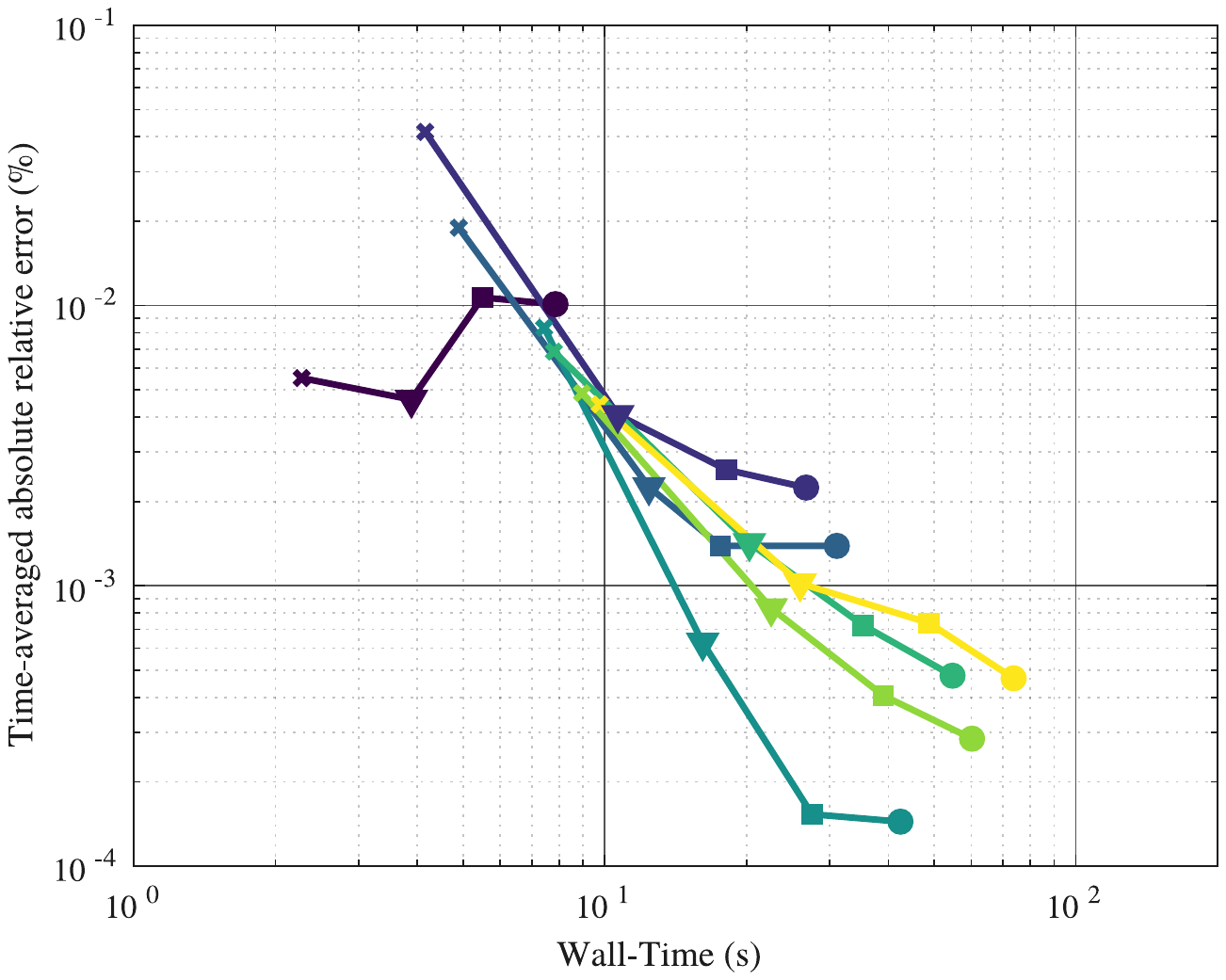}
		\caption[]%
		{Time-averaged $\textup{AE}_{H}$}
		\label{AEH_narrow}
	\end{subfigure}
		\caption[]
	{QoI results for narrow-width neighborhood hyper-parameter settings (a) Time-averaged $\textup{MAE}_D$ versus wall-time; (b) Time-averaged $\textup{AE}_H$ versus wall-time. Color scale indicates the number of particles in the domain and shapes correspond the bases rank listed Table \ref{PTROM_NarrowWidthHyperParameters}}  
	\label{PTROM_NarrowWidthResults}
\end{figure*}

Figure \ref{PTROM_NarrowWidthResults} presents the QoI reproductive results generated by the hyper-parameter settings of Table \ref{PTROM_NarrowWidthHyperParameters}. The time-averaged $\textup{MAE}_{D}$ results shown in Fig.~\ref{MAED_narrow} illustrate sub 0.1\% errors across all cases and particle domain sizes. It is important to point out that as the particle count increases the error decreases,  which is due to the PTROM being capable of accurately reproducing the growing number of particle trajectory paths, as reflected in the error quantification via Eq.~\ref{MAE_trajectory}. The time-averaged $\textup{AE}_{H}$ results shown in Fig.~\ref{AEH_narrow} also show sub 0.1\% errors across all cases and particle domain sizes. Both QoIs errors presented in Fig.~\ref{PTROM_NarrowWidthResults} reflect an increase in wall-time as the tolerance is tightened. It is interesting to note that for the numerical experiment corresponding to $N=100$, the QoI errors increase as the tolerance is tightened and more bases are added. It is unknown where this increase in error due to tightened tolerance stems from, but the hypothesis is that narrow neighborhood clustering generates poorer source approximations as more bases are added to the source surrogate POD matrix, $\tilde{\bm{\Phi}}$, for $N=100$.

Next, Table \ref{PTROM_ModerateWidthHyperParameters} lists the bases rank variations, $M$, width parameters, $p_c$, and resulting number of POD source clusters, $N$, for moderate neighborhood widths, i.e. the number of neighboring particles have increased from the narrow width case and the number of cluster sources have increased. The neighborhood width remained constant through out all hyper-parameter settings, i.e.  $p_c=1$ for all cases. However, it was seen that in the case of $N=100$ for Bases Case 4, the POD basis rank was increased from the prior setting listed in Table \ref{PTROM_NarrowWidthHyperParameters} to meet the max iteration criteria.

\begin{table}[h!]
    \caption{PTROM hyper-parameter settings for moderate-width neighborhood clustering.}
\makebox[\textwidth][c]{
    \begin{tabular}{cccccccccc}
\toprule
      &       &  & \multicolumn{7}{c}{Number of particles, $N$, in the domain} \\
\cmidrule(lr){4-10}
& tol & Hyper-parameter & 100  & 500 & 1000  & 2000  & 3000 & 4000 & 5000  \\ \midrule \midrule 
\multirow{3}{*} {Bases Case 1 } & \multirow{3}{*} {$10^{-1}$} & $M$ & 13 & 21 & 21  & 21  & 23  & 23  & 23  \\
\hhline{~~~~~}  &   & $ p_c$ & 1 & 1 & 1  & 1  & 1  & 1  & 1 \\
\hhline{~~~~~}  &   & $N_c$ & 73 & 122 & 139  & 123  & 135  & 137  & 149\vspace{0.25cm} \\
\multirow{3}{*} {Bases Case 2 } & \multirow{3}{*} {$10^{-2}$} & $M$ & 14 & 22 & 22  & 22  & 23  & 23  & 24  \\
\hhline{~~~~~}  &   & $ p_c$ & 1 & 1 & 1  & 1  & 1  & 1  & 1 \\
\hhline{~~~~~}  &   & $N_c$ & 79 & 130 & 142  & 149  & 137  & 150  & 144 \vspace{0.25cm} \\
\multirow{3}{*} {Bases Case 3 } & \multirow{3}{*} {$10^{-3}$} & $M$ & 14 & 23 & 23  & 23  & 25  & 25  & 26  \\
\hhline{~~~~~}  &   & $ p_c$ & 1 & 1 & 1  & 1  & 1  & 1  & 1 \\
\hhline{~~~~~}  &   & $N_c$ & 79 & 127 & 133  & 132  & 146  & 153  & 149 \vspace{0.25cm} \\
\multirow{3}{*} {Bases Case 4 } & \multirow{3}{*} {$10^{-4}$} & $M$ & 18 & 24 & 24  & 26  & 26  & 26  & 26  \\
\hhline{~~~~~}  &   & $ p_c$ & 1 & 1 & 1  & 1  & 1  & 1  & 1 \\
\hhline{~~~~~}  &   & $N_c$ & 84 & 131 & 144  & 144  & 147  & 159  & 143 \vspace{0.25cm} \\
\bottomrule
\end{tabular}%
}
\label{PTROM_ModerateWidthHyperParameters}%
\end{table}%

\begin{figure*}[h!]
	\centering
	\begin{subfigure}[b]{0.475\textwidth}
		\centering
		\includegraphics[scale=0.575, trim=4cm 8cm 50 8cm]{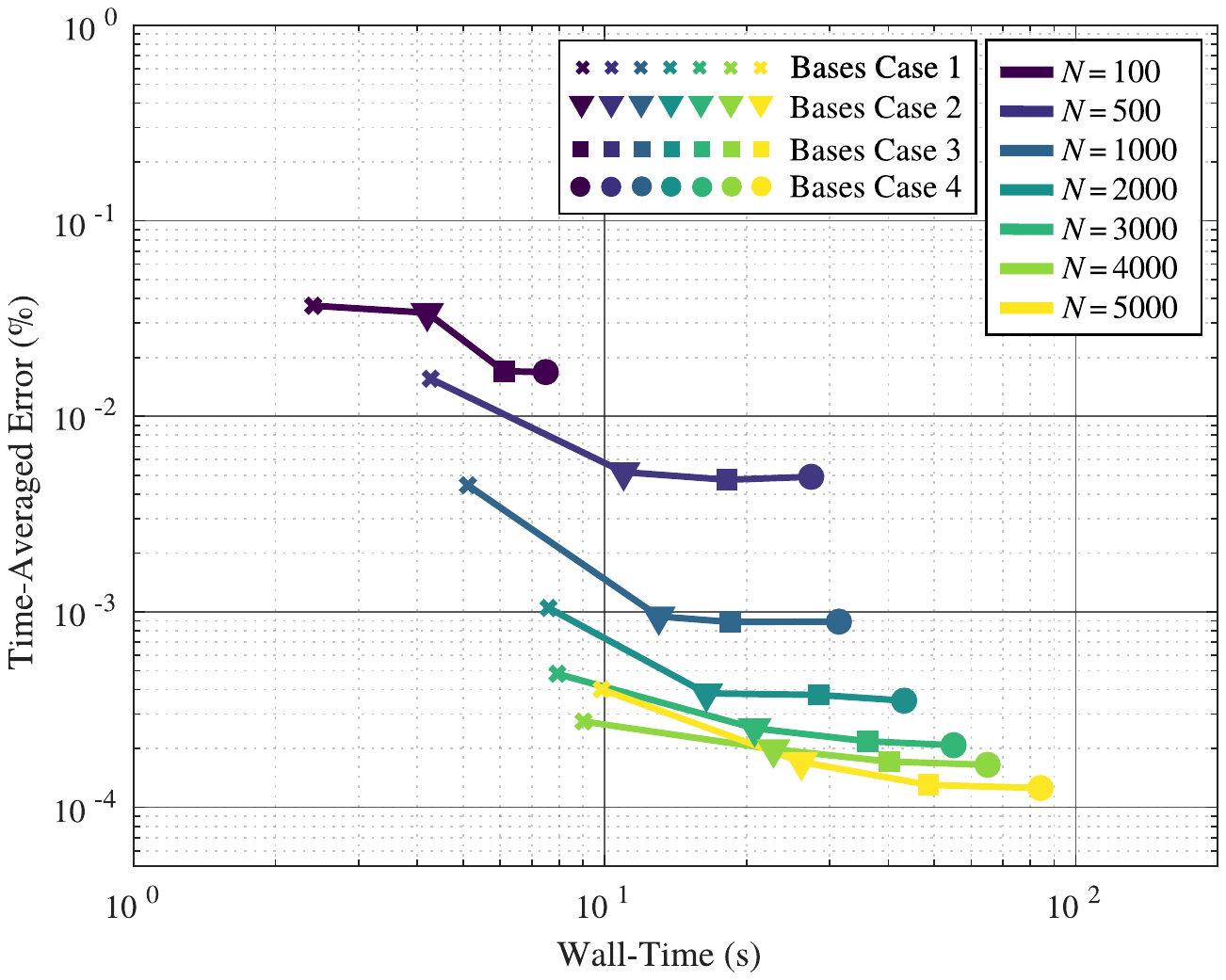}
		\caption[]%
		{Time-averaged $\textup{MAE}_{D}$}    
		\label{MAED_moderate}
	\end{subfigure}
	\hfill
	\begin{subfigure}[b]{0.475\textwidth}  
		\centering 
		\includegraphics[scale=0.575, trim=5cm 8cm 25 8cm]{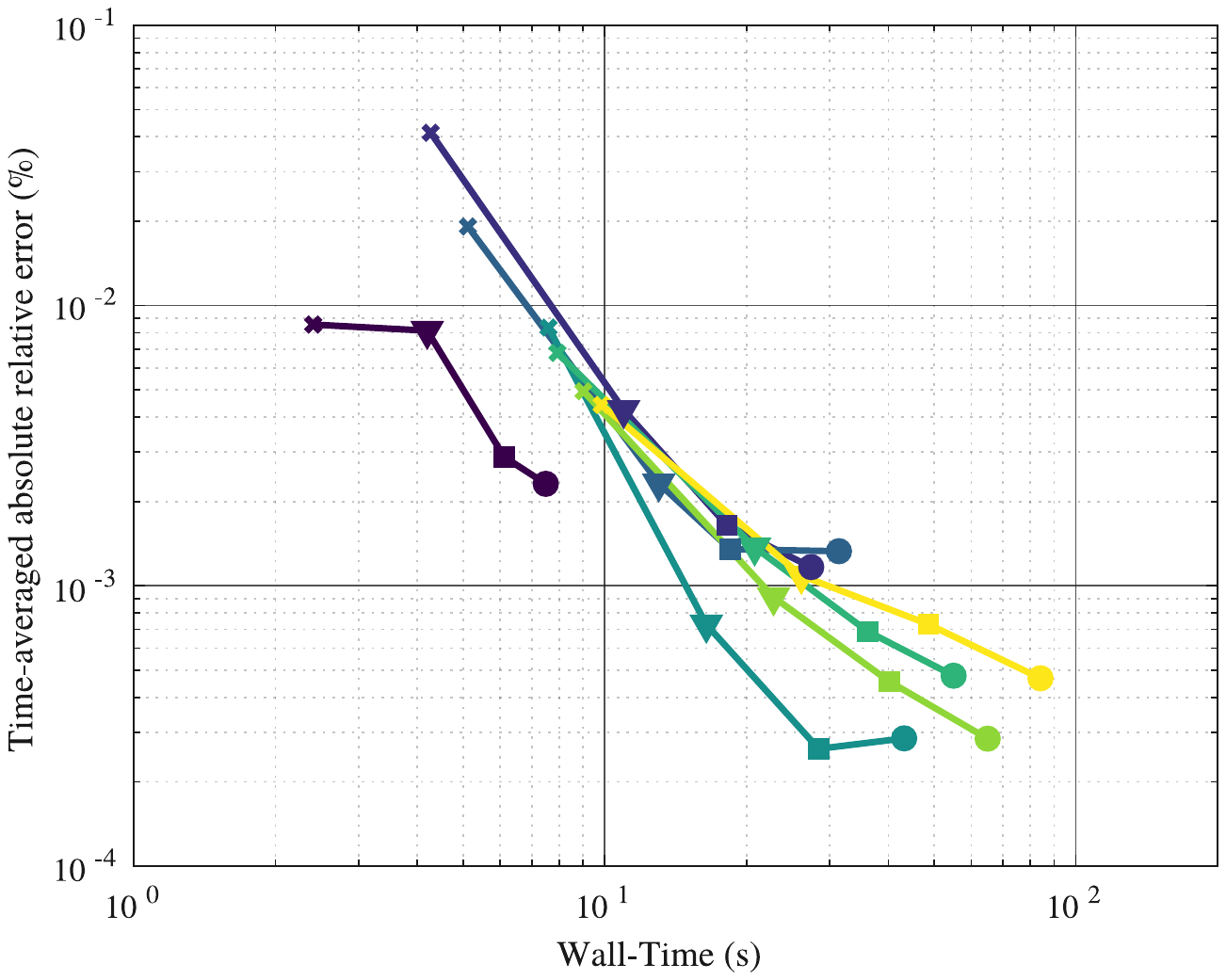}
		\caption[]%
		{Time-averaged $\textup{AE}_{H}$}
		\label{AEH_moderate}
	\end{subfigure}
		\caption[]
	{QoI results for moderate-width neighborhood hyper-parameter settings (a) Time-averaged $\textup{MAE}_D$ versus wall-time; (b) Time-averaged $\textup{AE}_H$ versus wall-time. Color scale indicates the number of particles in the domain and shapes correspond the bases rank listed Table \ref{PTROM_ModerateWidthHyperParameters}.}  
	\label{PTROM_ModerateWidthResults}
\end{figure*}

Figure \ref{PTROM_ModerateWidthResults} presents the QoI reproductive results generated by the hyper-parameter settings of Table \ref{PTROM_ModerateWidthHyperParameters}. The time-averaged $\textup{MAE}_{D}$ results shown in Fig.~\ref{MAED_moderate} illustrate sub 0.1\% errors across all cases and particle domain sizes. Similar to results generated by prior hyper-parameter settings in Table \ref{PTROM_NarrowWidthHyperParameters}, as the particle count increases the error decreases, which is due to the PTROM being capable of accurately reproducing the growing number of particle trajectory paths, as reflected in the error quantification via Eq.~\ref{MAE_trajectory}. The time-averaged $\textup{AE}_{H}$ results shown in Fig.~\ref{AEH_moderate} also show sub 0.1\% errors across all cases and particle domain sizes. Both QoIs errors presented in Fig.~\ref{PTROM_ModerateWidthResults} reflect an increase in wall-time as the tolerance is tightened and errors decreased. In addition, the increase in neighborhood width has resulted in increased wall-times relative to the hyper-parameter settings for the narrow-width neighborhood settings listed in Table \ref{PTROM_NarrowWidthHyperParameters}.

The final hyper-parametric settings that were tested in this reproductive experiment are presented in Table \ref{PTROM_WideWidthHyperParameters}, where the neighborhood width was increased to $p_c=2$ to increase the number of unclustered neighbors and overall source points. The neighborhood width remained constant through out all hyper-parameter settings, i.e.  $p_c=2$ for all cases. However, it was seen that in Bases Case 4 for $N=4000,$ and $5000$ the POD basis rank was increased from the prior setting listed in Table \ref{PTROM_ModerateWidthHyperParameters} to meet the max iteration criteria. It has been shown so far, with Tables \ref{PTROM_NarrowWidthHyperParameters} - \ref{PTROM_WideWidthHyperParameters} that hyper-parameter settings impact convergence. Future work will focus on attempting to unveil how the current hyper-parametric settings impact convergence.

\begin{table}[t]
  \caption{PTROM hyper-parameter settings for wide-width neighborhood clustering.}
\makebox[\textwidth][c]{
    \begin{tabular}{cccccccccc}
\toprule
      &       &  & \multicolumn{7}{c}{Number of particles, $N$, in the domain} \\
\cmidrule(lr){4-10}
& tol & Hyper-parameter & 100  & 500 & 1000  & 2000  & 3000 & 4000 & 5000 \\ \midrule \midrule
\multirow{3}{*} {Bases Case 1 } & \multirow{3}{*} {$10^{-1}$} & $M$ & 13 & 21 & 21  & 21  & 23  & 23  & 23  \\
\hhline{~~~~~}  &   & $ p_c$ & 2 & 2 & 2  & 2  & 2  & 2  & 2 \\
\hhline{~~~~~}  &   & $ N_c$ & 79 & 133 & 135  & 143  & 161  & 183  & 154 \vspace{0.25cm}\\
\multirow{3}{*} {Bases Case 2 } & \multirow{3}{*} {$10^{-2}$} & $M$ & 14 & 22 & 22  & 22  & 23  & 23  & 24  \\
\hhline{~~~~~}  &   & $ p_c$ & 2 & 2 & 2  & 2  & 2  & 2  & 2 \\
\hhline{~~~~~}  &   & $ N_c$ & 80 & 130 & 154  & 153  & 164  & 155  & 159 \vspace{0.25cm}\\
\multirow{3}{*} {Bases Case 3 } & \multirow{3}{*} {$10^{-3}$} & $M$ & 14 & 23 & 23  & 23  & 25  & 25  & 26  \\
\hhline{~~~~~}  &   & $ p_c$ & 2 & 2 & 2  & 2  & 2  & 2 & 2  \\
\hhline{~~~~~}  &   & $ N_c$ & 80 & 136 & 146  & 165  & 178  & 173  & 173 \vspace{0.25cm}\\
\multirow{3}{*} {Bases Case 4 } & \multirow{3}{*} {$10^{-4}$} & $M$ & 18 & 24 & 24  & 26  & 26  & 32  & 33  \\
\hhline{~~~~~}  &   & $ p_c$ & 2 & 2 & 2  & 2  & 2  & 2  & 2  \\
\hhline{~~~~~}  &   & $ N_c$ & 113 & 150 & 143  & 167  & 194  & 216  & 213 \vspace{0.25cm}\\
\bottomrule
\end{tabular}%
}
\label{PTROM_WideWidthHyperParameters}%
\end{table}%

\begin{figure*}[t]
	\centering
	\begin{subfigure}[b]{0.475\textwidth}
		\centering
		\includegraphics[scale=0.575, trim=4cm 8cm 50 8cm]{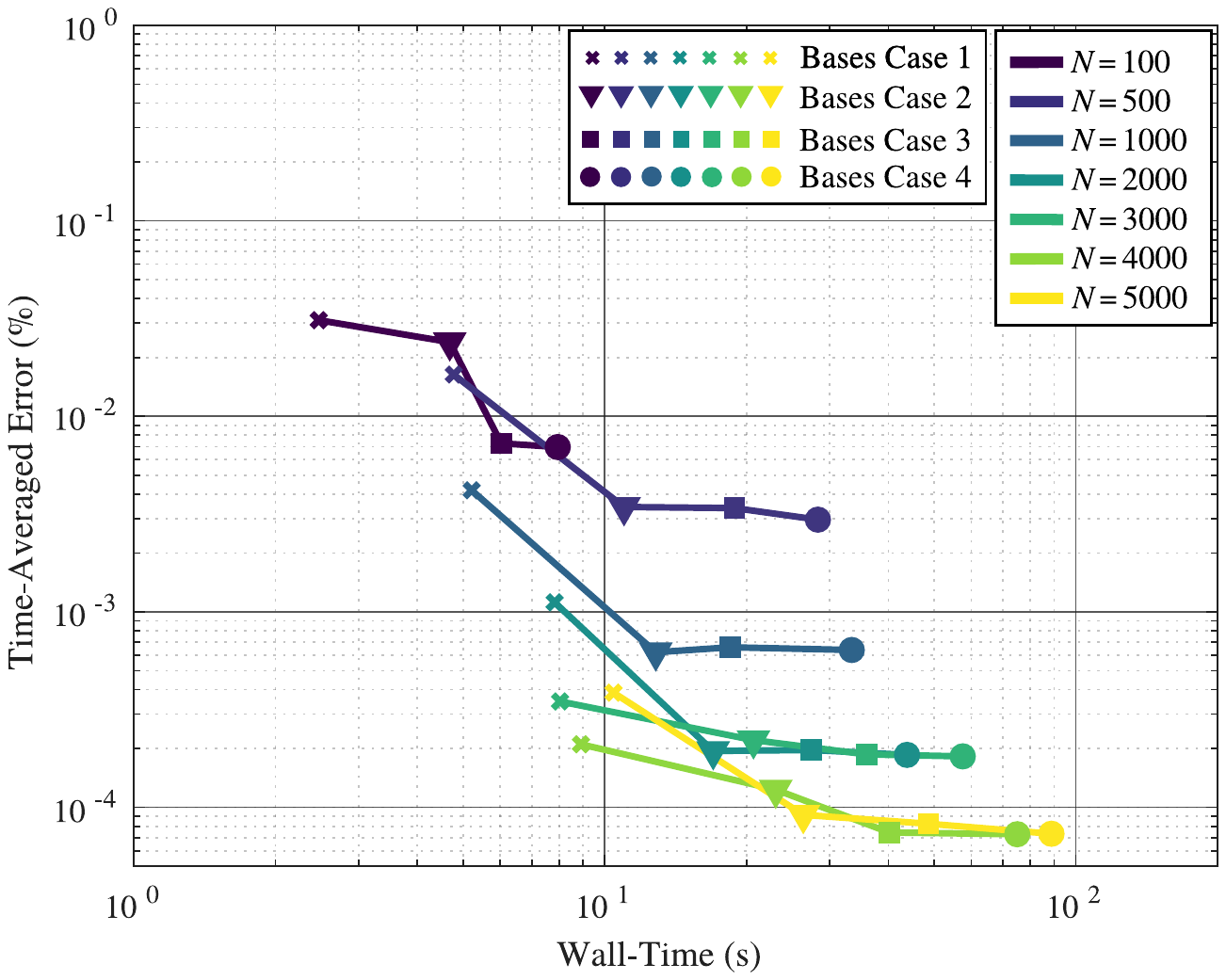}
		\caption[]%
		{Time-averaged $\textup{MAE}_{D}$}    
		\label{MAED_wide}
	\end{subfigure}
	\hfill
	\begin{subfigure}[b]{0.475\textwidth}  
		\centering 
		\includegraphics[scale=0.575, trim=5cm 8cm 25 8cm]{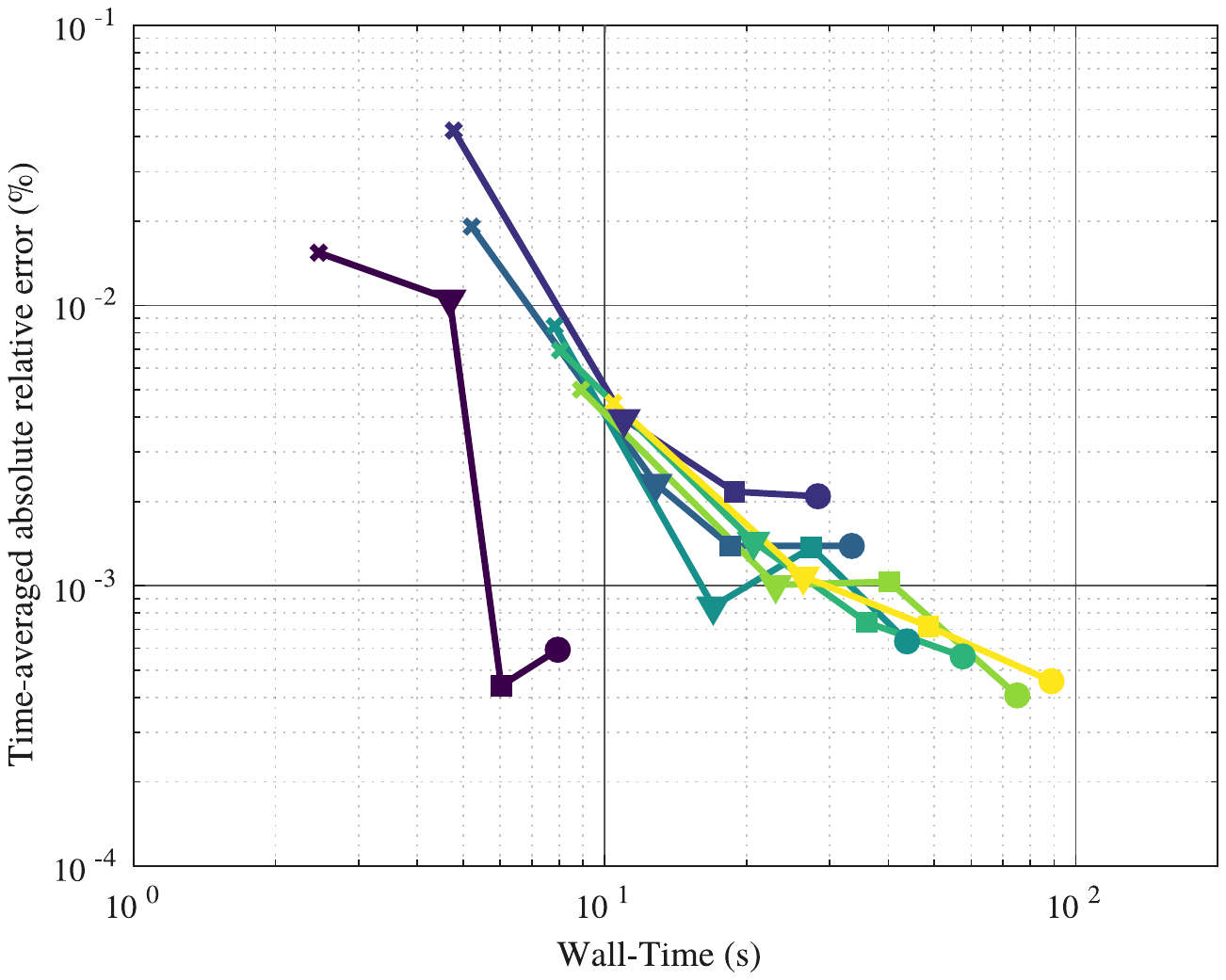}
		\caption[]%
		{Time-averaged $\textup{AE}_{H}$}
		\label{AE_wide}
	\end{subfigure}
		\caption[]
	{QoI results for wide-width neighborhood hyper-parameter settings (a) Time-averaged $\textup{MAE}_D$ versus wall-time; (b) Time-averaged $\textup{AE}_H$ versus wall-time. Color scale indicates the number of particles in the domain and shapes correspond the bases rank listed Table \ref{PTROM_WideWidthHyperParameters}.}
	\label{PTROM_WideWidthResults}
\end{figure*}

Figure \ref{PTROM_WideWidthResults} presents the QoI reproductive results generated by the hyper-parameter settings of Table \ref{PTROM_WideWidthHyperParameters}. The time-averaged $\textup{MAE}_{D}$ results shown in Fig.~\ref{MAED_wide} illustrate sub 0.1\% errors across all cases and particle domain sizes. Similar to results generated by prior hyper-parameter settings in Table \ref{PTROM_ModerateWidthHyperParameters}, as the particle count increases the error decreases, which is due to the PTROM being capable of accurately reproducing the growing number of particle trajectory paths, as reflected in the error quantification via Eq.~\ref{MAE_trajectory}. The time-averaged $\textup{AE}_{H}$ results shown in Fig.~\ref{AE_wide} also show sub 0.1\% errors across all cases and particle domain sizes. Both QoIs errors presented in Fig.~\ref{PTROM_WideWidthResults} reflect an increase in wall-time as the tolerance is tightened and errors decreased. In addition, the increase in neighborhood width has resulted in the highest wall-times incurred relative to the hyper-parameter settings for the narrow-width and moderate-width neighborhood settings listed in Table \ref{PTROM_NarrowWidthHyperParameters} and \ref{PTROM_ModerateWidthHyperParameters}.

\subsubsection{GNAT reproductive results}

The GNAT method hyper-parameter selection follows the PTROM hyper-parameter settings of Table \ref{PTROM_NarrowWidthHyperParameters}, i.e.~the coarsest neighbor-width of the PTROM. However, additional bases were added to the GNAT method if it did not meet the aforementioned max iteration criteria of $k\le 100$. Recall that the GNAT method employs no source clustering to the hyper-reduced set of residual computations and computes the influence of all $N$ sources. So with this in mind the GNAT method could be thought of as the PTROM method with an infinitely wide neighborhood width, such that there is no clustering of the sources. As a result, only one table of hyper-parameter settings is presented for the GNAT method reproductive experiment in Table \ref{GNAT_HyperParameters}, where the hyper-parameter settings for the residual POD basis and sampled residuals still hold, i.e. $M_r=2M$ and $\breve{n}=M_r$.

\begin{table}[t]
  \caption{GNAT hyper-parameter settings.}
\makebox[\textwidth][c]{
    \begin{tabular}{cccccccccc}
\toprule
      &       &  & \multicolumn{7}{c}{Number of particles, $N$, in the domain} \\
\cmidrule(lr){4-10}
& tol & Hyper-parameter & 100  & 500 & 1000  & 2000  & 3000 & 4000 & 5000 \\ \midrule \midrule
Bases Case 1 &  $10^{-1}$ & $M$ & 13 & 21 & 21  & 21  & 23  & 23  & 23  \\
Bases Case 2 &  $10^{-2}$ & $M$ & 14 & 22 & 22  & 22  & 24  & 24  & 25  \\
Bases Case 3 &  $10^{-3}$ & $M$ & 14 & 24 & 24  & 24  & 25  & 25  & 26  \\
Bases Case 4  & $10^{-4}$ & $M$ & 16 & 25 & 25  & 26  & 26  & 26  & 27  \\
\bottomrule
\end{tabular}%
}
\label{GNAT_HyperParameters}%
\end{table}%

\begin{figure*}[t]
	\centering
	\begin{subfigure}[b]{0.475\textwidth}
		\centering
		\includegraphics[scale=0.575, trim=4cm 8cm 50 9cm]{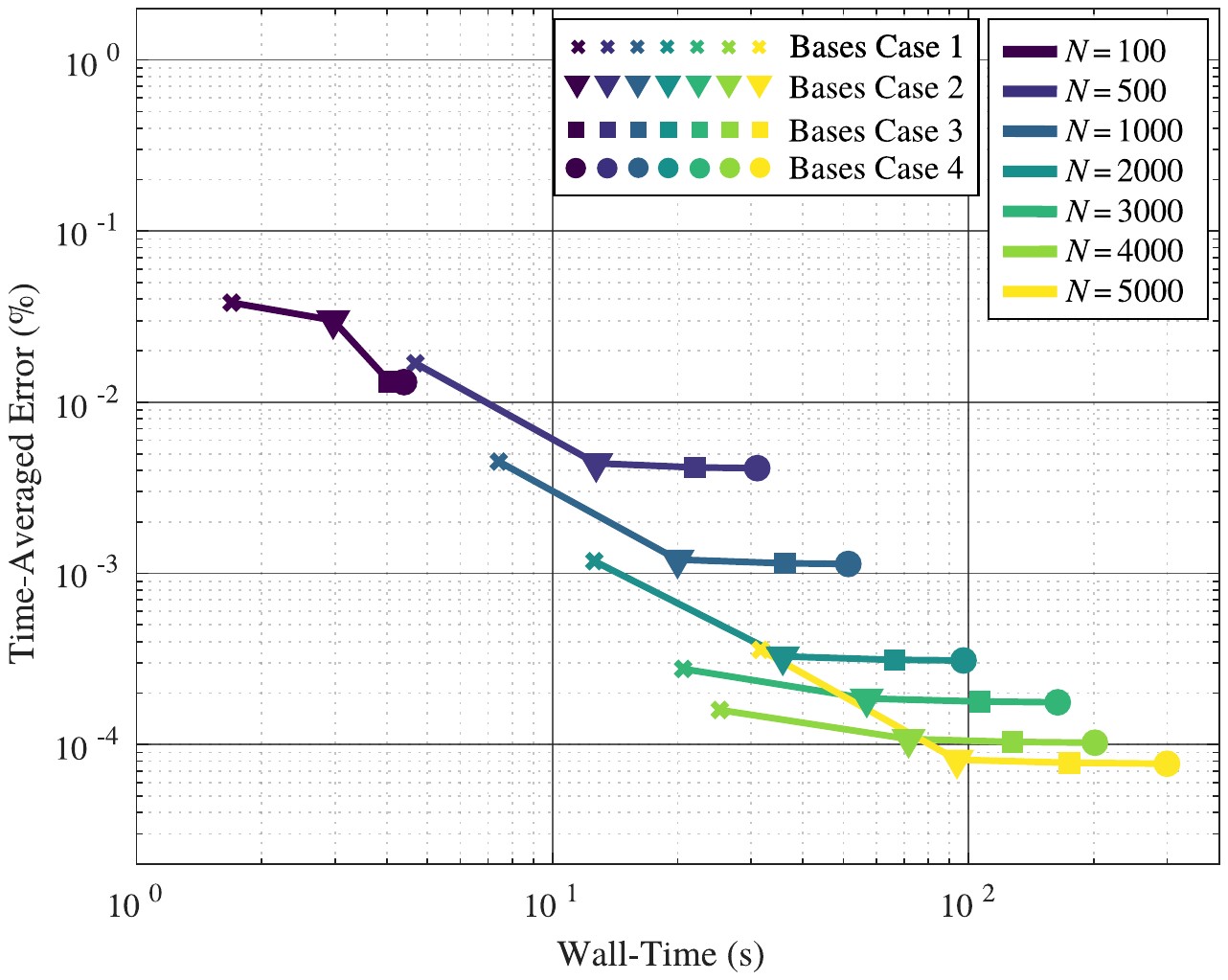}
		\caption[]%
		{Time-averaged $\textup{MAE}_{D}$}    
		\label{MAED_GNAT}
	\end{subfigure}
	\hfill
	\begin{subfigure}[b]{0.475\textwidth}  
		\centering 
		\includegraphics[scale=0.575, trim=5cm 8cm 25 9cm]{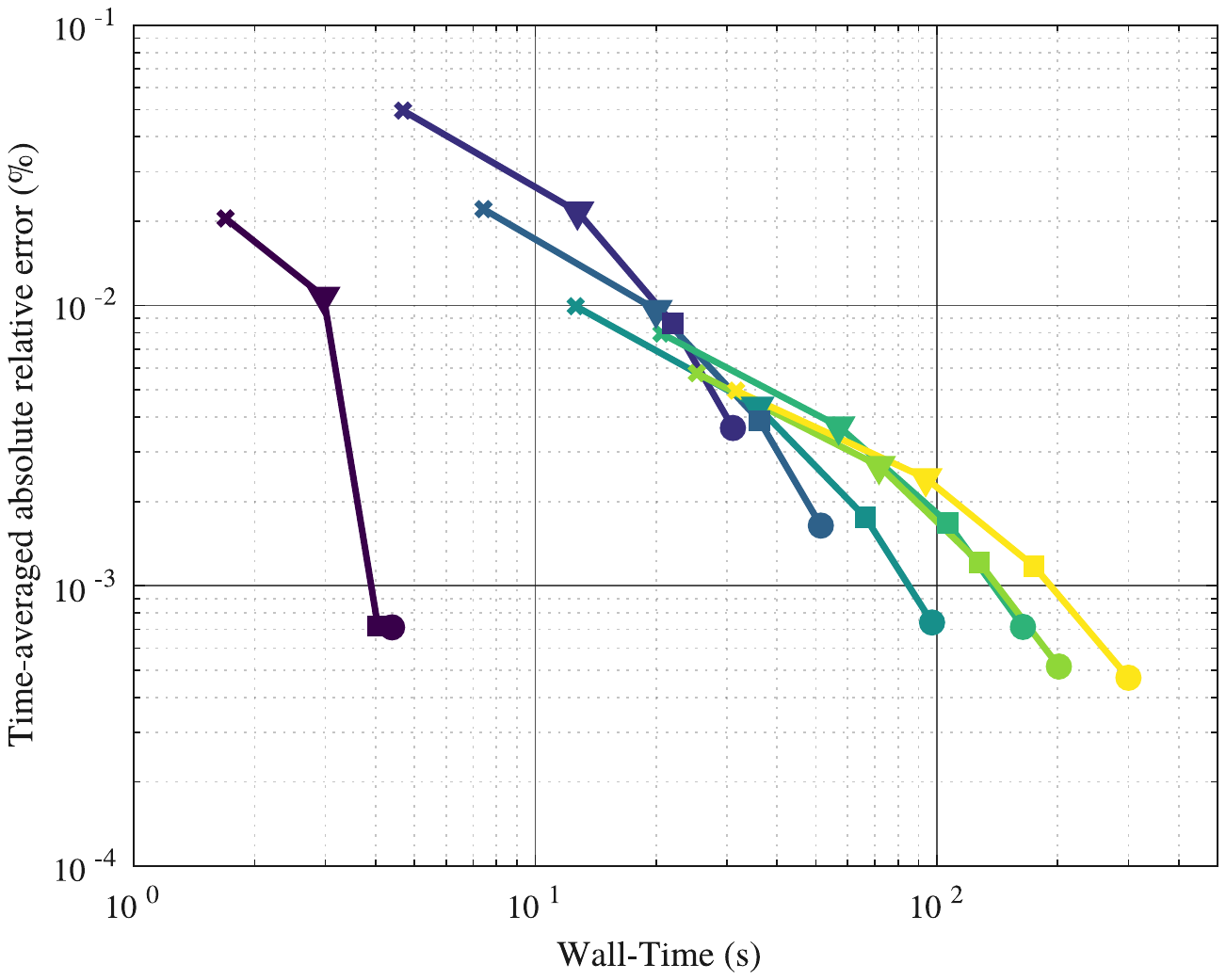}
		\caption[]%
		{Time-averaged $\textup{AE}_{H}$}
		\label{AE_GNAT}
	\end{subfigure}
	\caption[]
	{QoI results for GNAT hyper-parameter settings (a) Time-averaged $\textup{MAE}_D$ versus wall-time; (b) Time-averaged $\textup{AE}_H$ versus wall-time. Color scale indicates the number of particles in the domain and shapes correspond the bases rank listed Table \ref{GNAT_HyperParameters}.}  
	\label{GNAT_Results}
\end{figure*}

Figure \ref{GNAT_Results} presents the QoI reproductive results generated by the GNAT hyper-parameter settings of Table \ref{GNAT_HyperParameters}. The time-averaged $\textup{MAE}_{D}$ results shown in Fig.~\ref{MAED_GNAT} illustrate sub 0.1\% errors across all cases and particle domain sizes. Similar to results generated by prior PTROM reproductive experiments, as the particle count increases the error decreases due to the GNAT method's ability to accurately reproduce the growing number of particle trajectory paths. The time-averaged $\textup{AE}_{H}$ results shown in Fig.~\ref{AE_GNAT} also show sub 0.1\% errors across all cases and particle domain sizes. Both QoIs errors presented in Fig.~\ref{GNAT_Results} reflect an increase in wall-time as the tolerance is tightened and errors decreased. It is important to note that the QoI errors generated by the GNAT method are comparable to the PTROM errors, which highlight the minimal impact on QoI accuracy due to source clustering of POD basis in the PTROM method.

\subsubsection{Barnes--Hut results} \label{BarnesHutResultSubSection}
Next, results of the reproductive experiments with traditional hierarchical decomposition via the Barnes--Hut method with Barnes--Hut clustering and neighbor clustering are presented. It is important to note that to generate optimal hierarchical decomposition and clustering results, an additional parametric study was performed on the maximum number of particles contained in each leaf node. Ideally, the optimal choice would be to choose one particle per leaf node as was done in the PTROM. However, the traditional hierarchical decomposition requires online updates and rebuilds of the data-structure. These data-structure rebuilds incur more cost and wall-time if only one particle per leaf node is chosen due to the number of level traversals the algorithm has to go through in its clustering search. The results presented correspond to the simulations that incurred the lowest wall-time in the preliminary hyper-parametric search for the optimal number of particles in each leaf node. Results of this preliminary study are presented in Appendix \ref{appendix}. In addition, the number of source clusters is not fixed as it was for the PTROM due to the data-structure updates online and the number of clusters varies per target. As a result the number of clusters are not reported per each hyper-parameter setting for the hierarchical decomposition results.

First, hyper-parameters of the Barnes--Hut hierarchical decomposition with Barnes--Hut clustering are presented in Table \ref{BH_clustering_HyperParameters}. Here, $\theta=2$ generates the lowest number of clusters sources and $\theta = 0.5$ provides the largest number of cluster sources. Note that if $\theta = 0$ the original FOM would be retrieved. 

\begin{table}[t]
  \caption{Barnes--Hut hierarchical decomposition with Barnes--Hut clustering hyper-parameter settings.}
\makebox[\textwidth][c]{
    \begin{tabular}{ccccccccc}
\toprule
      &       &   \multicolumn{6}{c}{Number of particles, $N$, in the domain} \\
\cmidrule(lr){3-9}
 & Hyper-parameter & 100  & 500 & 1000  & 2000  & 3000 & 4000 & 5000 \\ \midrule \midrule
Barnes--Hut Case 1 & $\theta$ & 2 & 2 & 2  & 2  & 2  & 2  & 2  \\
Barnes--Hut Case 2 & $\theta$ & 1 & 1 & 1 & 1 & 1  & 1  & 1  \\
Barnes--Hut Case 3 & $\theta$ & 0.5 & 0.5 & 0.5  & 0.5 & 0.5  & 0.5 & 0.5  \\
\bottomrule
\end{tabular}%
}
\label{BH_clustering_HyperParameters}%
\end{table}%

\begin{figure*}[t]
	\centering
	\begin{subfigure}[b]{0.475\textwidth}
		\centering
		\includegraphics[scale=0.575, trim=4cm 8cm 50 9cm]{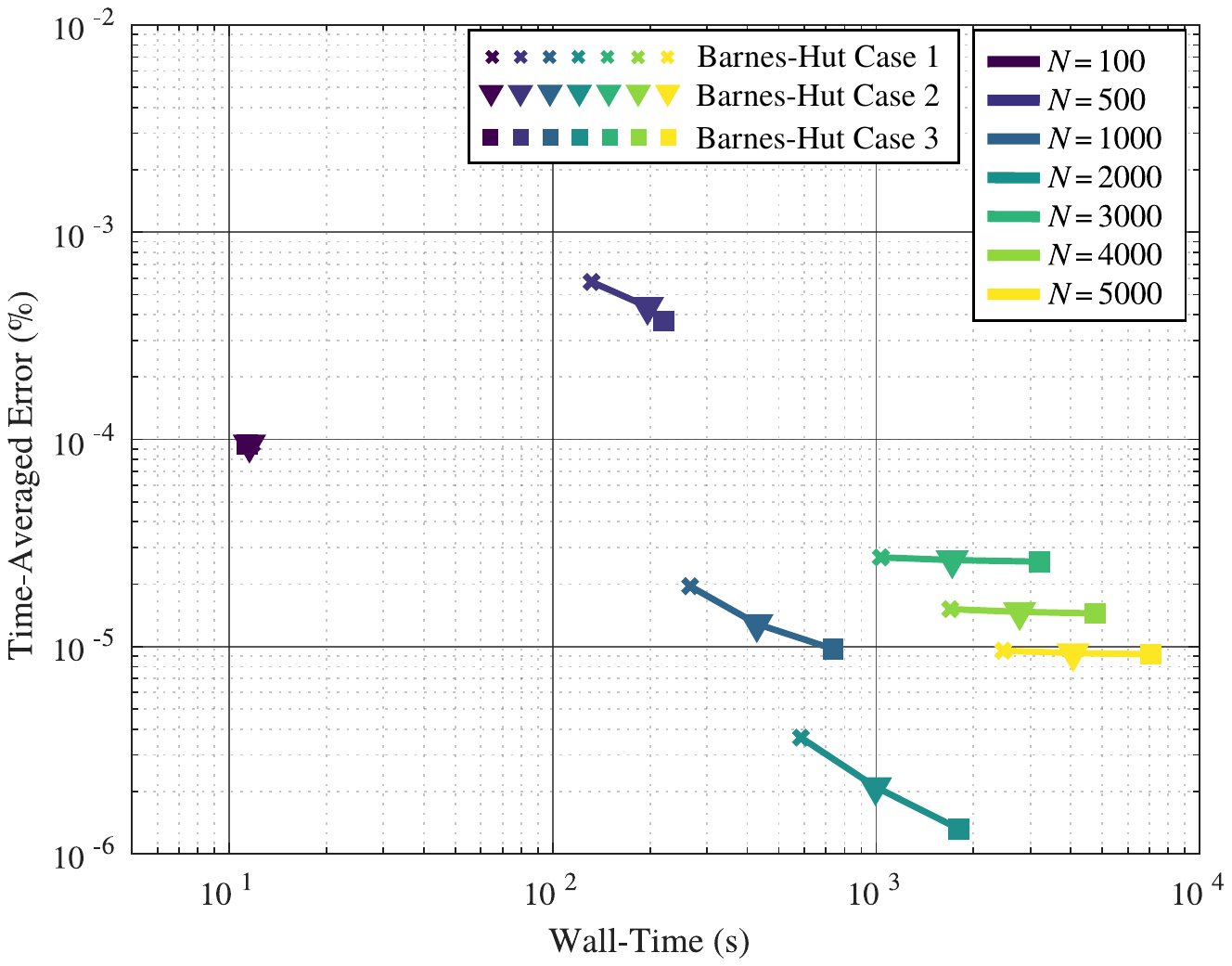}
		\caption[]%
		{Time-averaged $\textup{MAE}_{D}$}    
		\label{MAED_BHBH}
	\end{subfigure}
	\hfill
	\begin{subfigure}[b]{0.475\textwidth}  
		\centering 
		\includegraphics[scale=0.575, trim=5cm 8cm 25 9cm]{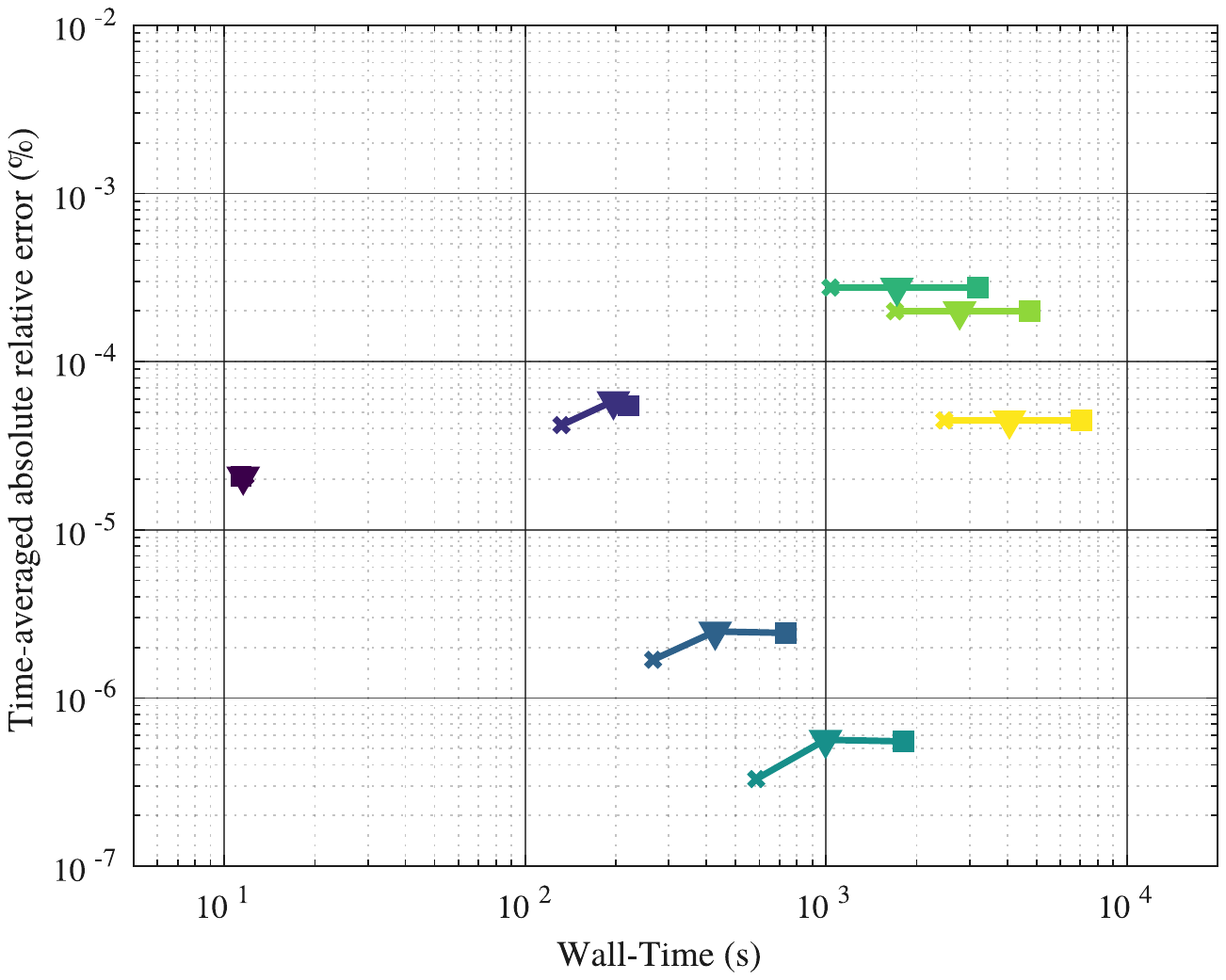}
		\caption[]%
		{Time-averaged $\textup{AE}_{H}$}
		\label{AE_BHBH}
	\end{subfigure}
	\caption[]
	{QoI results for the Barnes--Hut method with Barnes--Hut clustering hyper-parameter settings (a) Time-averaged $\textup{MAE}_D$ versus wall-time; (b) Time-averaged $\textup{AE}_H$ versus wall-time. Color scale indicates the number of particles in the domain and shapes correspond the clustering criteria listed Table \ref{BH_clustering_HyperParameters}}  
	\label{BHBH_Results}
\end{figure*}

Figure \ref{BHBH_Results} presents the QoI reproductive results generated by the Barnes--Hut hyper-parameter settings listed in Table \ref{BH_clustering_HyperParameters}. The time-averaged $\textup{MAE}_{D}$ results shown in Fig.~\ref{MAED_BHBH} illustrate improved reproductive errors, with respect to both PTROM and GNAT methods, that are all sub 0.001\% errors across all cases and particle domain sizes. Similar to results generated by prior PTROM and GNAT reproductive experiments, as the particle count increases the error decreases due to the hierarchical decomposition method's ability to accurately reproduce the growing number of particle trajectory paths. The time-averaged $\textup{AE}_{H}$ results shown in Fig.~\ref{AE_BHBH} also show an improved error over PTROM and GNAT method with sub 0.001\% error across all cases and particle domain sizes. The hierarchical decomposition method even reaches errors down to $10^{-7}\%$. However, it is important to note that the decrease in errors come with an increased wall-time of about two orders of magnitude with respect to the PTROM results in Fig.~\ref{PTROM_NarrowWidthResults}. A more detailed discussion of wall-time time performance across all methods will be given in Section \ref{computational_performance}

Next, hyper-parameters of the Barnes--Hut hierarchical decomposition with neighbor search clustering are presented in Table \ref{NN_clustering_HyperParameters}. Here, $p_c=0$ generates the lowest number of clusters sources and $p_c=2$ generates the largest number of cluster sources. Note that as $p_c \rightarrow \infty$ the original FOM would be retrieved.

\begin{table}[h!]
  \caption{Barnes--Hut hierarchical decomposition with neighbor search clustering hyper-parameter settings.}
\makebox[\textwidth][c]{
    \begin{tabular}{ccccccccc}
\toprule
      &       &   \multicolumn{6}{c}{Number of particles, $N$, in the domain} \\
\cmidrule(lr){3-9}
 & Hyper-parameter & 100  & 500 & 1000  & 2000  & 3000 & 4000 & 5000 \\ \midrule \midrule
Nearest-Neighbor Case 1 & $p_c$ & 0 & 0 & 0  & 0  & 0  & 0  & 0  \\
Nearest-Neighbor Case 2 & $p_c$ & 1 & 1 & 1 & 1  & 1  & 1  & 1  \\
Nearest-Neighbor Case 3 & $p_c$ & 2 & 2 & 2  & 2  & 2  & 2  & 2  \\
\bottomrule
\end{tabular}%
}
\label{NN_clustering_HyperParameters}%
\end{table}%

\begin{figure*}[h!]
	\centering
	\begin{subfigure}[b]{0.475\textwidth}
		\centering
		\includegraphics[scale=0.575, trim=4cm 8cm 50 9cm]{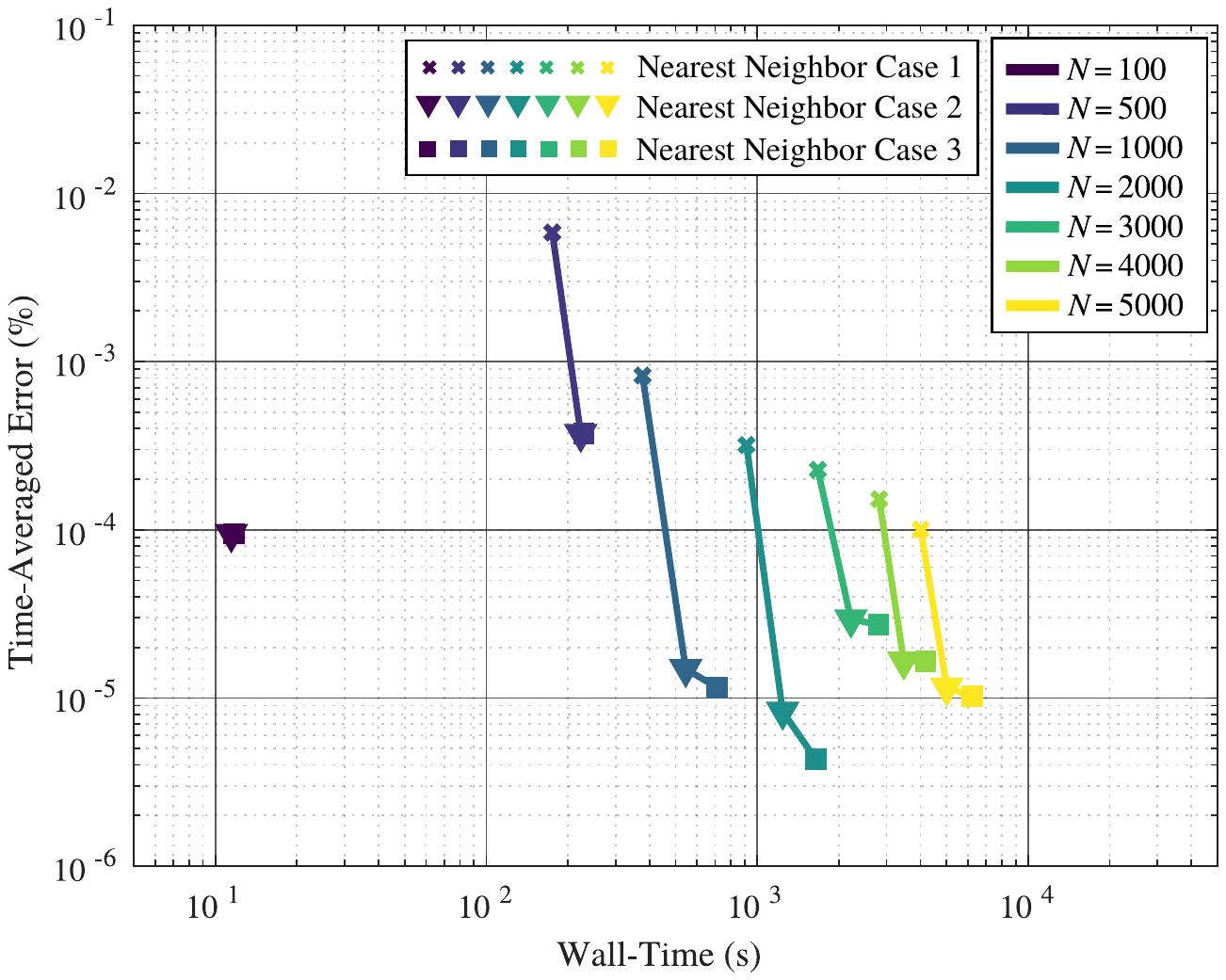}
		\caption[]%
		{Time-averaged $\textup{MAE}_{D}$}    
		\label{MAED_BHNN}
	\end{subfigure}
	\hfill
	\begin{subfigure}[b]{0.475\textwidth}  
		\centering 
		\includegraphics[scale=0.575, trim=5cm 8cm 25 9cm]{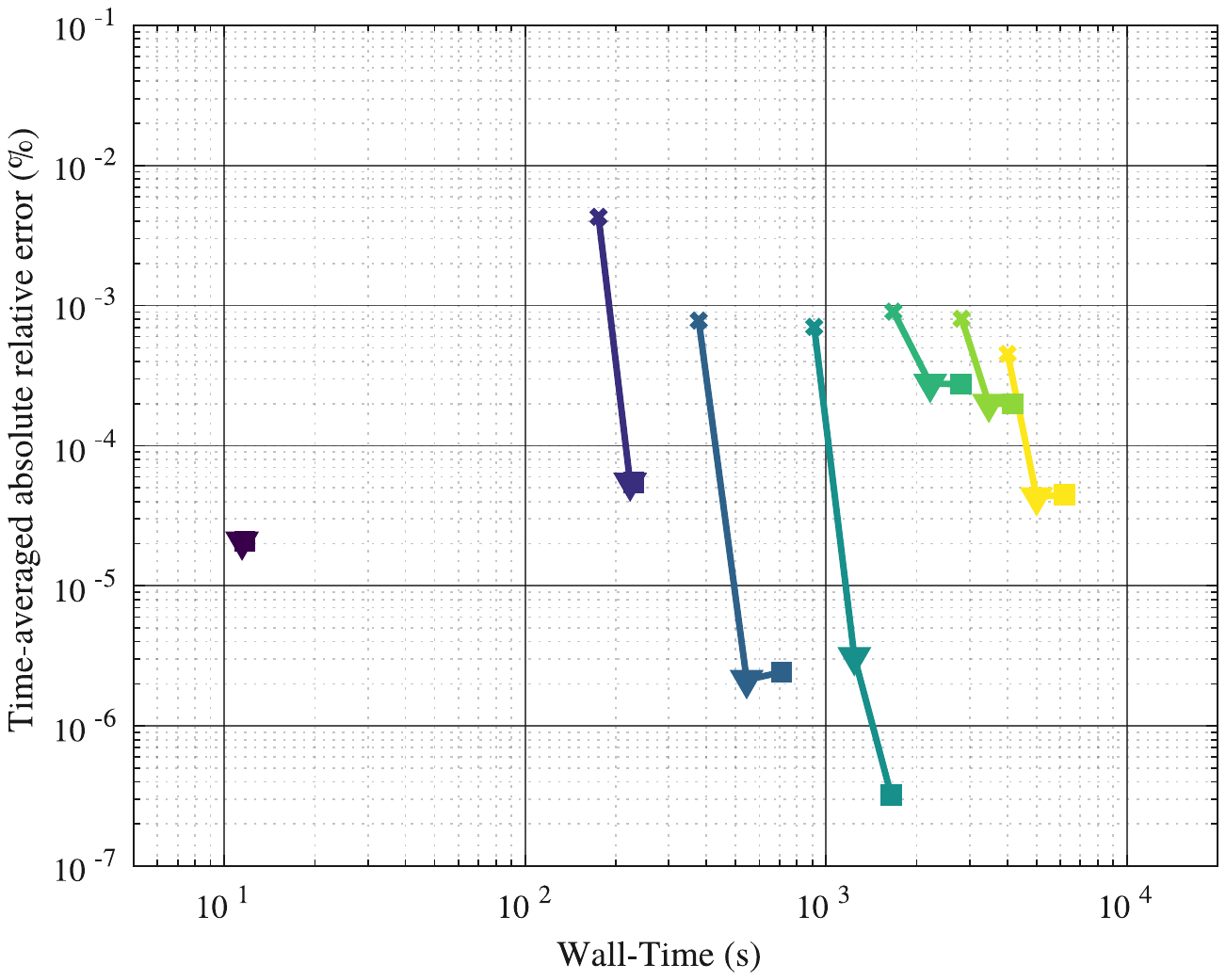}
		\caption[]%
		{Time-averaged $\textup{AE}_{H}$}
		\label{AE_BHNN}
	\end{subfigure}
	\caption[]
		{QoI results for the Barnes--Hut method with neighbor search clustering hyper-parameter settings (a) Time-averaged $\textup{MAE}_D$ versus wall-time; (b) Time-averaged $\textup{AE}_H$ versus wall-time. Color scale indicates the number of particles in the domain and shapes correspond the clustering criteria listed Table \ref{BH_clustering_HyperParameters}.}  
	\label{BHNN_Results}
\end{figure*}

Figure \ref{BHNN_Results} presents the QoI reproductive results generated by the Barnes--Hut hyper-parameter settings listed in Table \ref{NN_clustering_HyperParameters}. The time-averaged $\textup{MAE}_{D}$ results shown in Fig.~\ref{MAED_BHNN} illustrate improved reproductive errors, with respect to both PTROM and GNAT methods, that are all sub 0.01\% errors across all cases and particle domain sizes. Similar to results generated by prior PTROM and GNAT reproductive experiments, as the particle count increases the error decreases due to the hierarchical decomposition method's ability to accurately reproduce the growing number of particle trajectory paths. The time-averaged $\textup{AE}_{H}$ results shown in Fig.~\ref{AE_BHNN} also show an improved error over PTROM and GNAT method with sub 0.01\% error across all cases and particle domain sizes. The hierarchical decomposition method even reaches errors down to $10^{-7}\%$. However, it is important to note that the decrease in errors come with an increased wall-time of about two orders of magnitude with respect to the PTROM results in Fig.~\ref{PTROM_NarrowWidthResults}. A more detailed discussion of wall-time time performance across all methods will is now given.

\subsubsection{Computational savings and performance} \label{computational_performance}

\begin{figure*}[h!]
	\centering
	\begin{subfigure}[b]{0.475\textwidth}
		\centering
		\includegraphics[scale=0.575, trim=4cm 8cm 50 8cm]{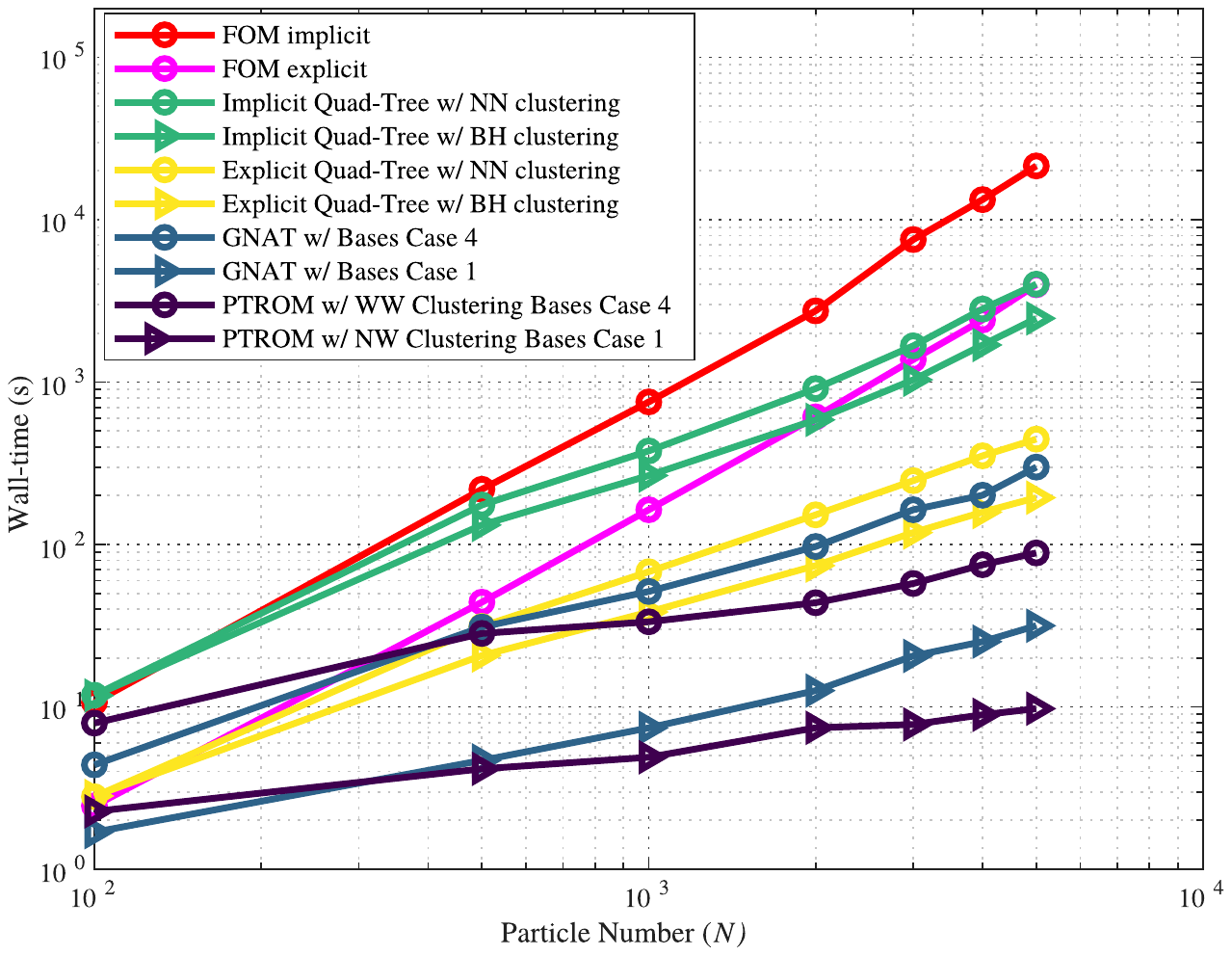}
		\caption[]%
		{Incurred wall-time}    
		\label{wall-time}
	\end{subfigure}
	\hfill
	\begin{subfigure}[b]{0.475\textwidth}  
		\centering 
		\includegraphics[scale=0.575, trim=5cm 8cm 25 8cm]{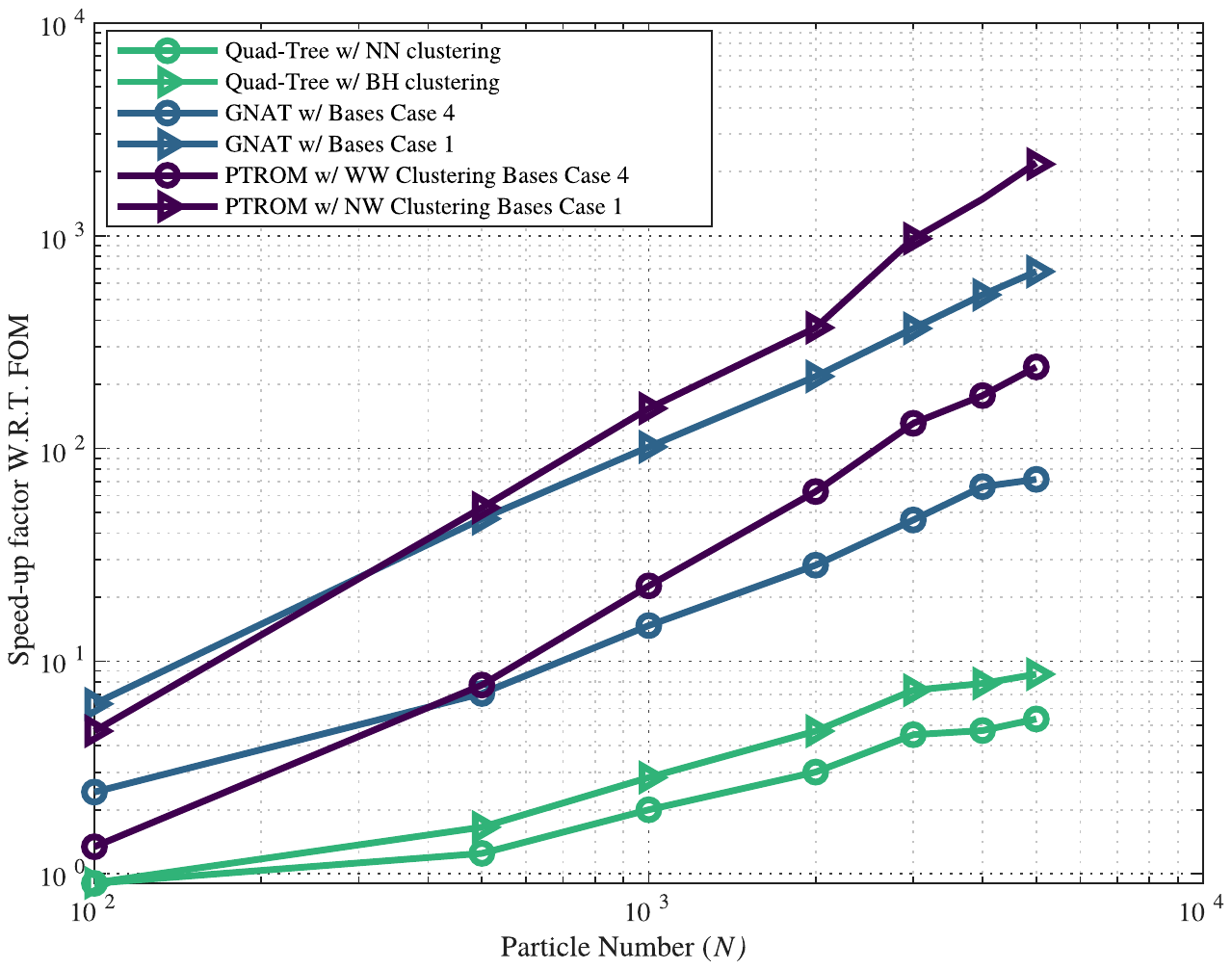}
		\caption[]%
		{Speed-up factors}
		\label{speed-up}
	\end{subfigure}
	\caption[]
	{Left: Wall-time assessment of FOM implicit and explicit, Barnes-Hut method with neighbor and Barnes-Hut clustering, GNAT with a POD bases of rank and rank, and PTROM with a POD bases of rank and rank. Right: Speed-up factors corresponding to the wall-time assessment with respect to the FOM  wall-time.}  
	\label{Savings_Results}
\end{figure*}

Wall-time and corresponding speed-up of the PTROM are presented in Fig.~\ref{Savings_Results}. Specifically, the PTROM narrow-width bases case 1 (highest PTROM computational savings) and narrow-width bases case 4 (lowest PTROM computational savings) are compared against GNAT bases case 1 and 4, and against the Barnes--Hut hierarchical decomposition with Barnes--Hut clustering set to $\theta=2$ and neighbor search clustering with $p_c=0$. In addition, the FOM with an explicit predictor-corrector modified Euler time integration is compared with the PTROM. Finally, the modified Euler's technique is equipped with hiearchical decomposition is also compared, such that Barnes--Hut clustering is set to $\theta=2$ and neighbor search clustering is set to $p_c=0$.

From an overhead view, results presented in Fig.~\ref{Savings_Results} show that the PTROM equipped with narrow-width neighbor search clustering and bases case 1, operates logarithmically efficient and out-performs all time-integration methods as the number of $particles$ increase. However, at $N=100$, the GNAT equipped with bases case 1 out-performs the PTROM due to lower number of iterations incurred during the Gauss-Newton loop. However, as the number of particles increases the GNAT cannot out-perform the PTROM even if less Gauss-Newton iterations are incurred, as the GNAT pair-wise interaction loop requires back-projection from the low-dimensional embedding back to the high-dimensional space to perform summation over all $N$ at the sampled residuals. Recall, the PTROM hyper-reduced pair-wise interaction does not require back-projection or summation over all $N$ sources. Similarly, at $N=100$ and $500$ the explicit hierarchical decomposition out-performs the PTROM method with wide-width neighbor search clustering and Bases Case 4, which can be attributed to the rapid explicit nature of the time integration which requires no Newton iterations. However, as the number of particles increases the explicit time-integration cannot out-perform the PTROM method even with a tight tolerance and relatively conservative hyper-parameters, i.e. Bases Case 4 with wide-width neighbor search clustering. 

Figure \ref{speed-up} highlights the computational performances of the implicit hierarchical decomposition, GNAT, and PTROM methods in terms of a speed-up factor as presented in Eq.~\ref{SF_equation}. It is shown in Fig.~\ref{speed-up} that the PTROM can reach a speed-up factor of up to 2199, i.e. the PTROM can generate results over 2000 times faster than the FOM and deliver sub 0.1 \% QoI reproductive erros for $N=5000$. The GNAT performance for Bases Case 1 provides a speed-up factor of 678 for $N=5000$. Barnes--Hut and narrow-neighbor search clustering for the hierarchical decomposition approach can at best provide a speed-up factor of 8.68 and 5.35 respectively. 

\begin{figure*}[h!]
	\centering
	\includegraphics[scale=0.575, trim=2cm 4cm 50 3cm]{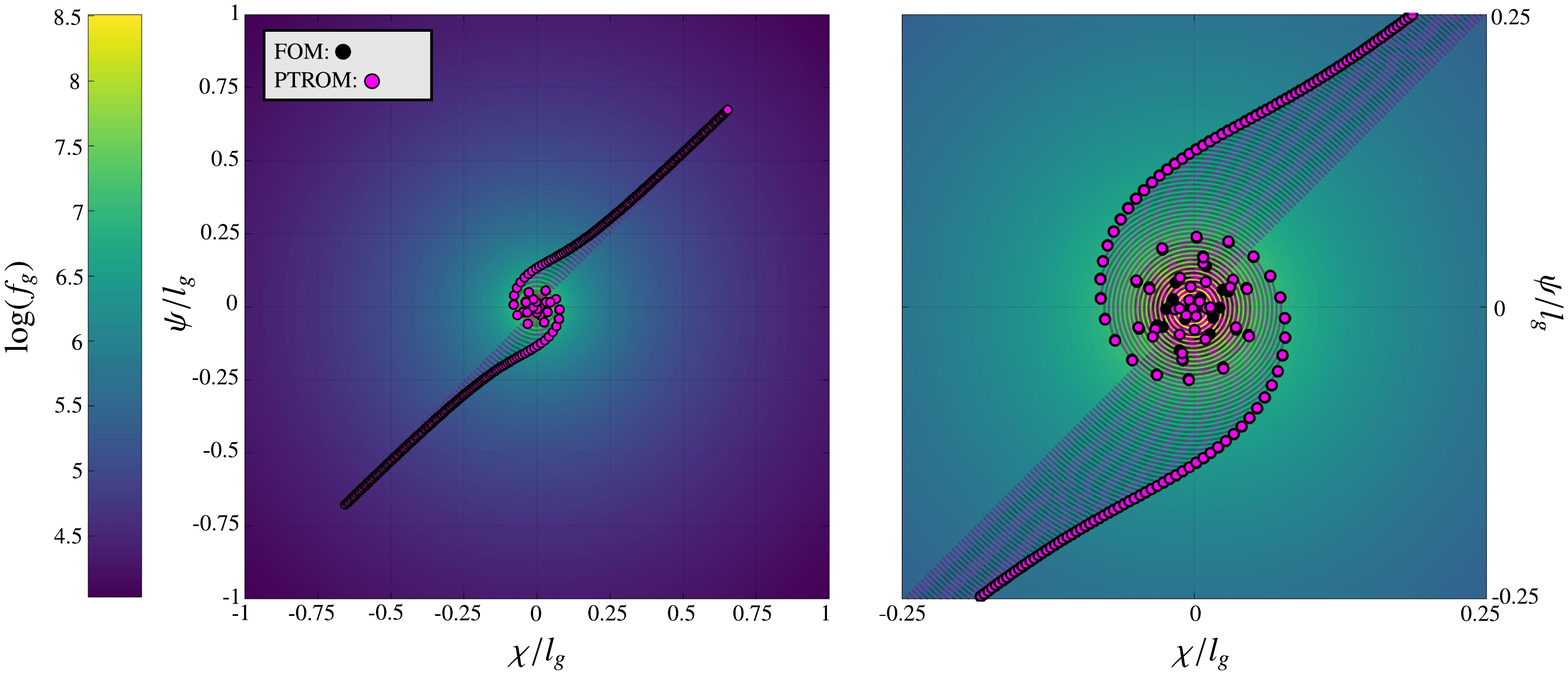}
		\caption[]
	{PTROM and FOM simulation snapshot at $t=5$ with the corresponding FOM velocity field. PTROM hyper-parameter settings correspond to the Bases Case 1 for a narrow-width neighbor search.  Left: Complete view of the simulation snapshot. Right: Magnified view of particles near the center. }  
	\label{PTROM_loose}
\end{figure*}

\begin{figure*}[h!]
	\centering
	\includegraphics[scale=0.575, trim=2cm 4cm 50 3cm]{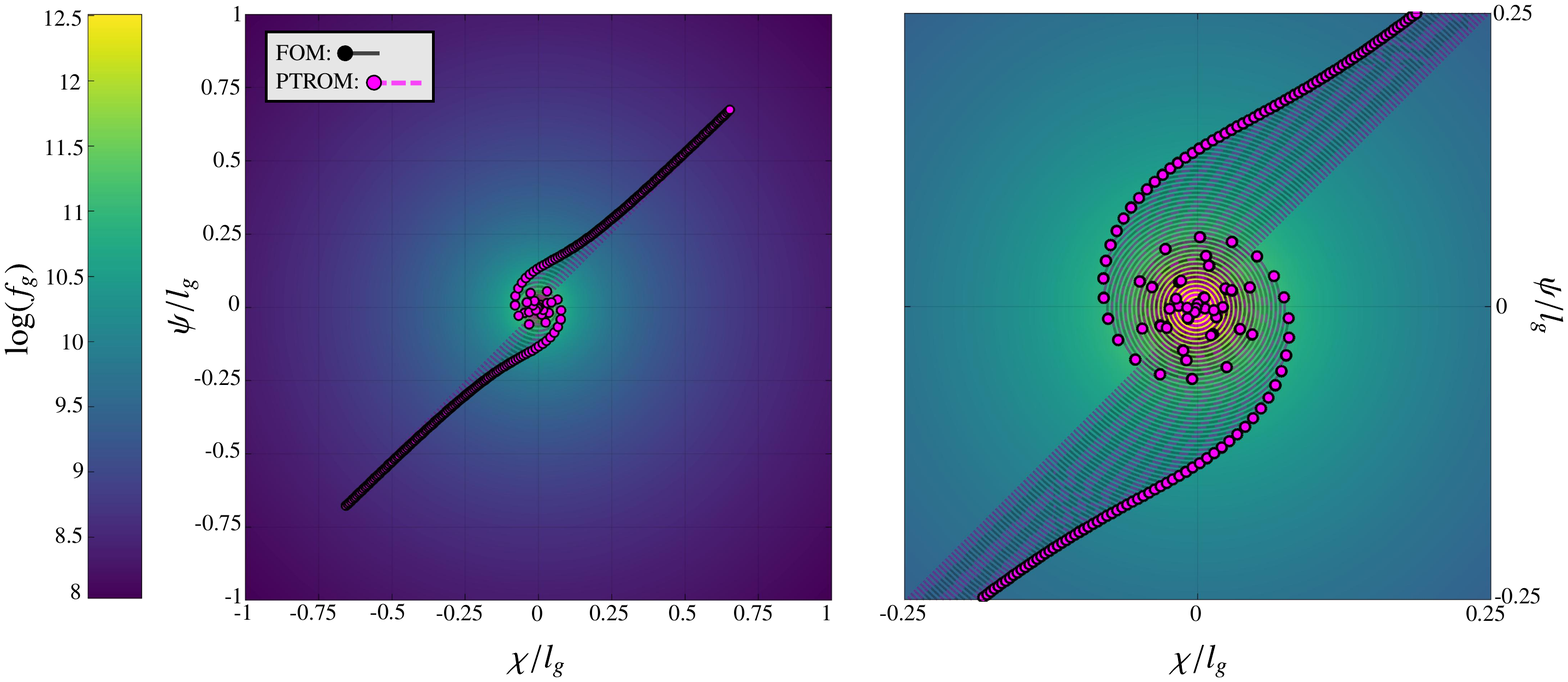}
		\caption[]
{PTROM and FOM simulation snapshot at $t=5$ with the corresponding FOM velocity field. PTROM hyper-parameter settings correspond to the Bases Case 4 for a wide-width neighbor search.  Left: Complete view of the simulation snapshot. Right: Magnified view of particles near the center. }  
	\label{PTROM_tight}
\end{figure*}

Figures \ref{PTROM_loose} and \ref{PTROM_tight} provide qualitative results of the PTROM  for $N=500$ using Bases Case 1 for a narrow-width neighbor search and Bases Case 4 for a wide-width neighbor search, respectively. Results show that the PTROM is capable of reproducing the FOM particle trajectories to a high level of accuracy away from the center particle with strong circulation. In the case of the PTROM with hyper-parameter settings of Bases Case 1 and a narrow-width neighbor search, results show a deviation in particle path trajectory near the center particle where the particle velocities increase inversely proportional to the squared distance from the center particle, as quantified by the Biot-Savart law in Eq.~\ref{BS_kernel}. However, in the case of the PTROM with Bases Case 4 for a wide-width neighbor search, the additional bases and sources along with a tightened tolerance has significantly improved the trajectory reconstruction near the center particle.

\section{Conclusions} \label{Section_Conclusion}

In this work, the projection-tree reduced order modeling (PTROM) technique is presented. The PTROM provides a new perspective in accelerating $N$-body problems by bridging traditional hierarchical decomposition methods with recent advancements in projection-based reduced order modeling to overcome the pairwise interaction problem and achieve $N$-independent operational count complexity. The PTROM is based on performing projection-based model reduction via the Gauss--Newton with approximated tensors (GNAT) approach, and hierarchical decomposition via the Barnes--Hut tree method. The effectiveness of the PTROM was tested on parametric and reproductive problems. In the parametric experiments, the PTROM was tested on vortex simulations with 1000 degrees-of-freedom where variable vortex circulation over a predefined space corresponded to parametric inputs. It was shown that the PTROM can deliver on average QoI errors between 0.0135-0.0634\% while delivering 1.7-3.94$\times$ computational speed-up, all with respect to the FOM. In the reproductive experiments, the PTROM was tested on a vortex simulation with degrees-of-freedom ranging from 200 to 10000 and was compared against stand-alone hiearchical decomposition methods, stand-alone projection-based model reduction, and explicit time integration with and without hierarchical decomposition. A range of PTROM hyper-parameters were varied and it was found that the PTROM was capable of delivering sub-0.1\% errors while delivering over a 2000$\times$ speed-up and out-performed explicit integration, the GNAT method, Barnes-Hut hierarchical decomposition with implicit and explicit integration.

Future work will involve investigating the impact hyper-parameters have on the PTROM convergence, and performing parallelized simulations with a large-scale particle count to assess more rigorously computional savings, core-hours, and resource expense. In addition, future work will include equipping the PTROM with more recent advances in model-reduction that include integrating nonlinear manifolds discovered by convolutional neural networks as the underlying reduced basis for dimensional compression. Future work will also focus on more physics driven applications of the PTROM, such as modeling heat-deposition of additive manufacturing, and heat/fluid transport phenomena.

\section*{Acknowledgements}

The inception of this work was developed while S.~N.~Rodriguez held a visiting scientist position at the University of Washington, Seattle. S.~N.~Rodriguez would like to acknowledge the NRL Isabella and Jerome Karle's Distinguished Scholar Fellowship and Dr.~Nathan F. Wagenhoffer for his expertise and insightful conversations on hierarchical decomposition and $N$-body acceleration methods. S.~L.~Brunton  would like to acknowledge support from the ARO PECASE (W911NF-19-1-0045). A.~P.~Iliopoulos, J.~C.~ Steuben, and J.~G.~Michopoulos would like to acknowledge support from ONR through NRL core funding.

\appendix \label{appendix}
\section{Barnes--Hut search for maximum particles per leaf node }

\begin{figure*}[h!]
	\centering
	\begin{subfigure}[b]{0.475\textwidth}
		\centering
		\includegraphics[scale=0.55, trim=4cm 8cm 50 7cm]{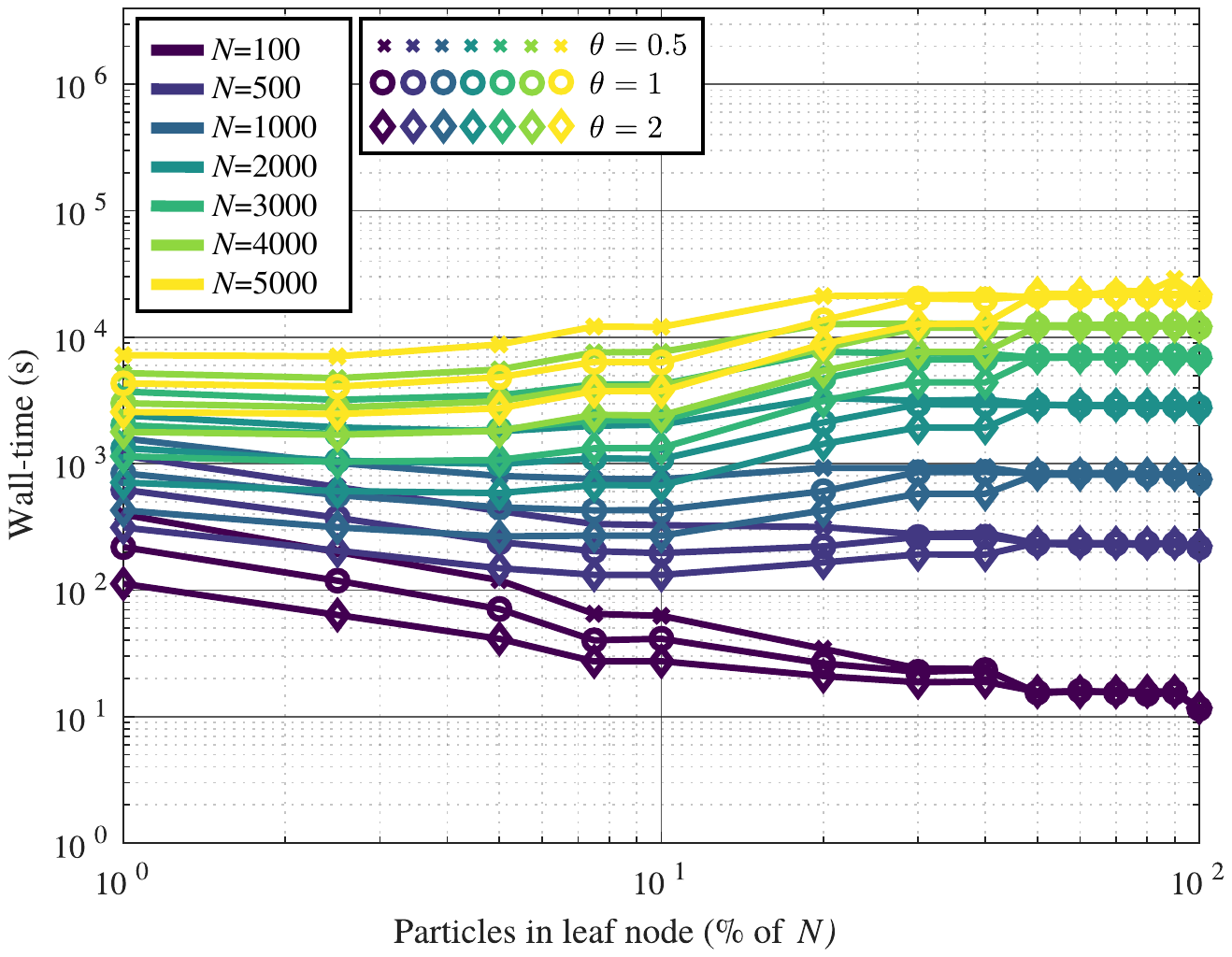}
		\caption[]%
		{Barnes--Hut clustering}    
		\label{}
	\end{subfigure}
	\hfill
	\begin{subfigure}[b]{0.475\textwidth}  
		\centering 
		\includegraphics[scale=0.55, trim=5cm 8cm 25 7cm]{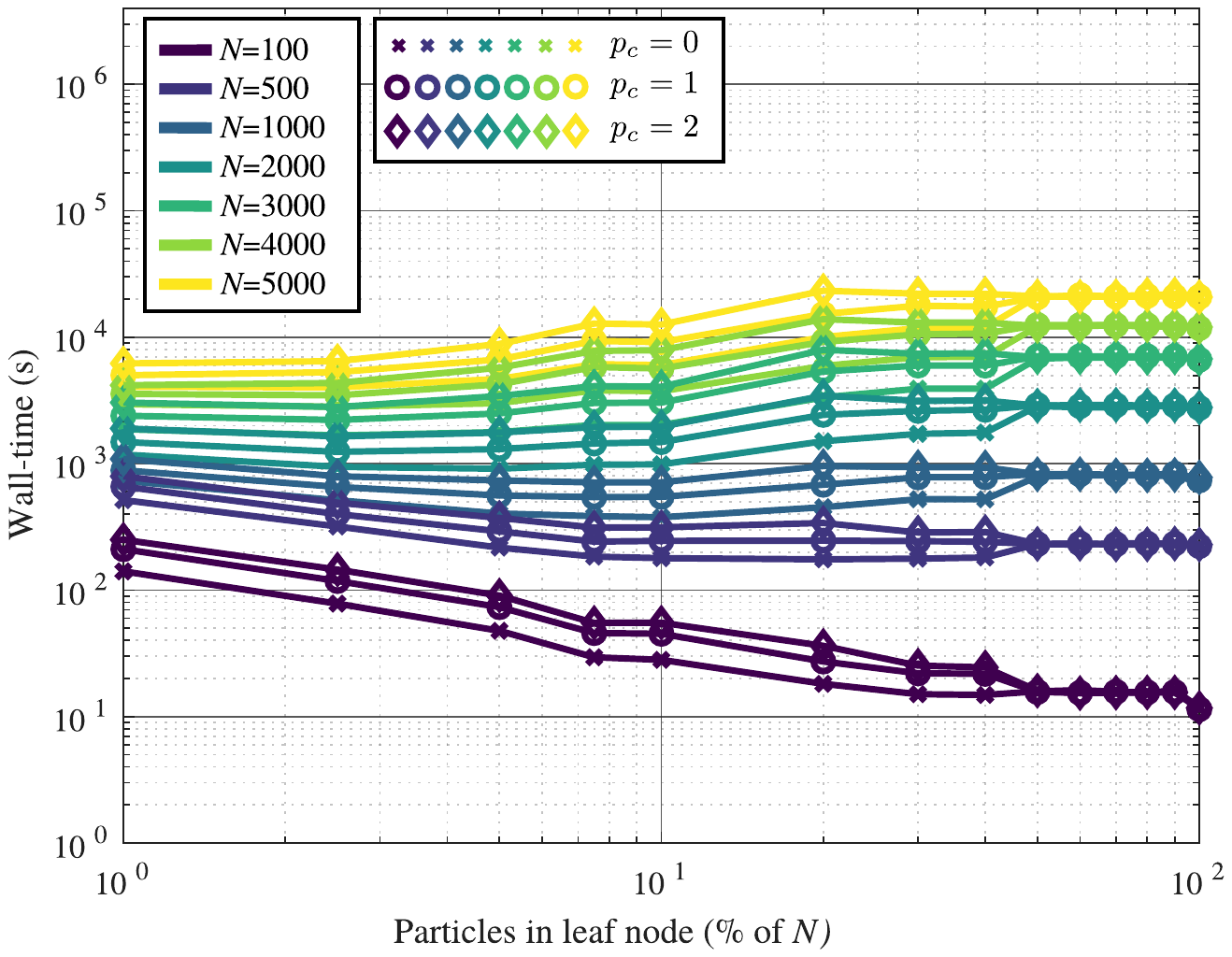}
		\caption[]%
		{Neighbor search clustering}
		\label{}
	\end{subfigure}
	\caption[]
	{Wall-times for implicit integration versus MMPL as a percent of $N$ rounded up.} 
	\label{Implicit_MaxPnts}
\end{figure*}

\begin{figure*}[h!]
	\centering
	\begin{subfigure}[b]{0.475\textwidth}
		\centering
		\includegraphics[scale=0.55, trim=4cm 8cm 50 8cm]{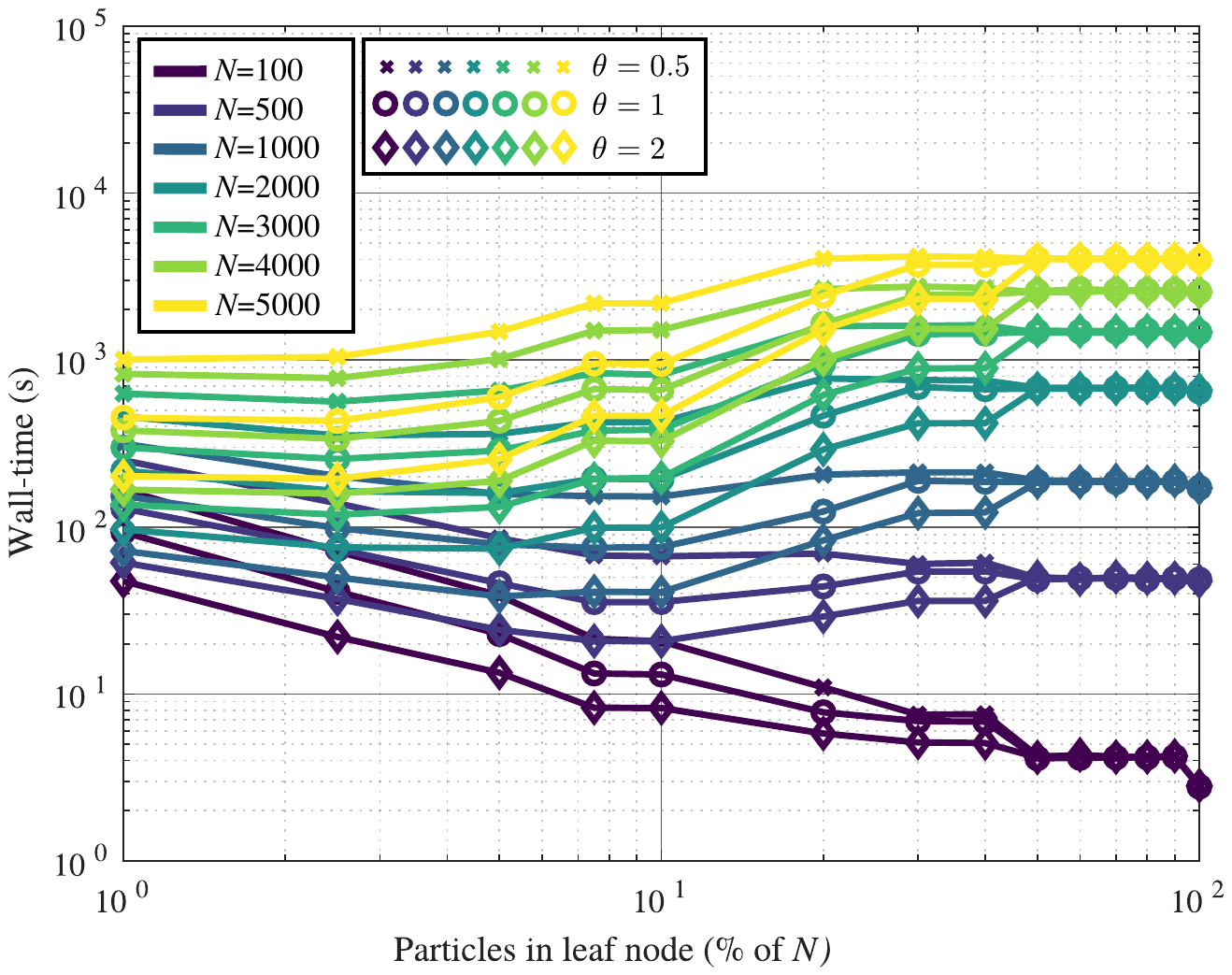}
		\caption[]%
		{Barnes--Hut clustering}    
		\label{}
	\end{subfigure}
	\hfill
	\begin{subfigure}[b]{0.475\textwidth}  
		\centering 
		\includegraphics[scale=0.55, trim=5cm 8cm 25 8cm]{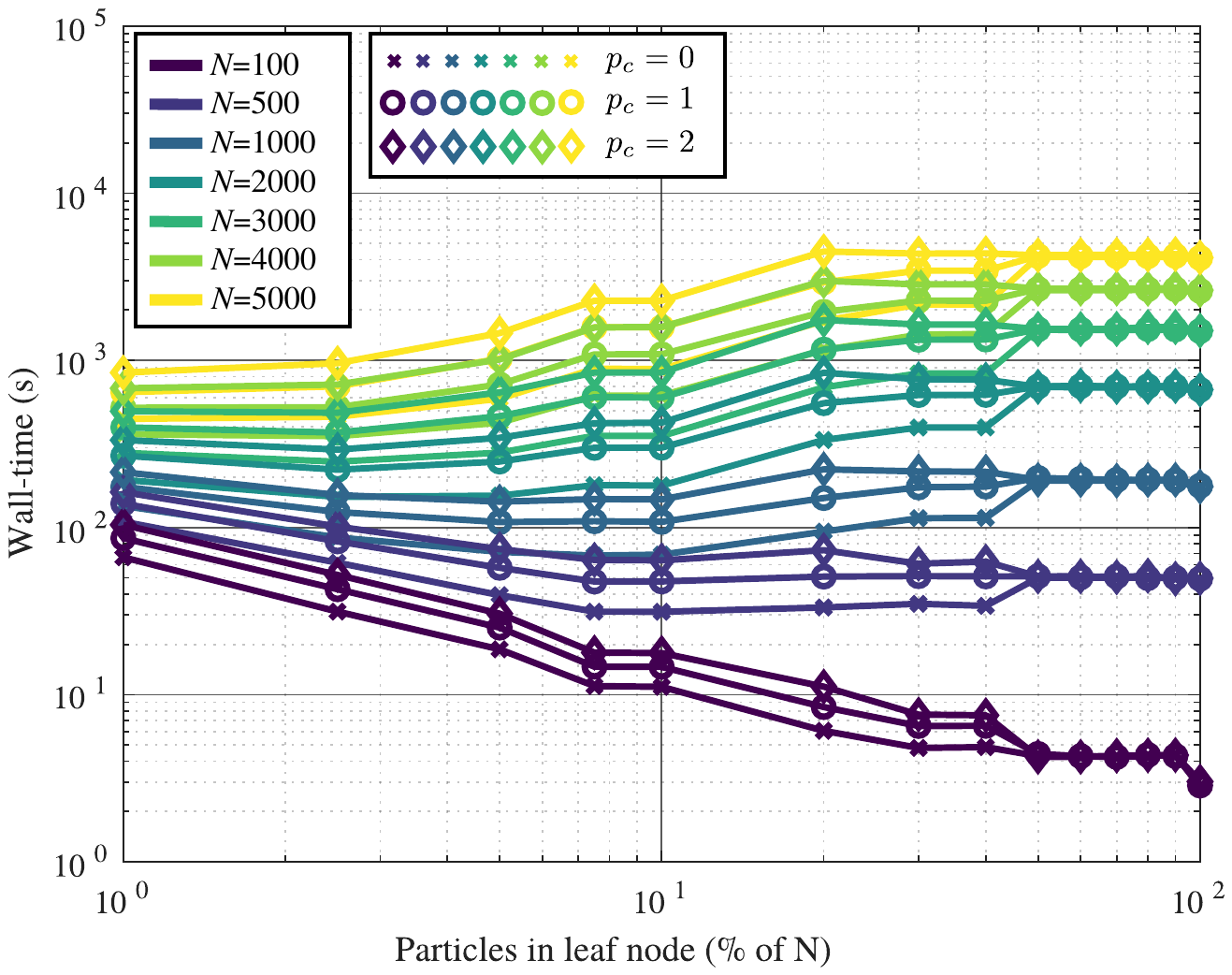}
		\caption[]%
	{Neighbor search clustering}
		\label{}
	\end{subfigure}
	\caption[]
	{ Wall-times for explicit integration versus MMPL as a percent of $N$ rounded up.}  
	\label{Explicit_MaxPnts}
\end{figure*}

As discussed in Section \ref{BarnesHutResultSubSection}, an additional parametric study was performed to find the maximum number of points per leaf nodes that would provide the most rapid results given the clustering hyper-parameters chosen in Tables \ref{BH_clustering_HyperParameters} and \ref{NN_clustering_HyperParameters}.  The maximum number of particles per leaf node (MMPL) were chosen by running the hyper-parameters in Tables \ref{BH_clustering_HyperParameters} and \ref{NN_clustering_HyperParameters} for a range of points per leaf node and chose the simulation with the lowest wall-time to compare against the PTROM. Results for the hierarchical decomposition with implicit time integration are presented in Fig.~\ref{Implicit_MaxPnts} and Fig.~\ref{Explicit_MaxPnts}.

 It was found that hierarchical decomposition did not improve pairwise interaction computations for lower particle count since the data structure build would incur more cost than computing the simulations directly. As a result the fastest run time for some particle counts were those that had the maximum number of particles per leaf node equal to the total number of particles in the domain. Tables \ref{Implicit_mxPntsList} and \ref{Explicit_mxPntsList} list the maximum number of points chosen for each clustering hyper-parameter and $N$ in the implicit and explicit integration cases, respectively.

\begin{table}[h!]
	\caption{Maximum number of particles per leaf node (MPPL) used in implicit time integration equipped with hierarchical decomposition.}
	\makebox[\textwidth][c]{
		\begin{tabular}{ccccccc}
			\toprule
			  &     \multicolumn{3}{c}{Barnes--Hut clustering MPPL}  &  \multicolumn{3}{c}{Neighbor search clustering MPPL} \\ \cmidrule(lr){2-4} \cmidrule(lr){5-7}
			$N$  &  $\theta = 0.5$ & $\theta = 1$  & $\theta = 2$  &  $p_c = 0$ & $p_c = 1$  &$p_c= 2$  \\ \midrule \midrule
			100     &  100 & 100 & 100 &  100 & 100 & 100 \\
			500    &  500 & 50 & 50  &  100 & 500 & 500\\
			1000   &  1000 & 75 & 50 &  100 & 75 & 75 \\
			2000    &  100 & 100 & 100  &  100 & 50 & 50\\
			3000    &  75 &  75 & 75  &  75 & 75 & 75 \\
			4000    &  100 & 100 & 100 &  100 & 100 & 40 \\
			5000    &  125 & 125 & 125 &  125 & 50 & 50 \\
			\bottomrule
		\end{tabular}%
	}
		\label{Implicit_mxPntsList}
\end{table}%

\begin{table}[h!]
	\caption{Maximum number of particles per leaf node (MPPL) used in explicit time integration equipped with hierarchical decomposition.}
	\makebox[\textwidth][c]{
		\begin{tabular}{ccccccc}
			\toprule
			&     \multicolumn{3}{c}{Barnes--Hut clustering MPPL}  &  \multicolumn{3}{c}{Neighbor search clustering MPPL} \\ \cmidrule(lr){2-4} \cmidrule(lr){5-7}
			$N$  &  $\theta = 0.5$ & $\theta = 1$  & $\theta = 2$  &  $p_c = 0$ & $p_c = 1$  &$p_c= 2$  \\ \midrule \midrule
			100     &  100 & 100 & 100  &  100  & 100 & 100 \\
			500    &  500 & 38   & 50   &  50   & 50 & 450\\
			1000   &  100 & 75   & 50    &  50   & 50   & 50 \\
			2000    &  50 & 100  & 100  &  100 & 100   & 50\\
			3000    &  75 &  75   & 75    &  75   & 75   & 75 \\
			4000    &  100 & 100 & 100  &  100  & 100 & 40 \\
			5000    &  50 & 125 & 125   &  125  & 125   & 50 \\
			\bottomrule
		\end{tabular}%
	}
	\label{Explicit_mxPntsList}
\end{table}%

\clearpage
\bibliography{references}

\begin{thebibliography}{74}
\expandafter\ifx\csname natexlab\endcsname\relax\def\natexlab#1{#1}\fi
\providecommand{\url}[1]{\texttt{#1}}
\providecommand{\href}[2]{#2}
\providecommand{\path}[1]{#1}
\providecommand{\DOIprefix}{doi:}
\providecommand{\ArXivprefix}{arXiv:}
\providecommand{\URLprefix}{URL: }
\providecommand{\Pubmedprefix}{pmid:}
\providecommand{\doi}[1]{\href{http://dx.doi.org/#1}{\path{#1}}}
\providecommand{\Pubmed}[1]{\href{pmid:#1}{\path{#1}}}
\providecommand{\bibinfo}[2]{#2}
\ifx\xfnm\relax \def\xfnm[#1]{\unskip,\space#1}\fi
\bibitem[{Afkham and Hesthaven(2017)}]{afkham2017structure}
\bibinfo{author}{Afkham, B.M.}, \bibinfo{author}{Hesthaven, J.S.},
  \bibinfo{year}{2017}.
\newblock \bibinfo{title}{Structure preserving model reduction of parametric
  {H}amiltonian systems}.
\newblock \bibinfo{journal}{SIAM Journal on Scientific Computing}
  \bibinfo{volume}{39}, \bibinfo{pages}{A2616--A2644}.
\bibitem[{Akoz and Moored(2018)}]{akoz2018unsteady}
\bibinfo{author}{Akoz, E.}, \bibinfo{author}{Moored, K.W.},
  \bibinfo{year}{2018}.
\newblock \bibinfo{title}{Unsteady propulsion by an intermittent swimming
  gait}.
\newblock \bibinfo{journal}{Journal of Fluid Mechanics} \bibinfo{volume}{834},
  \bibinfo{pages}{149}.
\bibitem[{Amsallem and Farhat(2012)}]{amsallem2012stabilization}
\bibinfo{author}{Amsallem, D.}, \bibinfo{author}{Farhat, C.},
  \bibinfo{year}{2012}.
\newblock \bibinfo{title}{Stabilization of projection-based reduced-order
  models}.
\newblock \bibinfo{journal}{International Journal for Numerical Methods in
  Engineering} \bibinfo{volume}{91}, \bibinfo{pages}{358--377}.
\bibitem[{Amsallem et~al.(2012)Amsallem, Zahr and
  Farhat}]{amsallem2012nonlinear}
\bibinfo{author}{Amsallem, D.}, \bibinfo{author}{Zahr, M.J.},
  \bibinfo{author}{Farhat, C.}, \bibinfo{year}{2012}.
\newblock \bibinfo{title}{Nonlinear model order reduction based on local
  reduced-order bases}.
\newblock \bibinfo{journal}{International Journal for Numerical Methods in
  Engineering} \bibinfo{volume}{92}, \bibinfo{pages}{891--916}.
\bibitem[{Antoulas(2005)}]{antoulas2005approximation}
\bibinfo{author}{Antoulas, A.C.}, \bibinfo{year}{2005}.
\newblock \bibinfo{title}{Approximation of large-scale dynamical systems}.
\newblock \bibinfo{publisher}{SIAM}.
\bibitem[{Barnes and Hut(1986)}]{barnes1986hierarchical}
\bibinfo{author}{Barnes, J.}, \bibinfo{author}{Hut, P.}, \bibinfo{year}{1986}.
\newblock \bibinfo{title}{A hierarchical o (n log n) force-calculation
  algorithm}.
\newblock \bibinfo{journal}{nature} \bibinfo{volume}{324},
  \bibinfo{pages}{446}.
\bibitem[{Barrault et~al.(2004)Barrault, Maday, Nguyen and
  Patera}]{barrault2004empirical}
\bibinfo{author}{Barrault, M.}, \bibinfo{author}{Maday, Y.},
  \bibinfo{author}{Nguyen, N.C.}, \bibinfo{author}{Patera, A.T.},
  \bibinfo{year}{2004}.
\newblock \bibinfo{title}{An ‘empirical interpolation’method: application
  to efficient reduced-basis discretization of partial differential equations}.
\newblock \bibinfo{journal}{Comptes Rendus Mathematique} \bibinfo{volume}{339},
  \bibinfo{pages}{667--672}.
\bibitem[{Benner et~al.(2017)Benner, Ohlberger, Cohen and
  Willcox}]{benner2017model}
\bibinfo{author}{Benner, P.}, \bibinfo{author}{Ohlberger, M.},
  \bibinfo{author}{Cohen, A.}, \bibinfo{author}{Willcox, K.},
  \bibinfo{year}{2017}.
\newblock \bibinfo{title}{Model reduction and approximation: theory and
  algorithms}.
\newblock \bibinfo{publisher}{SIAM}.
\bibitem[{Brown and Line(2005)}]{brown2005efficient}
\bibinfo{author}{Brown, R.E.}, \bibinfo{author}{Line, A.J.},
  \bibinfo{year}{2005}.
\newblock \bibinfo{title}{Efficient high-resolution wake modeling using the
  vorticity transport equation}.
\newblock \bibinfo{journal}{AIAA journal} \bibinfo{volume}{43},
  \bibinfo{pages}{1434--1443}.
\bibitem[{Brunton and Kutz(2019)}]{brunton2019data}
\bibinfo{author}{Brunton, S.L.}, \bibinfo{author}{Kutz, J.N.},
  \bibinfo{year}{2019}.
\newblock \bibinfo{title}{Data-driven science and engineering: Machine
  learning, dynamical systems, and control}.
\newblock \bibinfo{publisher}{Cambridge University Press}.
\bibitem[{Brunton and Noack(2015)}]{brunton2015closed}
\bibinfo{author}{Brunton, S.L.}, \bibinfo{author}{Noack, B.R.},
  \bibinfo{year}{2015}.
\newblock \bibinfo{title}{Closed-loop turbulence control: Progress and
  challenges}.
\newblock \bibinfo{journal}{Applied Mechanics Reviews} \bibinfo{volume}{67}.
\bibitem[{Buffa et~al.(2012)Buffa, Maday, Patera, Prud’homme and
  Turinici}]{buffa2012priori}
\bibinfo{author}{Buffa, A.}, \bibinfo{author}{Maday, Y.},
  \bibinfo{author}{Patera, A.T.}, \bibinfo{author}{Prud’homme, C.},
  \bibinfo{author}{Turinici, G.}, \bibinfo{year}{2012}.
\newblock \bibinfo{title}{A priori convergence of the greedy algorithm for the
  parametrized reduced basis method}.
\newblock \bibinfo{journal}{ESAIM: Mathematical Modelling and Numerical
  Analysis-Mod{\'e}lisation Math{\'e}matique et Analyse Num{\'e}rique}
  \bibinfo{volume}{46}, \bibinfo{pages}{595--603}.
\bibitem[{Bui-Thanh et~al.(2008)Bui-Thanh, Willcox and Ghattas}]{bui2008model}
\bibinfo{author}{Bui-Thanh, T.}, \bibinfo{author}{Willcox, K.},
  \bibinfo{author}{Ghattas, O.}, \bibinfo{year}{2008}.
\newblock \bibinfo{title}{Model reduction for large-scale systems with
  high-dimensional parametric input space}.
\newblock \bibinfo{journal}{SIAM Journal on Scientific Computing}
  \bibinfo{volume}{30}, \bibinfo{pages}{3270--3288}.
\bibitem[{Carlberg et~al.(2017)Carlberg, Barone and
  Antil}]{carlberg2017galerkin}
\bibinfo{author}{Carlberg, K.}, \bibinfo{author}{Barone, M.},
  \bibinfo{author}{Antil, H.}, \bibinfo{year}{2017}.
\newblock \bibinfo{title}{Galerkin v. least-squares {P}etrov--{G}alerkin
  projection in nonlinear model reduction}.
\newblock \bibinfo{journal}{Journal of Computational Physics}
  \bibinfo{volume}{330}, \bibinfo{pages}{693--734}.
\bibitem[{Carlberg et~al.(2015)Carlberg, Tuminaro and
  Boggs}]{carlberg2015preserving}
\bibinfo{author}{Carlberg, K.}, \bibinfo{author}{Tuminaro, R.},
  \bibinfo{author}{Boggs, P.}, \bibinfo{year}{2015}.
\newblock \bibinfo{title}{Preserving {L}agrangian structure in nonlinear model
  reduction with application to structural dynamics}.
\newblock \bibinfo{journal}{SIAM Journal on Scientific Computing}
  \bibinfo{volume}{37}, \bibinfo{pages}{B153--B184}.
\bibitem[{Carlberg et~al.(2013)Carlberg, Farhat, Cortial and
  Amsallem}]{carlberg2013gnat}
\bibinfo{author}{Carlberg, K.T.}, \bibinfo{author}{Farhat, C.},
  \bibinfo{author}{Cortial, J.}, \bibinfo{author}{Amsallem, D.},
  \bibinfo{year}{2013}.
\newblock \bibinfo{title}{The {GNAT} method for nonlinear model reduction:
  effective implementation and application to computational fluid dynamics and
  turbulent flows}.
\newblock \bibinfo{journal}{Journal of Computational Physics}
  \bibinfo{volume}{242}, \bibinfo{pages}{623--647}.
\bibitem[{Chaturantabut and Sorensen(2010)}]{chaturantabut2010nonlinear}
\bibinfo{author}{Chaturantabut, S.}, \bibinfo{author}{Sorensen, D.C.},
  \bibinfo{year}{2010}.
\newblock \bibinfo{title}{Nonlinear model reduction via discrete empirical
  interpolation}.
\newblock \bibinfo{journal}{SIAM Journal on Scientific Computing}
  \bibinfo{volume}{32}, \bibinfo{pages}{2737--2764}.
\bibitem[{Cipra(2000)}]{cipra2000best}
\bibinfo{author}{Cipra, B.A.}, \bibinfo{year}{2000}.
\newblock \bibinfo{title}{The best of the 20th century: Editors name top 10
  algorithms}.
\newblock \bibinfo{journal}{SIAM news} \bibinfo{volume}{33},
  \bibinfo{pages}{1--2}.
\bibitem[{Colmenares et~al.(2015)Colmenares, L{\'o}pez and
  Preidikman}]{colmenares2015computational}
\bibinfo{author}{Colmenares, J.D.}, \bibinfo{author}{L{\'o}pez, O.D.},
  \bibinfo{author}{Preidikman, S.}, \bibinfo{year}{2015}.
\newblock \bibinfo{title}{Computational study of a transverse rotor aircraft in
  hover using the unsteady vortex lattice method}.
\newblock \bibinfo{journal}{Mathematical Problems in Engineering}
  \bibinfo{volume}{2015}.
\bibitem[{Doerr et~al.(2016)Doerr, Harvey, No{\'e} and
  De~Fabritiis}]{doerr2016htmd}
\bibinfo{author}{Doerr, S.}, \bibinfo{author}{Harvey, M.},
  \bibinfo{author}{No{\'e}, F.}, \bibinfo{author}{De~Fabritiis, G.},
  \bibinfo{year}{2016}.
\newblock \bibinfo{title}{Htmd: high-throughput molecular dynamics for
  molecular discovery}.
\newblock \bibinfo{journal}{Journal of chemical theory and computation}
  \bibinfo{volume}{12}, \bibinfo{pages}{1845--1852}.
\bibitem[{Dongarra and Sullivan(2000)}]{dongarra2000guest}
\bibinfo{author}{Dongarra, J.}, \bibinfo{author}{Sullivan, F.},
  \bibinfo{year}{2000}.
\newblock \bibinfo{title}{Guest editors’ introduction: The top 10
  algorithms}.
\newblock \bibinfo{journal}{Computing in Science \& Engineering}
  \bibinfo{volume}{2}, \bibinfo{pages}{22}.
\bibitem[{Eldredge(2007)}]{eldredge2007numerical}
\bibinfo{author}{Eldredge, J.D.}, \bibinfo{year}{2007}.
\newblock \bibinfo{title}{Numerical simulation of the fluid dynamics of 2d
  rigid body motion with the vortex particle method}.
\newblock \bibinfo{journal}{Journal of Computational Physics}
  \bibinfo{volume}{221}, \bibinfo{pages}{626--648}.
\bibitem[{Eldredge et~al.(2002)Eldredge, Colonius and
  Leonard}]{eldredge2002vortex}
\bibinfo{author}{Eldredge, J.D.}, \bibinfo{author}{Colonius, T.},
  \bibinfo{author}{Leonard, A.}, \bibinfo{year}{2002}.
\newblock \bibinfo{title}{A vortex particle method for two-dimensional
  compressible flow}.
\newblock \bibinfo{journal}{Journal of Computational Physics}
  \bibinfo{volume}{179}, \bibinfo{pages}{371--399}.
\bibitem[{Erichson et~al.(2019)Erichson, Mathelin, Kutz and
  Brunton}]{erichson2019randomized}
\bibinfo{author}{Erichson, N.B.}, \bibinfo{author}{Mathelin, L.},
  \bibinfo{author}{Kutz, J.N.}, \bibinfo{author}{Brunton, S.L.},
  \bibinfo{year}{2019}.
\newblock \bibinfo{title}{Randomized dynamic mode decomposition}.
\newblock \bibinfo{journal}{SIAM Journal on Applied Dynamical Systems}
  \bibinfo{volume}{18}, \bibinfo{pages}{1867--1891}.
\bibitem[{Everson and Sirovich(1995)}]{everson1995karhunen}
\bibinfo{author}{Everson, R.}, \bibinfo{author}{Sirovich, L.},
  \bibinfo{year}{1995}.
\newblock \bibinfo{title}{Karhunen--{L}oeve procedure for gappy data}.
\newblock \bibinfo{journal}{JOSA A} \bibinfo{volume}{12},
  \bibinfo{pages}{1657--1664}.
\bibitem[{Farhat et~al.(2014)Farhat, Avery and Chapman}]{farhat2014dimensional}
\bibinfo{author}{Farhat, C.}, \bibinfo{author}{Avery, P.},
  \bibinfo{author}{Chapman, T.and~Cortial, J.}, \bibinfo{year}{2014}.
\newblock \bibinfo{title}{Dimensional reduction of nonlinear finite element
  dynamic models with finite rotations and energy-based mesh sampling and
  weighting for computational efficiency}.
\newblock \bibinfo{journal}{International Journal for Numerical Methods in
  Engineering} \bibinfo{volume}{98}, \bibinfo{pages}{625--662}.
\bibitem[{Farhat et~al.(2015)Farhat, Chapman and Avery}]{farhat2015structure}
\bibinfo{author}{Farhat, C.}, \bibinfo{author}{Chapman, T.},
  \bibinfo{author}{Avery, P.}, \bibinfo{year}{2015}.
\newblock \bibinfo{title}{Structure-preserving, stability, and accuracy
  properties of the energy-conserving sampling and weighting method for the
  hyper reduction of nonlinear finite element dynamic models}.
\newblock \bibinfo{journal}{International Journal for Numerical Methods in
  Engineering} \bibinfo{volume}{102}, \bibinfo{pages}{1077--1110}.
\bibitem[{Gaertner and Lackner(2015)}]{gaertner2015modeling}
\bibinfo{author}{Gaertner, E.M.}, \bibinfo{author}{Lackner, M.A.},
  \bibinfo{year}{2015}.
\newblock \bibinfo{title}{Modeling dynamic stall for a free vortex wake model}.
\newblock \bibinfo{journal}{Wind Engineering} \bibinfo{volume}{39},
  \bibinfo{pages}{675--691}.
\bibitem[{Gnedin(2019)}]{gnedin2019hierarchical}
\bibinfo{author}{Gnedin, N.Y.}, \bibinfo{year}{2019}.
\newblock \bibinfo{title}{Hierarchical particle mesh: An {FFT}-accelerated fast
  multipole method}.
\newblock \bibinfo{journal}{The Astrophysical Journal Supplement Series}
  \bibinfo{volume}{243}, \bibinfo{pages}{19}.
\bibitem[{Greengard and Rokhlin(1987)}]{greengard1987fast}
\bibinfo{author}{Greengard, L.}, \bibinfo{author}{Rokhlin, V.},
  \bibinfo{year}{1987}.
\newblock \bibinfo{title}{A fast algorithm for particle simulations}.
\newblock \bibinfo{journal}{Journal of Computational Physics}
  \bibinfo{volume}{73}, \bibinfo{pages}{325--348}.
\bibitem[{Grepl et~al.(2007)Grepl, Maday, Nguyen and
  Patera}]{grepl2007efficient}
\bibinfo{author}{Grepl, M.A.}, \bibinfo{author}{Maday, Y.},
  \bibinfo{author}{Nguyen, N.C.}, \bibinfo{author}{Patera, A.T.},
  \bibinfo{year}{2007}.
\newblock \bibinfo{title}{Efficient reduced-basis treatment of nonaffine and
  nonlinear partial differential equations}.
\newblock \bibinfo{journal}{ESAIM: Mathematical Modelling and Numerical
  Analysis} \bibinfo{volume}{41}, \bibinfo{pages}{575--605}.
\bibitem[{Guo and Hesthaven(2019)}]{guo2019data}
\bibinfo{author}{Guo, M.}, \bibinfo{author}{Hesthaven, J.S.},
  \bibinfo{year}{2019}.
\newblock \bibinfo{title}{Data-driven reduced order modeling for time-dependent
  problems}.
\newblock \bibinfo{journal}{Computer methods in applied mechanics and
  engineering} \bibinfo{volume}{345}, \bibinfo{pages}{75--99}.
\bibitem[{Hansson et~al.(2002)Hansson, Oostenbrink and van
  Gunsteren}]{hansson2002moleculardynamics}
\bibinfo{author}{Hansson, T.}, \bibinfo{author}{Oostenbrink, C.},
  \bibinfo{author}{van Gunsteren, W.F.}, \bibinfo{year}{2002}.
\newblock \bibinfo{title}{Molecular dynamics simulations}.
\newblock \bibinfo{journal}{Current opinion in structural biology}
  \bibinfo{volume}{12}, \bibinfo{pages}{190--196}.
\bibitem[{Holmes et~al.(2012)Holmes, Lumley, Berkooz and
  Rowley}]{holmes2012turbulence}
\bibinfo{author}{Holmes, P.}, \bibinfo{author}{Lumley, J.L.},
  \bibinfo{author}{Berkooz, G.}, \bibinfo{author}{Rowley, C.W.},
  \bibinfo{year}{2012}.
\newblock \bibinfo{title}{Turbulence, coherent structures, dynamical systems
  and symmetry}.
\newblock \bibinfo{publisher}{Cambridge university press}.
\bibitem[{Jeon et~al.(2014)Jeon, Lee and Lee}]{jeon2014unsteady}
\bibinfo{author}{Jeon, M.}, \bibinfo{author}{Lee, S.}, \bibinfo{author}{Lee,
  S.}, \bibinfo{year}{2014}.
\newblock \bibinfo{title}{Unsteady aerodynamics of offshore floating wind
  turbines in platform pitching motion using vortex lattice method}.
\newblock \bibinfo{journal}{Renewable Energy} \bibinfo{volume}{65},
  \bibinfo{pages}{207--212}.
\bibitem[{Jiang et~al.(2016)Jiang, Li, Zhao, Qin, Karpeev, Hernandez-Ortiz,
  de~Pablo and Heinonen}]{jiang2016n}
\bibinfo{author}{Jiang, X.}, \bibinfo{author}{Li, J.}, \bibinfo{author}{Zhao,
  X.}, \bibinfo{author}{Qin, J.}, \bibinfo{author}{Karpeev, D.},
  \bibinfo{author}{Hernandez-Ortiz, J.}, \bibinfo{author}{de~Pablo, J.J.},
  \bibinfo{author}{Heinonen, O.}, \bibinfo{year}{2016}.
\newblock \bibinfo{title}{An $o(n)$ and parallel approach to integral problems
  by a kernel-independent fast multipole method: Application to polarization
  and magnetization of interacting particles}.
\newblock \bibinfo{journal}{The Journal of Chemical Physics}
  \bibinfo{volume}{145}, \bibinfo{pages}{064307}.
\bibitem[{Jing and Stephansson(2007)}]{jing2007DEMfundamentals}
\bibinfo{author}{Jing, L.}, \bibinfo{author}{Stephansson, O.},
  \bibinfo{year}{2007}.
\newblock \bibinfo{title}{Fundamentals of discrete element methods for rock
  engineering: theory and applications}. volume~\bibinfo{volume}{85}.
\newblock \bibinfo{publisher}{Elsevier}.
\bibitem[{Kebbie-Anthony et~al.(2018)Kebbie-Anthony, Gumerov, Preidikman,
  Balachandran and Azarm}]{kebbie2018fast}
\bibinfo{author}{Kebbie-Anthony, A.B.}, \bibinfo{author}{Gumerov, N.},
  \bibinfo{author}{Preidikman, S.}, \bibinfo{author}{Balachandran, B.},
  \bibinfo{author}{Azarm, S.}, \bibinfo{year}{2018}.
\newblock \bibinfo{title}{Fast multipole method for nonlinear, unsteady
  aerodynamic simulations}, in: \bibinfo{booktitle}{2018 AIAA Modeling and
  Simulation Technologies Conference}, p. \bibinfo{pages}{1929}.
\bibitem[{Kutz et~al.(2016)Kutz, Brunton, Brunton and
  Proctor}]{kutz2016dynamic}
\bibinfo{author}{Kutz, J.N.}, \bibinfo{author}{Brunton, S.L.},
  \bibinfo{author}{Brunton, B.W.}, \bibinfo{author}{Proctor, J.L.},
  \bibinfo{year}{2016}.
\newblock \bibinfo{title}{Dynamic mode decomposition: data-driven modeling of
  complex systems}. volume \bibinfo{volume}{149}.
\newblock \bibinfo{publisher}{Siam}.
\bibitem[{Lee and Carlberg(2020)}]{lee2020model}
\bibinfo{author}{Lee, K.}, \bibinfo{author}{Carlberg, K.T.},
  \bibinfo{year}{2020}.
\newblock \bibinfo{title}{Model reduction of dynamical systems on nonlinear
  manifolds using deep convolutional autoencoders}.
\newblock \bibinfo{journal}{Journal of Computational Physics}
  \bibinfo{volume}{404}, \bibinfo{pages}{108973}.
\bibitem[{Leishman(2006)}]{leishman2006principles}
\bibinfo{author}{Leishman, G.J.}, \bibinfo{year}{2006}.
\newblock \bibinfo{title}{Principles of helicopter aerodynamics with CD extra}.
\newblock \bibinfo{publisher}{Cambridge university press}.
\bibitem[{Liu and Liu(2010)}]{liu2010smoothed}
\bibinfo{author}{Liu, M.B.}, \bibinfo{author}{Liu, G.R.}, \bibinfo{year}{2010}.
\newblock \bibinfo{title}{Smoothed particle hydrodynamics ({SPH}): an overview
  and recent developments}.
\newblock \bibinfo{journal}{Archives of computational methods in engineering}
  \bibinfo{volume}{17}, \bibinfo{pages}{25--76}.
\bibitem[{Martinsson(2015)}]{Martinsson2015}
\bibinfo{author}{Martinsson, P.G.}, \bibinfo{year}{2015}.
\newblock \bibinfo{title}{Fast Multipole Methods}. \bibinfo{publisher}{Springer
  Berlin Heidelberg}, \bibinfo{address}{Berlin, Heidelberg}.
\newblock pp. \bibinfo{pages}{498--508}.
\newblock \URLprefix \url{https://doi.org/10.1007/978-3-540-70529-1_448},
  \DOIprefix\doi{10.1007/978-3-540-70529-1_448}.
\bibitem[{Martinsson and Rokhlin(2007)}]{martinsson2007accelerated}
\bibinfo{author}{Martinsson, P.G.}, \bibinfo{author}{Rokhlin, V.},
  \bibinfo{year}{2007}.
\newblock \bibinfo{title}{An accelerated kernel-independent fast multipole
  method in one dimension}.
\newblock \bibinfo{journal}{SIAM Journal on Scientific Computing}
  \bibinfo{volume}{29}, \bibinfo{pages}{1160--1178}.
\bibitem[{Mocz and Succi(2015)}]{mocz2015numerical}
\bibinfo{author}{Mocz, P.}, \bibinfo{author}{Succi, S.}, \bibinfo{year}{2015}.
\newblock \bibinfo{title}{Numerical solution of the nonlinear schr{\"o}dinger
  equation using smoothed-particle hydrodynamics}.
\newblock \bibinfo{journal}{Physical Review E} \bibinfo{volume}{91},
  \bibinfo{pages}{053304}.
\bibitem[{Noack et~al.(2016)Noack, Stankiewicz, Morzynski and
  Schmid}]{noack2016recursive}
\bibinfo{author}{Noack, B.R.}, \bibinfo{author}{Stankiewicz, W.},
  \bibinfo{author}{Morzynski, M.}, \bibinfo{author}{Schmid, P.J.},
  \bibinfo{year}{2016}.
\newblock \bibinfo{title}{Recursive dynamic mode decomposition of transient and
  post-transient wake flows}.
\newblock \bibinfo{journal}{Journal of Fluid Mechanics} \bibinfo{volume}{809},
  \bibinfo{pages}{843}.
\bibitem[{Nocedal and Wright(2006)}]{nocedal2006numerical}
\bibinfo{author}{Nocedal, J.}, \bibinfo{author}{Wright, S.},
  \bibinfo{year}{2006}.
\newblock \bibinfo{title}{Numerical optimization}.
\newblock \bibinfo{publisher}{Springer Science \& Business Media}.
\bibitem[{No{\'e} and Clementi(2015)}]{noe2015kinetic}
\bibinfo{author}{No{\'e}, F.}, \bibinfo{author}{Clementi, C.},
  \bibinfo{year}{2015}.
\newblock \bibinfo{title}{Kinetic distance and kinetic maps from molecular
  dynamics simulation}.
\newblock \bibinfo{journal}{Journal of Chemical Theory and Computation}
  \bibinfo{volume}{11}, \bibinfo{pages}{5002--5011}.
\bibitem[{Parish and Carlberg(2020)}]{parish2020time}
\bibinfo{author}{Parish, E.J.}, \bibinfo{author}{Carlberg, K.T.},
  \bibinfo{year}{2020}.
\newblock \bibinfo{title}{Time-series machine-learning error models for
  approximate solutions to parameterized dynamical systems}.
\newblock \bibinfo{journal}{Computer Methods in Applied Mechanics and
  Engineering} \bibinfo{volume}{365}, \bibinfo{pages}{112990}.
\bibitem[{Peng and Mohseni(2016)}]{peng2016symplectic}
\bibinfo{author}{Peng, L.}, \bibinfo{author}{Mohseni, K.},
  \bibinfo{year}{2016}.
\newblock \bibinfo{title}{Symplectic model reduction of {H}amiltonian systems}.
\newblock \bibinfo{journal}{SIAM Journal on Scientific Computing}
  \bibinfo{volume}{38}, \bibinfo{pages}{A1--A27}.
\bibitem[{Pfalzner and Gibbon(2005)}]{pfalzner2005many}
\bibinfo{author}{Pfalzner, S.}, \bibinfo{author}{Gibbon, P.},
  \bibinfo{year}{2005}.
\newblock \bibinfo{title}{Many-body tree methods in physics}.
\newblock \bibinfo{publisher}{Cambridge University Press}.
\bibitem[{Pfrommer et~al.(2006)Pfrommer, Springel, En{\ss}lin and
  Jubelgas}]{pfrommer2006detecting}
\bibinfo{author}{Pfrommer, C.}, \bibinfo{author}{Springel, V.},
  \bibinfo{author}{En{\ss}lin, T.A.}, \bibinfo{author}{Jubelgas, M.},
  \bibinfo{year}{2006}.
\newblock \bibinfo{title}{Detecting shock waves in cosmological smoothed
  particle hydrodynamics simulations}.
\newblock \bibinfo{journal}{Monthly Notices of the Royal Astronomical Society}
  \bibinfo{volume}{367}, \bibinfo{pages}{113--131}.
\bibitem[{Rodriguez(2018)}]{rodriguez2018phd}
\bibinfo{author}{Rodriguez, S.N.}, \bibinfo{year}{2018}.
\newblock \bibinfo{title}{Stability and Dynamic Properties of Tip Vortices Shed
  from Flexible Rotors of Floating Offshore Wind Turbines}.
\newblock Ph.D. thesis. Lehigh University.
\bibitem[{Rodriguez et~al.(2020a)Rodriguez, Iliopoulos, Michopoulos and
  Jaworski}]{rodriguez2020investigating}
\bibinfo{author}{Rodriguez, S.N.}, \bibinfo{author}{Iliopoulos, A.P.},
  \bibinfo{author}{Michopoulos, J.G.}, \bibinfo{author}{Jaworski, J.W.},
  \bibinfo{year}{2020}a.
\newblock \bibinfo{title}{Investigating the coupled effects between rotor-blade
  aeroelasticity and tip vortex stability}, in:
  \bibinfo{booktitle}{International Design Engineering Technical Conferences
  and Computers and Information in Engineering Conference},
  \bibinfo{organization}{American Society of Mechanical Engineers}. p.
  \bibinfo{pages}{V009T09A001}.
\bibitem[{Rodriguez and Jaworski(2017)}]{rodriguez2017jert}
\bibinfo{author}{Rodriguez, S.N.}, \bibinfo{author}{Jaworski, J.W.},
  \bibinfo{year}{2017}.
\newblock \bibinfo{title}{Toward identifying aeroelastic mechanisms in
  near-wake instabilities of floating offshore wind turbines}.
\newblock \bibinfo{journal}{Journal of Energy Resources Technology, Special
  Issue: Wind Energy} \bibinfo{volume}{139}, \bibinfo{pages}{051203}.
\bibitem[{Rodriguez and Jaworski(2019)}]{rodriguez2019strongly}
\bibinfo{author}{Rodriguez, S.N.}, \bibinfo{author}{Jaworski, J.W.},
  \bibinfo{year}{2019}.
\newblock \bibinfo{title}{Strongly-coupled aeroelastic free-vortex wake
  framework for floating offshore wind turbine rotors. part 1: Numerical
  framework}.
\newblock \bibinfo{journal}{Renewable Energy} \bibinfo{volume}{141},
  \bibinfo{pages}{1127--1145}.
\bibitem[{Rodriguez and Jaworski(2020)}]{rodriguez2020strongly}
\bibinfo{author}{Rodriguez, S.N.}, \bibinfo{author}{Jaworski, J.W.},
  \bibinfo{year}{2020}.
\newblock \bibinfo{title}{Strongly-coupled aeroelastic free-vortex wake
  framework for floating offshore wind turbine rotors. part 2: Application}.
\newblock \bibinfo{journal}{Renewable Energy} \bibinfo{volume}{149},
  \bibinfo{pages}{1018--1031}.
\bibitem[{Rodriguez et~al.(2020b)Rodriguez, Jaworski and
  Michopoulos}]{rodriguez2020stability}
\bibinfo{author}{Rodriguez, S.N.}, \bibinfo{author}{Jaworski, J.W.},
  \bibinfo{author}{Michopoulos, J.G.}, \bibinfo{year}{2020}b.
\newblock \bibinfo{title}{Stability of helical vortex structures shed from
  flexible rotors}.
\newblock \bibinfo{journal}{arXiv preprint arXiv:2008.08969} .
\bibitem[{Rowley(2005)}]{rowley2005model}
\bibinfo{author}{Rowley, C.W.}, \bibinfo{year}{2005}.
\newblock \bibinfo{title}{Model reduction for fluids, using balanced proper
  orthogonal decomposition}.
\newblock \bibinfo{journal}{International Journal of Bifurcation and Chaos}
  \bibinfo{volume}{15}, \bibinfo{pages}{997--1013}.
\bibitem[{Rowley et~al.(2004)Rowley, Colonius and Murray}]{rowley2004model}
\bibinfo{author}{Rowley, C.W.}, \bibinfo{author}{Colonius, T.},
  \bibinfo{author}{Murray, R.M.}, \bibinfo{year}{2004}.
\newblock \bibinfo{title}{Model reduction for compressible flows using {POD}
  and {G}alerkin projection}.
\newblock \bibinfo{journal}{Physica D: Nonlinear Phenomena}
  \bibinfo{volume}{189}, \bibinfo{pages}{115--129}.
\bibitem[{Russell et~al.(2018)Russell, Souto-Iglesias and
  Zohdi}]{russell2018numerical}
\bibinfo{author}{Russell, M.A.}, \bibinfo{author}{Souto-Iglesias, A.},
  \bibinfo{author}{Zohdi, T.}, \bibinfo{year}{2018}.
\newblock \bibinfo{title}{Numerical simulation of laser fusion additive
  manufacturing processes using the sph method}.
\newblock \bibinfo{journal}{Computer Methods in Applied Mechanics and
  Engineering} \bibinfo{volume}{341}, \bibinfo{pages}{163--187}.
\bibitem[{Schmid(2010)}]{schmid2010dynamic}
\bibinfo{author}{Schmid, P.J.}, \bibinfo{year}{2010}.
\newblock \bibinfo{title}{Dynamic mode decomposition of numerical and
  experimental data}.
\newblock \bibinfo{journal}{Journal of fluid mechanics} \bibinfo{volume}{656},
  \bibinfo{pages}{5--28}.
\bibitem[{Sebastian and Lackner(2012)}]{sebastian2012development}
\bibinfo{author}{Sebastian, T.}, \bibinfo{author}{Lackner, M.},
  \bibinfo{year}{2012}.
\newblock \bibinfo{title}{Development of a free vortex wake method code for
  offshore floating wind turbines}.
\newblock \bibinfo{journal}{Renewable Energy} \bibinfo{volume}{46},
  \bibinfo{pages}{269--275}.
\bibitem[{Shadloo et~al.(2016)Shadloo, Oger and
  Le~Touz{\'e}}]{shadloo2016smoothed}
\bibinfo{author}{Shadloo, M.S.}, \bibinfo{author}{Oger, G.},
  \bibinfo{author}{Le~Touz{\'e}, D.}, \bibinfo{year}{2016}.
\newblock \bibinfo{title}{Smoothed particle hydrodynamics method for fluid
  flows, towards industrial applications: Motivations, current state, and
  challenges}.
\newblock \bibinfo{journal}{Computers \& Fluids} \bibinfo{volume}{136},
  \bibinfo{pages}{11--34}.
\bibitem[{Sirovich(1987)}]{sirovich1987turbulence}
\bibinfo{author}{Sirovich, L.}, \bibinfo{year}{1987}.
\newblock \bibinfo{title}{Turbulence and the dynamics of coherent structures.
  i. coherent structures}.
\newblock \bibinfo{journal}{Quarterly of applied mathematics}
  \bibinfo{volume}{45}, \bibinfo{pages}{561--571}.
\bibitem[{Steuben et~al.(2016)Steuben, Iliopoulos and
  Michopoulos}]{steuben2016discrete}
\bibinfo{author}{Steuben, J.C.}, \bibinfo{author}{Iliopoulos, A.P.},
  \bibinfo{author}{Michopoulos, J.G.}, \bibinfo{year}{2016}.
\newblock \bibinfo{title}{Discrete element modeling of particle-based additive
  manufacturing processes}.
\newblock \bibinfo{journal}{Computer Methods in Applied Mechanics and
  Engineering} \bibinfo{volume}{305}, \bibinfo{pages}{537--561}.
\bibitem[{Taira et~al.(2017)Taira, Brunton, Dawson, Rowley, Colonius, McKeon,
  Schmidt, Gordeyev, Theofilis and Ukeiley}]{taira2017modal}
\bibinfo{author}{Taira, K.}, \bibinfo{author}{Brunton, S.L.},
  \bibinfo{author}{Dawson, S.T.M.}, \bibinfo{author}{Rowley, C.W.},
  \bibinfo{author}{Colonius, T.}, \bibinfo{author}{McKeon, B.J.},
  \bibinfo{author}{Schmidt, O.T.}, \bibinfo{author}{Gordeyev, S.},
  \bibinfo{author}{Theofilis, V.}, \bibinfo{author}{Ukeiley, L.S.},
  \bibinfo{year}{2017}.
\newblock \bibinfo{title}{Modal analysis of fluid flows: {A}n overview}.
\newblock \bibinfo{journal}{Aiaa Journal} \bibinfo{volume}{55},
  \bibinfo{pages}{4013--4041}.
\bibitem[{Tartakovsky and Meakin(2005)}]{tartakovsky2005smoothed}
\bibinfo{author}{Tartakovsky, A.M.}, \bibinfo{author}{Meakin, P.},
  \bibinfo{year}{2005}.
\newblock \bibinfo{title}{A smoothed particle hydrodynamics model for miscible
  flow in three-dimensional fractures and the two-dimensional rayleigh--taylor
  instability}.
\newblock \bibinfo{journal}{Journal of Computational Physics}
  \bibinfo{volume}{207}, \bibinfo{pages}{610--624}.
\bibitem[{Tiso and Rixen(2013)}]{tiso2013discrete}
\bibinfo{author}{Tiso, P.}, \bibinfo{author}{Rixen, D.J.},
  \bibinfo{year}{2013}.
\newblock \bibinfo{title}{Discrete empirical interpolation method for finite
  element structural dynamics}, in: \bibinfo{booktitle}{Topics in Nonlinear
  Dynamics, Volume 1}. \bibinfo{publisher}{Springer}, pp.
  \bibinfo{pages}{203--212}.
\bibitem[{Tu et~al.(2013)Tu, Rowley, Luchtenburg, Brunton and
  Kutz}]{tu2013dynamic}
\bibinfo{author}{Tu, J.H.}, \bibinfo{author}{Rowley, C.W.},
  \bibinfo{author}{Luchtenburg, D.M.}, \bibinfo{author}{Brunton, S.L.},
  \bibinfo{author}{Kutz, J.N.}, \bibinfo{year}{2013}.
\newblock \bibinfo{title}{On dynamic mode decomposition: theory and
  applications}.
\newblock \bibinfo{journal}{Journal of Nonlinear Science} \bibinfo{volume}{22},
  \bibinfo{pages}{887--915}.
\bibitem[{Wang et~al.(2019)Wang, Olsson, Wehmeyer, P{\'e}rez, Charron,
  De~Fabritiis, No{\'e} and Clementi}]{wang2019machine}
\bibinfo{author}{Wang, J.}, \bibinfo{author}{Olsson, S.},
  \bibinfo{author}{Wehmeyer, C.}, \bibinfo{author}{P{\'e}rez, A.},
  \bibinfo{author}{Charron, N.E.}, \bibinfo{author}{De~Fabritiis, G.},
  \bibinfo{author}{No{\'e}, F.}, \bibinfo{author}{Clementi, C.},
  \bibinfo{year}{2019}.
\newblock \bibinfo{title}{Machine learning of coarse-grained molecular dynamics
  force fields}.
\newblock \bibinfo{journal}{ACS central science} \bibinfo{volume}{5},
  \bibinfo{pages}{755--767}.
\bibitem[{Willcox and Peraire(2002)}]{willcox2002balanced}
\bibinfo{author}{Willcox, K.}, \bibinfo{author}{Peraire, J.},
  \bibinfo{year}{2002}.
\newblock \bibinfo{title}{Balanced model reduction via the proper orthogonal
  decomposition}.
\newblock \bibinfo{journal}{AIAA journal} \bibinfo{volume}{40},
  \bibinfo{pages}{2323--2330}.
\bibitem[{Ying(2006)}]{ying2006kernel}
\bibinfo{author}{Ying, L.}, \bibinfo{year}{2006}.
\newblock \bibinfo{title}{A kernel independent fast multipole algorithm for
  radial basis functions}.
\newblock \bibinfo{journal}{Journal of Computational Physics}
  \bibinfo{volume}{213}, \bibinfo{pages}{451--457}.
\bibitem[{Ying et~al.(2004)Ying, Biros and Zorin}]{ying2004kernel}
\bibinfo{author}{Ying, L.}, \bibinfo{author}{Biros, G.},
  \bibinfo{author}{Zorin, D.}, \bibinfo{year}{2004}.
\newblock \bibinfo{title}{A kernel-independent adaptive fast multipole
  algorithm in two and three dimensions}.
\newblock \bibinfo{journal}{Journal of Computational Physics}
  \bibinfo{volume}{196}, \bibinfo{pages}{591--626}.

\end{thebibliography}

\end{document}